\documentclass[a4paper,11pt]{article}%
\usepackage{jheppub}
\usepackage{amsmath}
\usepackage{amsfonts}
\usepackage{amssymb}
\usepackage{graphicx}%
\providecommand{\U}[1]{\protect\rule{.1in}{.1in}}
\pdfoutput=1
\abstract{It is well known that high-energy scattering of a meson from some hadronic target can be described by the 
interaction of that target  with a color dipole formed by two Wilson lines corresponding to fast quark-antiquark pair.
Moreover,  the energy dependence of the scattering amplitude is governed by the evolution equation of this color dipole with respect to rapidity.
Similarly, the energy dependence of scattering of a baryon can be described in terms of evolution of a three-Wilson-lines operator 
with respect to the rapidity of the Wilson lines. We calculate the evolution of the 3-quark Wilson loop operator  in the next-to-leading order
(NLO) and present a quasi-conformal evolution equation for a composite 3-Wilson-lines operator. We also obtain the linearized version 
of that evolution equation describing the amplitude of the odderon exchange at high energies.}
\arxivnumber{}
\affiliation{$^{*}$ Physics Dept., Old Dominion University, Norfolk VA 23529
and Theory Group, JLAB, 12000 Jefferson Ave, Newport News, VA 23606}
\affiliation{$^{\dag}$ Budker Institute of Nuclear Physics and Novosibirsk State University,  630090 Novosibirsk, Russia}
\emailAdd{balitsky@jlab.org}
\emailAdd{A.V.Grabovsky@inp.nsk.su}
\begin{document}

\title{\boldmath NLO evolution of 3-quark Wilson loop operator}
\author{I. Balitsky$^{*}$ and A. V. Grabovsky$^{\dag}$}
\maketitle

\subheader{BUDKER INP-2013-25}

\flushbottom

\section{Introduction}

The description of the proton scattering in the framework of $k_{T}%
$-factorization can be addressed within the high energy operator expansion
developed in \cite{Balitsky96}. In that paper this method was applied to
derive the full leading order (LO) hierarchy of the low-x evolution equations
for Wilson lines with arbitrary indices and to the most important case of the
color dipole. In the next to leading order (NLO) the evolution equation for
the color dipole was derived in \cite{Balitsky:2006wa},
\cite{Balitsky:2008zza}, \cite{Balitsky:2009xg} and the connected evolution of
the 3 Wilson lines was found in \cite{Grabovsky:2013mba}. Finally the full NLO
hierarchy of the low-x evolution equations was written in
\cite{Balitsky:2013fea} and the JIMWLK hamiltonian equivalent to it in
\cite{Kovner:2013ona}.

In this framework the amplitude in the Regge limit can be written as a
convolution of the impact factors and the matrix elements of the Wilson line
operators. The impact factors consist of the wavefunctions of the incoming and
outgoing particles, which describe their splitting into the quarks and gluons
propagating through the shockwave formed by the other particle. It is well
known that the propagation of the fast particle is described by a Wilson line
- infinite gauge link ordered along the classical trajectory of the fast
particle. For the virtual photon or meson scattering the relevant
two-Wilson-lines operator is a color dipole. In the proton case assuming SU(3)
symmetry it is the baryon or 3-quark Wilson loop (3QWL) $\varepsilon
^{i^{\prime}j^{\prime}h^{\prime}}\varepsilon_{ijh}U_{1i^{\prime}}%
^{i}U_{2j^{\prime}}^{j}U_{3h^{\prime}}^{h}$. Its leading order linear
evolution equation was studied in the C-odd case within the JIMWLK formalism
and proved equivalent to the C-odd BKP equation \cite{Bartels:1980pe}%
-\cite{Kwiecinski:1980wb} in \cite{Hatta:2005as} and its nonlinear evolution
equation was derived within Wilson line approach \cite{Balitsky96} in
\cite{Gerasimov:2012bj}. The connected contribution to the NLO kernel of the
equation was calculated in \cite{Grabovsky:2013mba}. In the momentum
representation the evolution of this operator was first studied in
\cite{Praszalowicz:1997nf}, and the nonlinear equation was worked out in
\cite{Bartels:2007aa}. In the C-odd case the linear NLO evolution equation for
the odderon Green function was obtained in \cite{Bartels:2012sw}.

Here the NLO evolution equation for the 3QWL operator is presented. Then as in
\cite{Balitsky:2008zza}, we construct the composite 3QWL operator obeying the
quasi-conformal evolution equation and after that give its linearized kernel
in the 3-gluon approximation. In this approximation we also linearize the BK
equation and show that it contains the nondipole 3QWL operators.

After completion of this paper we were informed about JIMWLK calculation of
the NLO evolution of 3-Wilson-line operator \cite{Kovner:2014lca}. Both
evolution kernels reproduce NLO BK in the dipole limit $\vec{r}_{1}%
\rightarrow\vec{r}_{2}$ and survive other checks but the detailed comparison
of these two results is beyond the scope of present paper. We also wish to
compare our results with the results of S. Caron-Huot \cite{Simon}.

The paper is organized as follows. In Section 2 we remind the general logic of
high-energy OPE. Sections 3 and 4 present the derivation of the NLO kernel for
3QWL operator and Section 5 describes the calculation of the quasi-conformal
kernel for the composite 3QWL operator. The linearized kernel is given in
Section 6. The main results of the paper are listed in Sect. 7 and conclusions
in Sect. 8. Appendices comprise the necessary technical details.

\section{Rapidity factorization and evolution of Wilson lines}

Let us consider the proton scattering off a hadron target like another proton
or nucleus. First, we assume that due to saturation the characteristic
transverse momenta of exchanged and produced gluons are relatively high
($Q_{s}\sim2-3$ GeV for $pA$ scattering at LHC) so the application of
perturbation theory is justified. Alternatively, one may think about
high-energy scattering of a charmed baryon.

If pQCD is applicable, in accordance with general logic of high-energy OPE we
factorize all amplitudes in rapidity. First, we integrate over gluons with
rapidity $Y$ close to that of the projectile proton $Y_{p}$. To this end we
introduce the rapidity divide $\eta\leq Y_{p}$ which separates the
\textquotedblleft fast\textquotedblright\ gluons from the \textquotedblleft
slow\textquotedblright\ ones.

\begin{figure}[tbh]
\includegraphics[width=44mm]{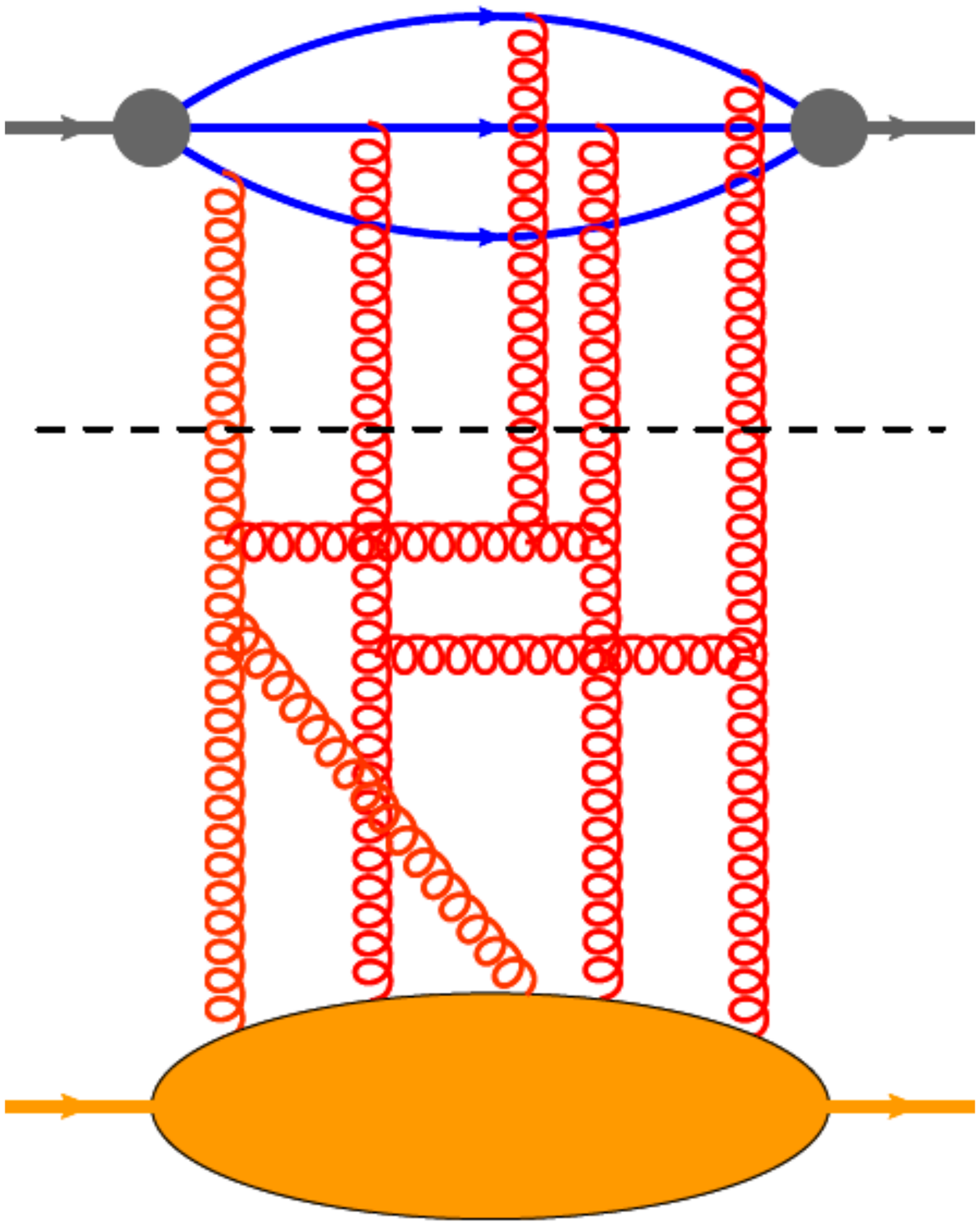} \hspace{1cm}
\includegraphics[width=22mm]{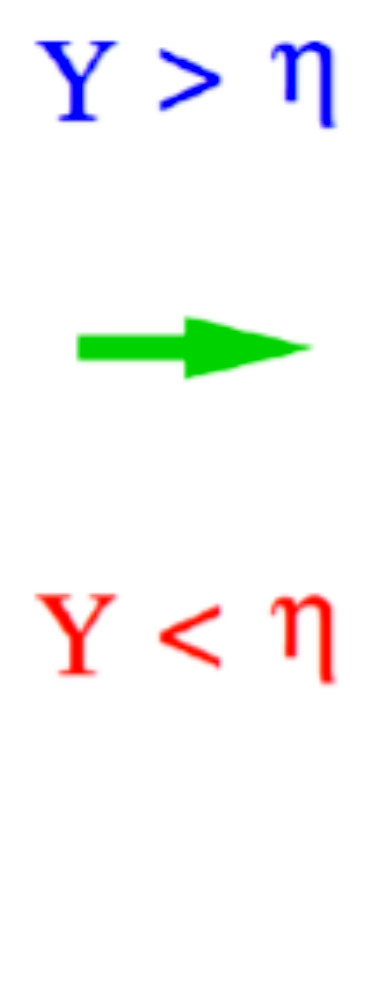} \hspace{1cm}
\includegraphics[width=44mm]{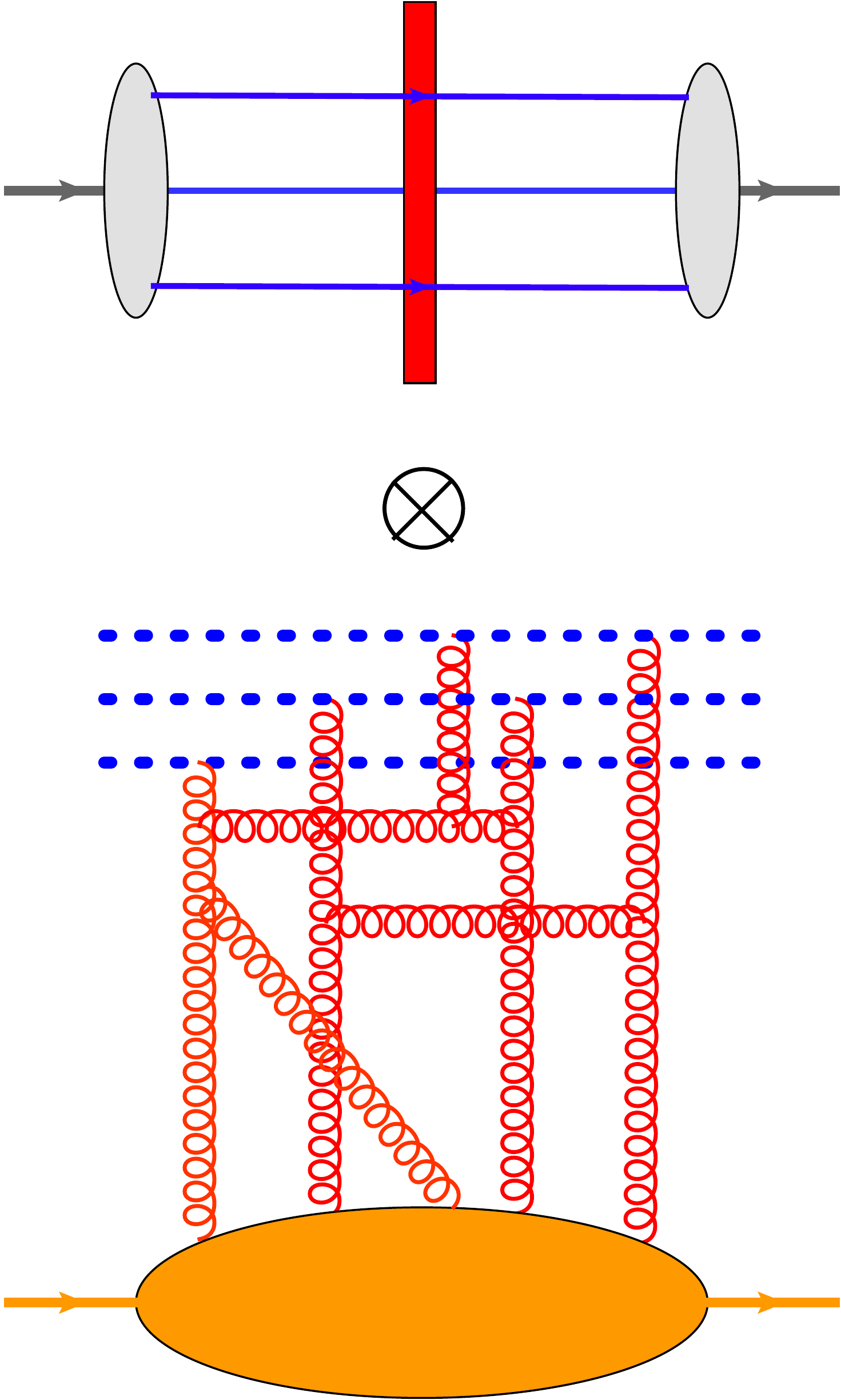}\caption{Rapidity factorization. The
impact factors with $Y>\eta$ are given by diagrams in the shock-wave
background. Wilson-line operators with $Y<\eta$ are denoted by dotted lines.}%
\label{fig:rapidityfac}%
\end{figure}

It is convenient to use the background field formalism: we integrate over
gluons with $\alpha>\sigma=e^{\eta}$ and leave gluons with $\alpha<\sigma$ as
a background field, to be integrated over later. Since the rapidities of the
background gluons are very different from the rapidities of gluons in our
Feynman diagrams, the background field can be taken in the form of a shock
wave due to the Lorentz contraction. To derive the expression of a quark (or
gluon) propagator in this shock-wave background we represent the propagator as
a path integral over various trajectories, each of them weighed with the gauge
factor Pexp$(ig\int\! dx_{\mu}A^{\mu})$ ordered along the propagation path.
Now, since the shock wave is very thin, quarks (or gluons) do not have time to
deviate in transverse direction so their trajectory inside the shock wave can
be approximated by a segment of the straight line. Moreover, since there is no
external field outside the shock wave, the integral over the segment of
straight line can be formally extended to $\pm\infty$ limits yielding the
Wilson-line gauge factor
\begin{align}
& \hspace{-0mm} U^{\eta}_{x}~=~\mathrm{Pexp}\Big[ig\!\int_{-\infty}^{\infty
}\!\! du ~p_{1}^{\mu}A^{\sigma}_{\mu}(up_{1}+x_{\perp})\Big],\nonumber\\
& \hspace{-0mm} A^{\eta}_{\mu}(x)~=~\int\!d^{4} k ~\theta(e^{\eta}-|\alpha
_{k}|)e^{ik\cdot x} A_{\mu}(k)\label{cutoff}%
\end{align}
\begin{equation}
U\left(  \vec{r},\eta\right)  =Pe^{ig\int_{-\infty}^{+\infty}A_{\eta}%
^{-}(r^{+},\vec{r})dr^{+}}, \label{WL}%
\end{equation}
where $A_{\eta}^{-}$ is the external shock wave field built from only slow
gluons
\begin{equation}
A_{\eta}^{-}=\int\frac{d^{4}p}{\left(  2\pi\right)  ^{4}}e^{-ipz}A^{-}\left(
p\right)  \theta(e^{\eta}-p^{+}).
\end{equation}
(Our light-cone conventions are listed in the Appendix A). The propagation of
a quark (or gluon) in the shock-wave background is then described by free
propagation to a point of interaction with the shock wave, Wilson line $U$ at
the point of interaction, and free propagation to the final point.

Thus, the result of the integration over rapidities $Y>\eta$ gives the proton
impact factor proportional to product of two proton wavefunctions integrated
over longitudinal momenta. This impact factor is multiplied by a ``color
tripole'' - 3QWL operator made of three (light-like) Wilson lines with
rapidities up to $\eta$:
\begin{equation}
B_{123}\equiv\varepsilon^{i^{\prime}j^{\prime}h^{\prime}}\varepsilon
_{ijh}U_{1i^{\prime}}^{i}U_{2j^{\prime}}^{j}U_{3h^{\prime}}^{h}\equiv
U_{1}\cdot U_{2}\cdot U_{3},\label{tripole}%
\end{equation}
where $U_{i}\equiv U(\vec{r}_{i},\eta)$. (As discussed in Refs.
\cite{Balitsky96,Balitsky:2013fea}, these Wilson lines should be connected by
appropriate gauge links at infinity making the operator (\ref{tripole}) and
similar many-Wilson-lines operators below gauge invariant.) It should be noted
that the proton impact factor is non-perturbative, so at this point it can be
calculated only using some models of proton wavefunctions.

At the second step the integrals over gluons with rapidity $Y<\eta$ give
matrix element of triple Wilson-line operator $B_{123}$ between target states.
The \textquotedblleft rapidity cutoff\textquotedblright\ $\eta$ is arbitrary
(between rapidities of projectile and target) but it is convenient to choose
it in such a way that the impact factor does not scale with energy so all
energy dependence is shifted to the matrix element of triple Wilson-line
operator (see the discussion in Ref. \cite{Balitsky:2013if}). In the leading
order the rapidity evolution of this operator was calculated in Ref.
\cite{Gerasimov:2012bj}, \cite{Grabovsky:2013mba} while in this paper we
present the result for the NLO evolution.

\section{NLO evolution of triple-Wilson-line operator}

In this Section we outline the calculation of the NLO kernel for the rapidity
evolution of 3QWL operator (\ref{tripole}). In accordance with general logic
of high-energy OPE in order to find the evolution of the Wilson-line operators
with respect to rapidity cutoff we consider matrix element of operators with
rapidities up to $\eta_{1}$ and we integrate over the region of rapidities
$\Delta\eta=\eta_{1}-\eta_{2}$ (where $\eta_{1}>\eta_{2}>Y_{\mathrm{target}}%
$). Since particles with different rapidities perceive each other as Wilson
lines, the result of integration will be $\Delta\eta$ times kernel of the
rapidity evolution times Wilson lines with rapidities up to $\eta_{2}$.  As we
discussed in Sect. 2, it is convenient to use background-field formalism where
gluons with rapidities $Y<\eta_{2}$  form a narrow shockwave.

\begin{figure}[tbh]
\begin{center}
\includegraphics[width=88mm]{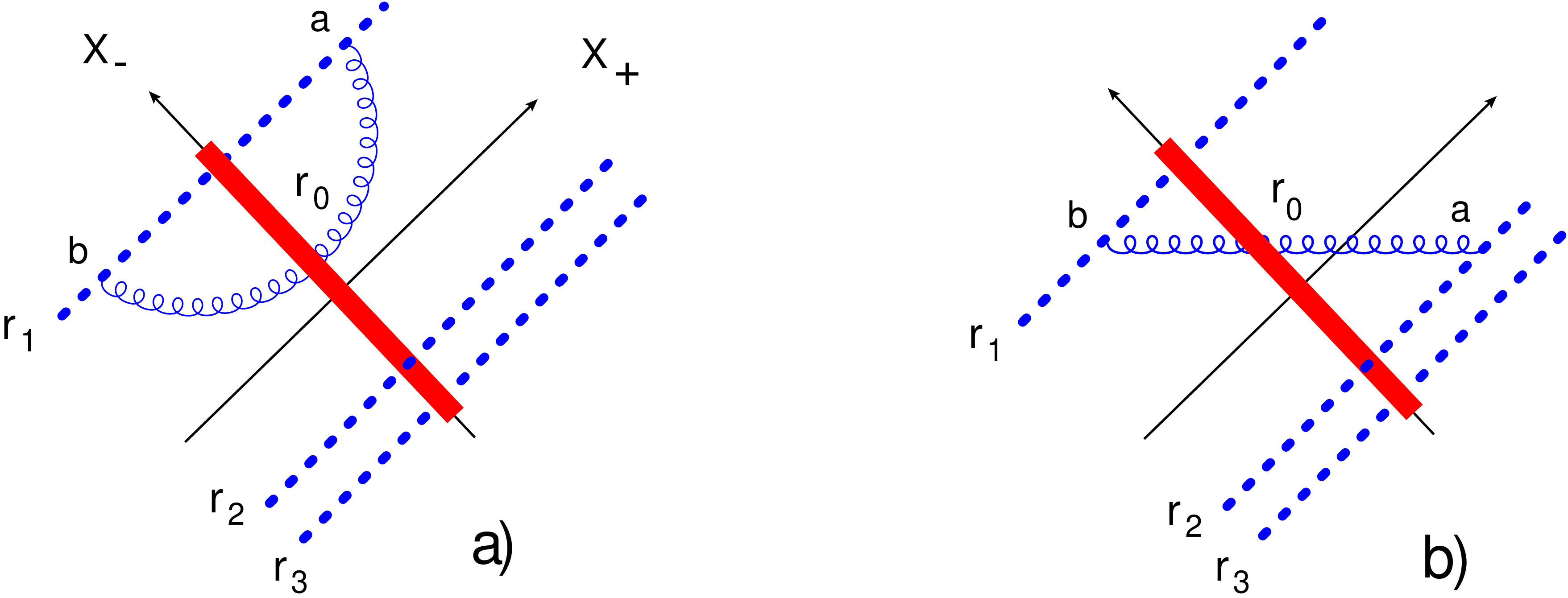}
\end{center}
\caption{Leading-order diagrams.}%
\label{fig:lo}%
\end{figure}

The typical leading-order diagrams are shown in Fig. \ref{fig:lo} and it is
clear that at this order the evolution equation for 3-line operator can be
restored from the evolution of two-line operators since all interactions are
either pairwise (Fig. \ref{fig:lo}b) or self-interactions (Fig. \ref{fig:lo}%
a). The result for the LO evolution has the form \cite{Gerasimov:2012bj}
\[
\frac{\partial B_{123}}{\partial\eta}=\frac{\alpha_{s}3}{4\pi^{2}}\int
d\vec{r}_{0}\left[  \frac{\vec{r}_{12}^{\,\,2}}{\vec{r}_{01}^{\,\,2}\vec
{r}_{02}^{\,\,2}}(-B_{123}+\frac{1}{6}(B_{100}B_{320}+B_{200}B_{310}%
-B_{300}B_{210}))\right.
\]%
\begin{equation}
\left.  \frac{{}}{{}}+(1\leftrightarrow3)+(2\leftrightarrow3)\right]
.\label{LO}%
\end{equation}
where $\vec{r}_{1},\vec{r}_{2},\vec{r}_{3}$ are the coordinates of the quark
Wilson lines within the 3QWL and $\vec{r}_{0}$ is the coordinate of gluon
Wilson line coming from the intersection with the shock wave. (In this paper
we set $N_{c}=3$ explicitly).

\begin{figure}[tbh]
\begin{center}
\includegraphics[width=66mm]{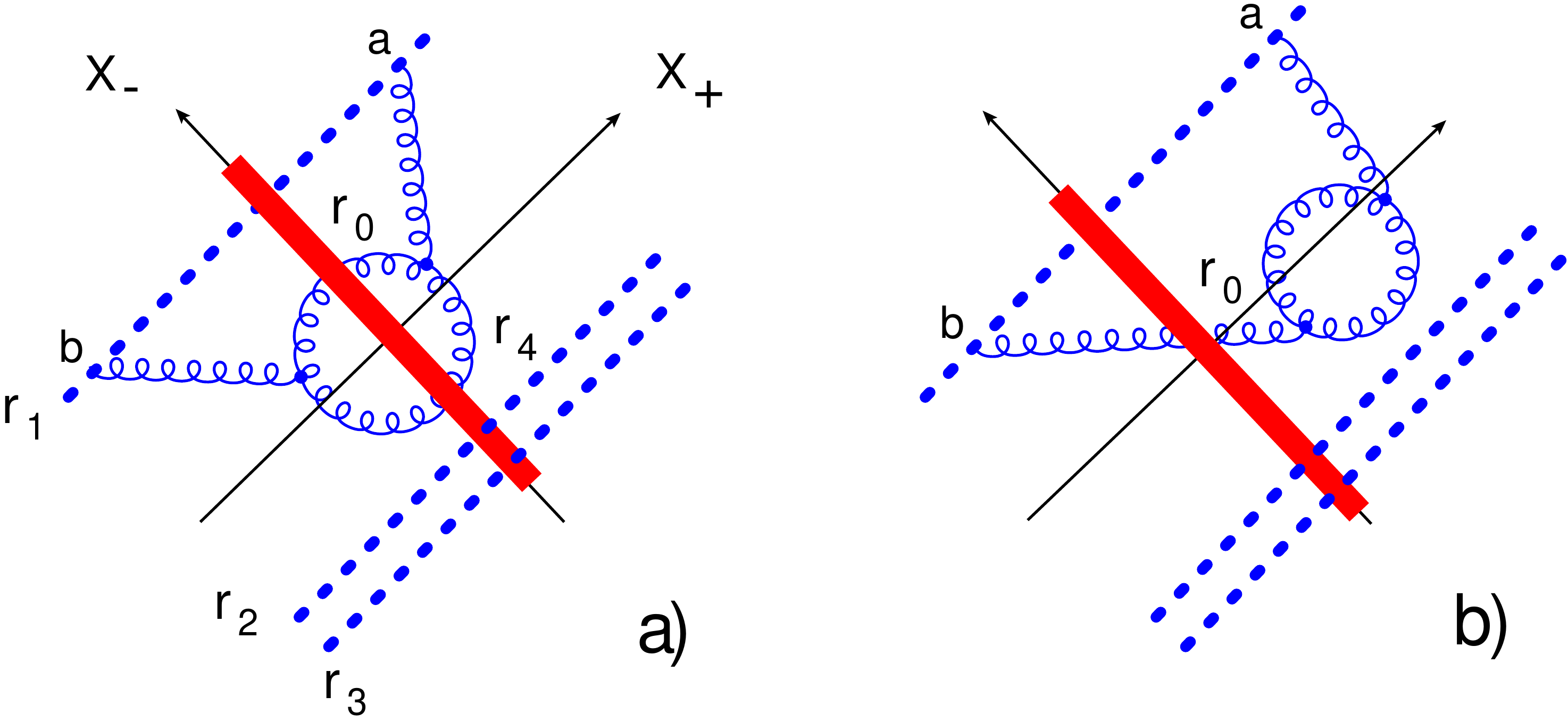} \hspace{11mm}
\includegraphics[width=70mm]{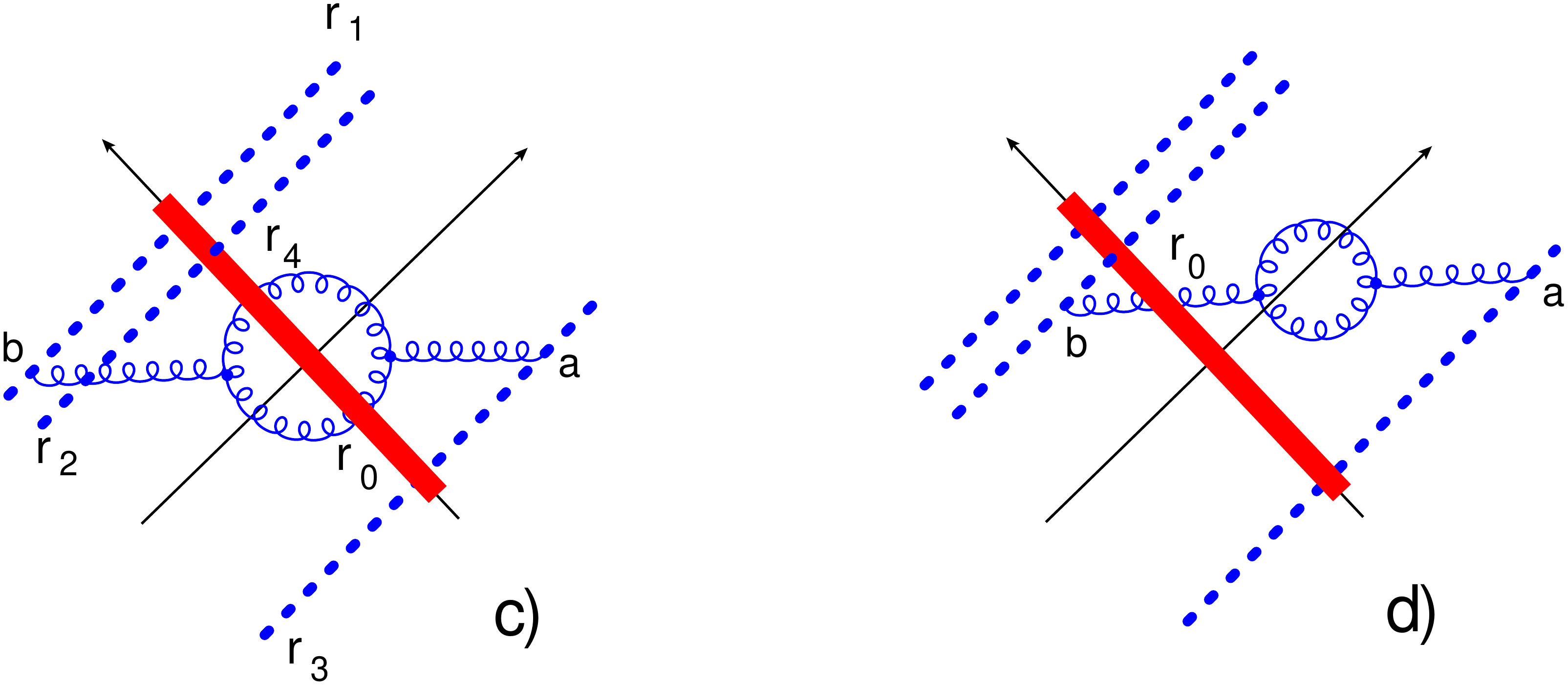}
\par
\vspace{7mm} \includegraphics[width=70mm]{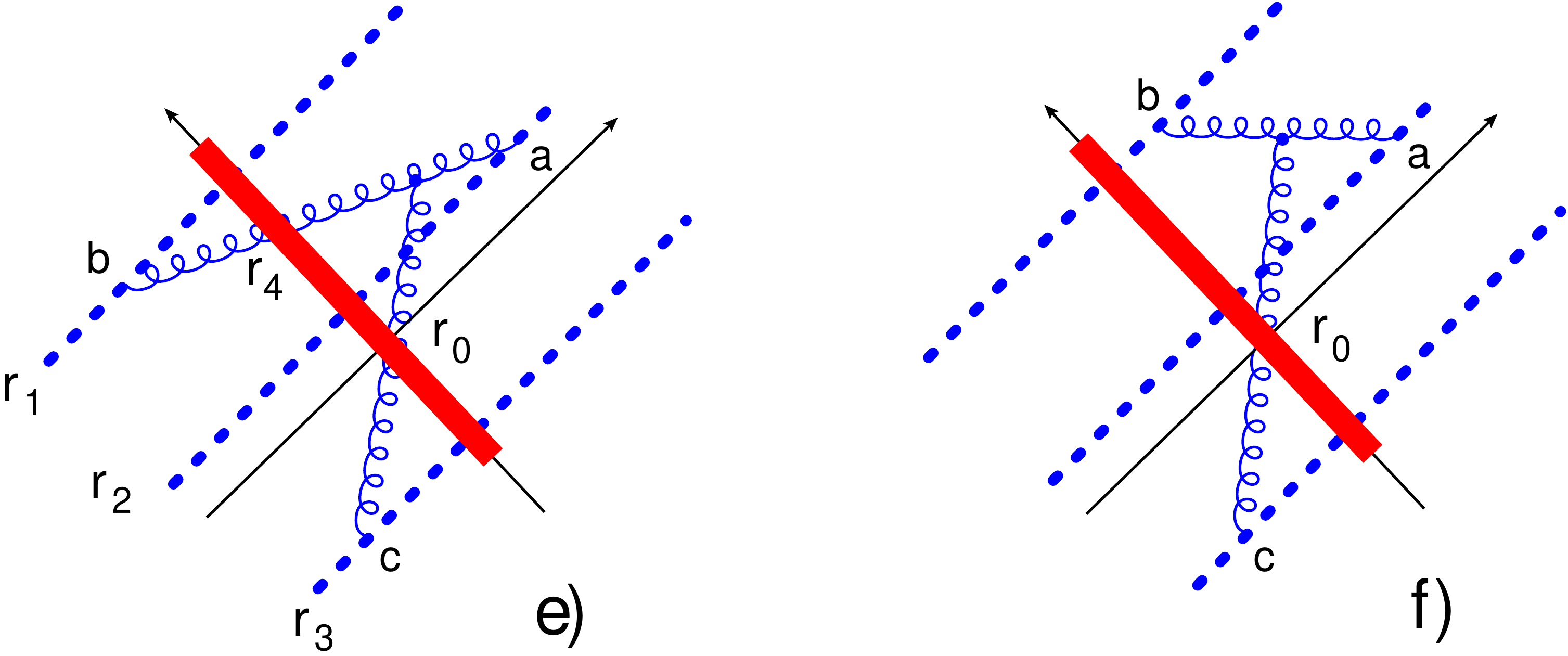}
\end{center}
\caption{Typical NLO diagrams.}%
\label{fig:typinlo}%
\end{figure}

The typical diagrams in the next-to-leading order are shown in Fig.
\ref{fig:typinlo} where $\vec{r}_{0}$ and $\vec{r}_{4}$ are the coordinates of
intersections with the shock wave. It is clear that at the NLO in addition to
self-interaction (Fig.\ref{fig:typinlo}a) and pairwise-interaction
(Fig.\ref{fig:typinlo}b) diagrams we have also the triple-interaction diagrams
of Fig.\ref{fig:typinlo}c type.
It should be emphasized that for the self-interaction and pairwise diagrams
one can use the results of Ref. \cite{Balitsky:2013fea} while the
triple-interaction diagrams were already calculated in Ref.
\cite{Grabovsky:2013mba}. In this paper we will combine these results to get a
concise expression for the evolution of three-Wilson-line operator
(\ref{tripole}).

Let us start with self- and pairwise-interactions of the type shown in Fig.
\ref{fig:typinlo} a-d. At $N_{c}=3$ one can use the $SU(3)$ identities
\begin{equation}
U_{4}^{ba}=2tr(t^{b}U_{4}t^{a}U_{4}^{\dag}),\quad(t^{a})_{i}^{j}(t^{a}%
)_{k}^{l}=\frac{1}{2}\delta_{i}^{l}\delta_{k}^{j}-\frac{1}{6}\delta_{i}%
^{j}\delta_{k}^{l} \label{Uadjoint}%
\end{equation}
to rewrite the result of \cite{Balitsky:2013fea}\ only through the Wilson
lines in the fundamental representation. For the contribution of the the
states with 2 gluons crossing the shockwave it reads
\begin{equation}
\langle K_{NLO}\otimes(U_{1})_{i_{1}}^{i_{3}}(U_{2})_{j_{1}}^{j_{3}}%
\rangle|_{2g}=-{\frac{\alpha_{s}^{2}}{8\pi^{4}}}\!\int\!d\vec{r}_{0}d\vec
{r}_{4}~\mathbf{G}_{12},
\end{equation}%
\[
\mathbf{G}_{12}=G_{1}\left\{  \left(  U_{1}U_{0}^{\dag}U_{4}\right)  _{i_{1}%
}^{i_{3}}\left(  U_{0}U_{4}^{\dag}U_{2}\right)  _{j_{1}}^{j_{3}}+\left(
U_{4}U_{0}^{\dag}U_{1}\right)  _{i_{1}}^{i_{3}}\left(  U_{2}U_{4}^{\dag}%
U_{0}\right)  _{j_{1}}^{j_{3}}\right\}
\]%
\[
+G_{2}(U_{1})_{i_{1}}^{i_{3}}\left\{  \left(  U_{0}U_{4}^{\dag}U_{2}%
U_{0}^{\dag}U_{4}\right)  _{j_{1}}^{j_{3}}+\left(  U_{4}U_{0}^{\dag}U_{2}%
U_{4}^{\dag}U_{0}\right)  _{j_{1}}^{j_{3}}\right\}
\]%
\[
+G_{3}(U_{2})_{j_{1}}^{j_{3}}\left\{  \left(  U_{0}U_{4}^{\dag}U_{1}%
U_{0}^{\dag}U_{4}\right)  _{i_{1}}^{i_{3}}+\left(  U_{4}U_{0}^{\dag}U_{1}%
U_{4}^{\dag}U_{0}\right)  _{i_{1}}^{i_{3}}\right\}
\]%
\[
+G_{4}\left\{  (U_{4})_{j_{1}}^{i_{3}}\left(  U_{1}U_{0}^{\dag}U_{2}\right)
_{i_{1}}^{j_{3}}+(U_{4})_{i_{1}}^{j_{3}}\left(  U_{2}U_{0}^{\dag}U_{1}\right)
_{j_{1}}^{i_{3}}\right\}  tr(U_{0}U_{4}^{\dag})
\]%
\[
+G_{5}(U_{2})_{j_{1}}^{j_{3}}(U_{4})_{i_{1}}^{i_{3}}tr(U_{0}U_{4}^{\dag
})tr(U_{0}^{\dag}U_{1})
\]%
\[
+G_{6}\left(  (U_{0})_{i_{1}}^{i_{3}}\left\{  \left(  U_{4}U_{0}^{\dag}%
U_{1}U_{4}^{\dag}U_{2}\right)  _{j_{1}}^{j_{3}}+\left(  U_{2}U_{4}^{\dag}%
U_{1}U_{0}^{\dag}U_{4}\right)  _{j_{1}}^{j_{3}}\right\}  \right.
\]%
\[
-\left.  \left\{  (U_{4})_{j_{1}}^{i_{3}}\left(  U_{0}U_{4}^{\dag}%
U_{2}\right)  _{i_{1}}^{j_{3}}+(U_{4})_{i_{1}}^{j_{3}}\left(  U_{2}U_{4}%
^{\dag}U_{0}\right)  _{j_{1}}^{i_{3}}\right\}  tr(U_{0}^{\dag}U_{1})\right)
\]%
\[
+G_{7}(U_{1})_{i_{1}}^{i_{3}}(U_{4})_{j_{1}}^{j_{3}}tr(U_{0}U_{4}^{\dag
})tr(U_{0}^{\dag}U_{2})
\]%
\[
+G_{8}\left(  (U_{4})_{j_{1}}^{j_{3}}\left\{  \left(  U_{0}U_{4}^{\dag}%
U_{2}U_{0}^{\dag}U_{1}\right)  _{i_{1}}^{i_{3}}+\left(  U_{1}U_{0}^{\dag}%
U_{2}U_{4}^{\dag}U_{0}\right)  _{i_{1}}^{i_{3}}\right\}  \right.
\]%
\begin{equation}
\left.  -\left\{  (U_{0})_{i_{1}}^{j_{3}}\left(  U_{4}U_{0}^{\dag}%
U_{1}\right)  _{j_{1}}^{i_{3}}+(U_{0})_{j_{1}}^{i_{3}}\left(  U_{1}U_{0}%
^{\dag}U_{4}\right)  _{i_{1}}^{j_{3}}\right\}  tr(U_{4}^{\dag}U_{2})\right)  .
\label{G12}%
\end{equation}
Note that as discussed in Ref. \cite{Balitsky:2013fea}, it is convenient to
present some of the terms with one intersection in the two-intersection form
with an additional integration over $\vec{r}_{4}$. In doing so, some $U_{4}$
and $U_{0}$ factors in Eq. (\ref{G12}) are replaced by $U_{4}-U_{i}$ and
$U_{4}-U_{i}$ ($i=1,2$ or 3). We do not write such subtraction terms here
since it is easier to make the subtraction after the color convolution. The
functions have the form
\[
G_{1}=-\left(  \frac{\vec{r}_{04}{}^{2}-2\vec{r}_{02}{}^{2}}{2\vec{r}_{02}%
{}^{2}\vec{r}_{04}{}^{2}\left(  \vec{r}_{24}{}^{2}-\vec{r}_{02}{}^{2}\right)
}-\frac{\vec{r}_{04}{}^{2}\vec{r}_{12}{}^{2}+\vec{r}_{02}{}^{2}\left(  \vec
{r}_{12}{}^{2}-\vec{r}_{14}{}^{2}\right)  +\left(  \vec{r}_{01}{}^{2}+\vec
{r}_{02}{}^{2}-\vec{r}_{12}{}^{2}\right)  \vec{r}_{24}{}^{2}}{2\vec{r}_{01}%
{}^{2}\vec{r}_{02}{}^{2}\vec{r}_{04}{}^{2}\vec{r}_{24}{}^{2}}\right.
\]%
\[
\left.  +\frac{1}{\vec{r}_{01}{}^{2}\vec{r}_{24}{}^{2}-\vec{r}_{02}{}^{2}%
\vec{r}_{14}{}^{2}}\left[  \frac{2\vec{r}_{12}{}^{2}}{\vec{r}_{04}{}^{2}%
}-\frac{\vec{r}_{12}{}^{4}{}}{2\vec{r}_{01}{}^{2}\vec{r}_{24}{}^{2}}%
-\frac{\vec{r}_{14}{}^{2}}{\vec{r}_{04}{}^{2}}-\frac{\left(  \vec{r}_{02}%
{}^{2}-\vec{r}_{04}{}^{2}\right)  \left(  \vec{r}_{14}{}^{2}-\vec{r}_{01}%
{}^{2}\right)  \vec{r}_{24}{}^{2}}{\vec{r}_{04}{}^{4}{}\left(  \vec{r}_{24}%
{}^{2}-\vec{r}_{02}{}^{2}\right)  }\right]  \right)
\]%
\begin{equation}
\times\ln\left(  \frac{\vec{r}_{02}{}^{2}}{\vec{r}_{24}{}^{2}}\right)
-\frac{1}{2\vec{r}_{04}{}^{4}}+\left(  0\leftrightarrow4,1\leftrightarrow
2\right)  .
\end{equation}%
\[
G_{2}=\left(  \frac{1}{\left(  \vec{r}_{02}{}^{2}-\vec{r}_{24}{}^{2}\right)
}\left[  \left(  \frac{1}{\vec{r}_{04}{}^{4}{}}+\frac{1}{2\vec{r}_{02}{}%
^{2}\vec{r}_{24}{}^{2}}\right)  \frac{\left(  \vec{r}_{02}{}^{2}+\vec{r}%
_{24}{}^{2}\right)  }{2}-\frac{2}{\vec{r}_{04}{}^{2}}\right]  -\frac{\vec
{r}_{02}{}^{2}-\vec{r}_{24}{}^{2}}{4\vec{r}_{02}{}^{2}\vec{r}_{04}{}^{2}%
\vec{r}_{24}{}^{2}}\right)
\]%
\begin{equation}
\times\ln\left(  \frac{\vec{r}_{02}{}^{2}}{\vec{r}_{24}{}^{2}}\right)
-\frac{1}{\vec{r}_{04}{}^{4}{}}.
\end{equation}%
\begin{equation}
G_{3}=G_{2}|_{1\leftrightarrow2}.
\end{equation}%
\[
G_{4}=\left(  \frac{\left(  \vec{r}_{02}{}^{2}-\vec{r}_{24}{}^{2}\right)
\left(  \vec{r}_{02}{}^{2}\vec{r}_{14}{}^{2}-\vec{r}_{01}{}^{2}\vec{r}_{24}%
{}^{2}\right)  }{2\vec{r}_{01}{}^{2}\vec{r}_{02}{}^{2}\vec{r}_{14}{}^{2}%
\vec{r}_{24}{}^{2}\vec{r}_{04}{}^{2}}-\frac{\vec{r}_{02}{}^{2}+\vec{r}_{24}%
{}^{2}}{2\vec{r}_{02}{}^{2}\vec{r}_{24}{}^{2}\vec{r}_{04}{}^{2}}+\frac
{1}{2\vec{r}_{01}{}^{2}\vec{r}_{14}{}^{2}}+\frac{\vec{r}_{12}{}^{4}{}}%
{2\vec{r}_{01}{}^{2}\vec{r}_{02}{}^{2}\vec{r}_{14}{}^{2}\vec{r}_{24}{}^{2}%
}\right.
\]%
\begin{equation}
\left.  +\left(  \frac{\left(  \vec{r}_{01}{}^{2}-\vec{r}_{14}{}^{2}\right)
\left(  \vec{r}_{02}{}^{2}-\vec{r}_{24}{}^{2}\right)  }{2\vec{r}_{01}{}%
^{2}\vec{r}_{02}{}^{2}\vec{r}_{04}{}^{2}\vec{r}_{14}{}^{2}\vec{r}_{24}{}^{2}%
}-\frac{\vec{r}_{02}{}^{2}+\vec{r}_{24}{}^{2}}{2\vec{r}_{01}{}^{2}\vec{r}%
_{02}{}^{2}\vec{r}_{14}{}^{2}\vec{r}_{24}{}^{2}}\right)  \vec{r}_{12}{}%
^{2}\right)  \ln\left(  \frac{\vec{r}_{01}{}^{2}}{\vec{r}_{14}{}^{2}}\right)
-G_{1}.
\end{equation}%
\begin{equation}
G_{5}=\frac{2}{\vec{r}_{04}{}^{4}{}}+\left(  \frac{1}{\vec{r}_{01}{}^{2}%
-\vec{r}_{14}{}^{2}}\left[  \frac{4}{\vec{r}_{04}{}^{2}}-\frac{\vec{r}_{01}%
{}^{2}+\vec{r}_{14}{}^{2}}{\vec{r}_{04}{}^{4}{}}-\frac{1}{\vec{r}_{01}{}^{2}%
}\right]  -\frac{1}{\vec{r}_{01}{}^{2}\vec{r}_{04}{}^{2}}\right)  \ln\left(
\frac{\vec{r}_{01}{}^{2}}{\vec{r}_{14}{}^{2}}\right)  .
\end{equation}%
\begin{equation}
G_{6}=\left(  \frac{\vec{r}_{12}{}^{2}-\vec{r}_{24}{}^{2}}{2\vec{r}_{01}{}%
^{2}\vec{r}_{14}{}^{2}\vec{r}_{24}{}^{2}}+\frac{\vec{r}_{14}{}^{2}\left(
\vec{r}_{24}{}^{2}-\vec{r}_{02}{}^{2}\right)  +\vec{r}_{01}{}^{2}\left(
\vec{r}_{14}{}^{2}-\vec{r}_{12}{}^{2}+\vec{r}_{24}{}^{2}\right)  }{2\vec
{r}_{01}{}^{2}\vec{r}_{04}{}^{2}\vec{r}_{14}{}^{2}\vec{r}_{24}{}^{2}}\right)
\ln\left(  \frac{\vec{r}_{01}{}^{2}}{\vec{r}_{14}{}^{2}}\right)  .
\end{equation}%
\begin{equation}
G_{7}=G_{5}|_{1\leftrightarrow2},\quad G_{8}=G_{6}|_{1\leftrightarrow
2,0\leftrightarrow4}.
\end{equation}
After the convolution with $\varepsilon^{i_{1}j_{1}h}\varepsilon_{i_{3}%
j_{3}h^{\prime}}(U_{3})_{h}^{h^{\prime}},$ (\ref{G12}) gives the contribution
of the 2-gluon states to the evolution of the 3QWL operator $U_{1}\cdot
U_{2}\cdot U_{3}$ describing the total interaction of Wilson lines 1 and 2,
leaving Wilson line 3 intact.%
\[
\mathbf{G}_{12}\varepsilon^{i_{1}j_{1}h}\varepsilon_{i_{3}j_{3}h^{\prime}%
}(U_{3})_{h}^{h^{\prime}}%
\]%
\[
=G_{1}\left(  \frac{{}}{{}}\left(  U_{0}U_{4}{}^{\dag}U_{2}\right)
\cdot\left(  U_{1}U_{0}{}^{\dag}U_{4}\right)  +\left(  U_{2}U_{4}{}^{\dag
}U_{0}\right)  \cdot\left(  U_{4}U_{0}{}^{\dag}U_{1}\right)  \right)  \cdot
U_{3}%
\]%
\[
+\left[  G_{2}\left(  U_{0}U_{4}{}^{\dag}U_{2}U_{0}{}^{\dag}U_{4}+U_{4}U_{0}%
{}^{\dag}U_{2}U_{4}{}^{\dag}U_{0}\frac{{}}{{}}\right)  \cdot U_{1}\cdot
U_{3}+\left(  1\leftrightarrow2\right)  \right]
\]%
\[
-G_{4}tr\left(  U_{0}U_{4}{}^{\dag}\right)  \left(  U_{1}U_{0}{}^{\dag}%
U_{2}+U_{2}U_{0}{}^{\dag}U_{1}\frac{{}}{{}}\right)  \cdot U_{3}\cdot U_{4}%
\]%
\[
+\left[  G_{5}\,\,U_{2}\cdot U_{3}\cdot U_{4}\,\,tr\left(  U_{0}{}^{\dag}%
U_{1}\right)  tr\left(  U_{0}U_{4}{}^{\dag}\right)  +\left(  1\leftrightarrow
2\right)  \frac{{}}{{}}\right]
\]%
\[
+\left[  G_{6}\left(  tr\left(  U_{0}{}^{\dag}U_{1}\right)  \left(  U_{0}%
U_{4}{}^{\dag}U_{2}+U_{2}U_{4}{}^{\dag}U_{0}\frac{{}}{{}}\right)  \cdot
U_{3}\cdot U_{4}\right.  \right.
\]%
\begin{equation}
\left.  \left.  +\left(  U_{2}U_{4}{}^{\dag}U_{1}U_{0}{}^{\dag}U_{4}%
+U_{4}U_{0}{}^{\dag}U_{1}U_{4}{}^{\dag}U_{2}\frac{{}}{{}}\right)  \cdot
U_{0}\cdot U_{3}\right)  +\left(  1\leftrightarrow2,0\leftrightarrow4\right)
\right]  . \label{G<12>3}%
\end{equation}
One can also write%
\begin{equation}
\langle K_{NLO}\otimes(U_{1}^{\dag})_{j_{1}}^{j_{3}}\rangle|_{2g}%
=-{\frac{\alpha_{s}^{2}}{8\pi^{4}}}\!\int\!d\vec{r}_{0}d\vec{r}_{4}%
~\mathbf{G}_{1^{\dag}},
\end{equation}%
\[
\mathbf{G}_{1^{\dag}}=G_{3}\left(  U_{4}{}^{\dag}U_{0}U_{1}{}^{\dag}U_{4}%
U_{0}{}^{\dag}+U_{0}{}^{\dag}U_{4}U_{1}{}^{\dag}U_{0}U_{4}{}^{\dag}\frac{{}%
}{{}}\right.
\]%
\[
\left.  -tr\left(  U_{1}{}^{\dag}U_{4}\right)  tr\left(  U_{4}{}^{\dag}%
U_{0}\right)  U_{0}{}^{\dag}-tr\left(  U_{0}{}^{\dag}U_{4}\right)  tr\left(
U_{1}{}^{\dag}U_{0}\right)  U_{4}{}^{\dag}\frac{{}}{{}}\right)  _{j_{1}%
}^{j_{3}}%
\]%
\begin{equation}
+G_{9}\left(  tr\left(  U_{1}{}^{\dag}U_{4}\right)  tr\left(  U_{4}{}^{\dag
}U_{0}\right)  U_{0}{}^{\dag}-tr\left(  U_{0}{}^{\dag}U_{4}\right)  tr\left(
U_{1}{}^{\dag}U_{0}\right)  U_{4}{}^{\dag}\frac{{}}{{}}\right)  _{j_{1}%
}^{j_{3}},
\end{equation}%
\begin{equation}
G_{9}=\frac{\vec{r}_{01}{}^{2}-\vec{r}_{04}{}^{2}+\vec{r}_{14}{}^{2}}{4\vec
{r}_{01}{}^{2}\vec{r}_{04}{}^{2}\vec{r}_{14}{}^{2}}\ln\left(  \frac{\vec
{r}_{01}{}^{2}}{\vec{r}_{14}{}^{2}}\right)  .
\end{equation}
And then take the convolution%
\[
\mathbf{G}_{\langle1^{\dag}\rangle3}=\mathbf{G}_{1^{\dag}}\left(
U_{3}\right)  _{j_{3}}^{j_{1}}=G_{3}\left(  tr\left(  U_{0}{}^{\dag}U_{3}%
U_{4}{}^{\dag}U_{0}U_{1}{}^{\dag}U_{4}\right)  +tr\left(  U_{0}{}^{\dag}%
U_{4}U_{1}{}^{\dag}U_{0}U_{4}{}^{\dag}U_{3}\right)  \frac{{}}{{}}\right.
\]%
\[
\left.  -tr\left(  U_{0}{}^{\dag}U_{3}\right)  tr\left(  U_{1}{}^{\dag}%
U_{4}\right)  tr\left(  U_{4}{}^{\dag}U_{0}\right)  -tr\left(  U_{0}{}^{\dag
}U_{4}\right)  tr\left(  U_{1}{}^{\dag}U_{0}\right)  tr\left(  U_{4}{}^{\dag
}U_{3}\right)  \frac{{}}{{}}\right)
\]%
\begin{equation}
+G_{9}\left(  tr\left(  U_{0}{}^{\dag}U_{3}\right)  tr\left(  U_{1}{}^{\dag
}U_{4}\right)  tr\left(  U_{4}{}^{\dag}U_{0}\right)  -tr\left(  U_{0}{}^{\dag
}U_{4}\right)  tr\left(  U_{1}{}^{\dag}U_{0}\right)  tr\left(  U_{4}{}^{\dag
}U_{3}\right)  \frac{{}}{{}}\right)  . \label{G<1dag>3}%
\end{equation}
For the elements of $SU(3)$ group one has the identity%
\begin{equation}
\varepsilon^{ijh}\varepsilon_{i^{\prime}j^{\prime}h^{\prime}}(U_{1}%
)_{i}^{i^{\prime}}(U_{1})_{j}^{j^{\prime}}=2(U_{1}^{\dag})_{h^{\prime}}%
^{h},\quad U_{1}\cdot U_{1}\cdot U_{3}=2tr(U_{1}^{\dag}U_{3}%
),\label{dipolimit}%
\end{equation}
Taking $\vec{r}_{2}=\vec{r}_{1}$ in (\ref{G<12>3}) one can check that it is
related to (\ref{G<1dag>3}) via this identity using the other $SU(3)$
identities (\ref{3qWlidentity}) and (\ref{IDENTITY}). Taking the conjugate of
$\mathbf{G}_{1^{\dag}},$ one gets
\begin{equation}
\langle K_{NLO}\otimes(U_{1})_{j}^{j^{\prime}}\rangle|_{2g}=-{\frac{\alpha
_{s}^{2}}{8\pi^{4}}}\!\int\!d\vec{r}_{0}d\vec{r}_{4}~\mathbf{G}_{1},
\end{equation}%
\[
\mathbf{G}_{1}=G_{3}\left(  U_{4}{}U_{0}^{\dag}U_{1}{}U_{4}^{\dag}U_{0}%
{}+U_{0}{}U_{4}^{\dag}U_{1}{}U_{0}^{\dag}U_{4}{}\frac{{}}{{}}\right.
\]%
\[
\left.  -tr(U_{1}{}U_{4}^{\dag})tr(U_{4}{}U_{0}^{\dag})U_{0}{}-tr(U_{0}{}%
U_{4}^{\dag})tr(U_{1}{}U_{0}^{\dag})U_{4}{}\frac{{}}{{}}\right)
_{j}^{j^{\prime}}%
\]%
\begin{equation}
+G_{9}\left(  tr(U_{1}{}U_{4}^{\dag})tr(U_{4}{}U_{0}^{\dag})U_{0}{}-tr(U_{0}%
{}U_{4}^{\dag})tr(U_{1}{}U_{0}^{\dag})U_{4}{}\frac{{}}{{}}\right)
_{j}^{j^{\prime}},
\end{equation}
The contribution of the evolution of only one line $U_{1}$ to the evolution of
the 3QWL reads%
\[
\mathbf{G}_{\langle1\rangle23}=\mathbf{G}_{1}\varepsilon^{ijh}\varepsilon
_{i^{\prime}j^{\prime}h^{\prime}}\left(  U_{2}\right)  _{i}^{i^{\prime}}%
(U_{3})_{h}^{h^{\prime}}=G_{3}\left(  U_{4}U_{0}{}^{\dag}U_{1}{}U_{4}{}^{\dag
}U_{0}+U_{0}{}U_{4}^{\dag}U_{1}{}U_{0}^{\dag}U_{4}{}\frac{{}}{{}}\right.
\]%
\[
\left.  -{}tr(U_{1}{}U_{4}^{\dag})tr(U_{4}{}U_{0}^{\dag})U_{0}-tr\left(
U_{0}{}^{\dag}U_{1}\right)  tr\left(  U_{4}{}^{\dag}U_{0}\right)  U_{4}{}%
\frac{{}}{{}}\right)  \cdot U_{2}\cdot U_{3}%
\]%
\begin{equation}
+G_{9}\left(  tr(U_{1}{}U_{4}^{\dag})tr(U_{4}{}U_{0}^{\dag})U_{0}-tr\left(
U_{0}{}^{\dag}U_{1}\right)  tr\left(  U_{4}{}^{\dag}U_{0}\right)  U_{4}{}%
\frac{{}}{{}}\right)  \cdot U_{2}\cdot U_{3}. \label{<1>23}%
\end{equation}
Then the connected contribution of the evolution of lines 1 and 2 reads
\[
\mathbf{G}_{\langle12\rangle3}=\frac{1}{2}[H_{1}-(1\leftrightarrow2)]
\]%
\[
\times\left[  \left(  U_{0}U_{4}{}^{\dag}U_{2}\right)  \cdot\left(  U_{1}%
U_{0}{}^{\dag}U_{4}\right)  \cdot U_{3}-\left(  U_{0}U_{4}{}^{\dag}%
U_{1}\right)  \cdot\left(  U_{2}U_{0}{}^{\dag}U_{4}\right)  \cdot
U_{3}-(4\leftrightarrow0)\right]
\]%
\[
+H_{2}\left[  tr\left(  U_{0}{}^{\dag}U_{1}\right)  \left(  U_{0}U_{4}{}%
^{\dag}U_{2}+U_{2}U_{4}{}^{\dag}U_{0}\right)  \cdot U_{3}\cdot U_{4}\right.
\]%
\[
\left.  -\left(  U_{2}U_{0}{}^{\dag}U_{1}U_{4}{}^{\dag}U_{0}+U_{0}U_{4}%
{}^{\dag}U_{1}U_{0}{}^{\dag}U_{2}\right)  \cdot U_{3}\cdot U_{4}%
-(4\leftrightarrow0)\right]
\]%
\[
+H_{3}\left[  tr\left(  U_{0}{}^{\dag}U_{1}\right)  \left(  U_{0}U_{4}{}%
^{\dag}U_{2}+U_{2}U_{4}{}^{\dag}U_{0}\right)  \cdot U_{3}\cdot U_{4}\right.
\]%
\[
\left.  +\left(  U_{2}U_{0}{}^{\dag}U_{1}U_{4}{}^{\dag}U_{0}+U_{0}U_{4}%
{}^{\dag}U_{1}U_{0}{}^{\dag}U_{2}\right)  \cdot U_{3}\cdot U_{4}%
+(4\leftrightarrow0)\right]
\]%
\[
+H_{4}[tr\left(  U_{0}U_{4}{}^{\dag}\right)  \left(  U_{1}U_{0}{}^{\dag}%
U_{2}+U_{2}U_{0}{}^{\dag}U_{1}\right)  \cdot U_{3}\cdot U_{4}%
\]%
\[
+\left(  U_{0}U_{4}{}^{\dag}U_{1}\right)  \cdot\left(  U_{2}U_{0}{}^{\dag
}U_{4}\right)  \cdot U_{3}+\left(  U_{0}U_{4}{}^{\dag}U_{2}\right)
\cdot\left(  U_{1}U_{0}{}^{\dag}U_{4}\right)  \cdot U_{3}+(4\leftrightarrow
0)]
\]%
\begin{equation}
+H_{1}[tr\left(  U_{4}U_{0}{}^{\dag}\right)  \left(  U_{1}U_{4}{}^{\dag}%
U_{2}+U_{2}U_{4}{}^{\dag}U_{1}\right)  \cdot U_{0}\cdot U_{3}%
-(4\leftrightarrow0)]+(1\leftrightarrow2). \label{<12>3}%
\end{equation}%
\[
H_{1}=\frac{1}{8}\left[  \frac{\left(  \vec{r}_{02}{}^{2}-\vec{r}_{12}{}%
^{2}\right)  \left(  \vec{r}_{14}{}^{2}\left(  \vec{r}_{02}{}^{2}-\vec{r}%
_{24}{}^{2}\right)  +\vec{r}_{04}{}^{2}\left(  \vec{r}_{24}{}^{2}-\vec{r}%
_{12}{}^{2}\right)  \right)  }{\vec{r}_{01}{}^{2}\vec{r}_{02}{}^{2}\vec
{r}_{04}{}^{2}\vec{r}_{14}{}^{2}\vec{r}_{24}{}^{2}}\right.
\]%
\begin{equation}
\left.  +\frac{\vec{r}_{12}{}^{2}-\vec{r}_{14}{}^{2}-\vec{r}_{24}{}^{2}}%
{\vec{r}_{04}{}^{2}\vec{r}_{14}{}^{2}\vec{r}_{24}{}^{2}}+\frac{\vec{r}_{24}%
{}^{2}-\vec{r}_{12}{}^{2}-\vec{r}_{14}{}^{2}}{\vec{r}_{02}{}^{2}\vec{r}_{04}%
{}^{2}\vec{r}_{14}{}^{2}}\right]  \ln\left(  \frac{\vec{r}_{01}{}^{2}}{\vec
{r}_{14}{}^{2}}\right)  .
\end{equation}%
\[
H_{2}=\frac{1}{8}\left[  \frac{\vec{r}_{12}{}^{2}}{\vec{r}_{01}{}^{2}\vec
{r}_{14}{}^{2}\vec{r}_{02}{}^{2}}+\frac{\vec{r}_{12}{}^{2}}{\vec{r}_{01}{}%
^{2}\vec{r}_{14}{}^{2}\vec{r}_{24}{}^{2}}+\frac{\vec{r}_{01}{}^{2}-\vec
{r}_{02}{}^{2}}{\vec{r}_{01}{}^{2}\vec{r}_{04}{}^{2}\vec{r}_{24}{}^{2}}%
+2\frac{\vec{r}_{14}{}^{2}-\vec{r}_{04}{}^{2}+\vec{r}_{01}{}^{2}}{\vec{r}%
_{01}{}^{2}\vec{r}_{04}{}^{2}\vec{r}_{14}{}^{2}}\right.
\]%
\begin{equation}
\left.  -\frac{\vec{r}_{12}{}^{2}}{\vec{r}_{04}{}^{2}\vec{r}_{14}{}^{2}\vec
{r}_{24}{}^{2}}+\frac{\vec{r}_{01}{}^{2}-\vec{r}_{12}{}^{2}}{\vec{r}_{01}%
{}^{2}\vec{r}_{04}{}^{2}\vec{r}_{02}{}^{2}}-\frac{\vec{r}_{24}{}^{2}}{\vec
{r}_{04}{}^{2}\vec{r}_{14}{}^{2}\vec{r}_{02}{}^{2}}\right]  \ln\left(
\frac{\vec{r}_{01}{}^{2}}{\vec{r}_{14}{}^{2}}\right)  .
\end{equation}%
\[
H_{3}=\frac{1}{8}\left[  \frac{\vec{r}_{01}{}^{2}-\vec{r}_{02}{}^{2}}{\vec
{r}_{01}{}^{2}\vec{r}_{04}{}^{2}\vec{r}_{24}{}^{2}}+\frac{\vec{r}_{12}{}^{2}%
}{\vec{r}_{01}{}^{2}\vec{r}_{14}{}^{2}\vec{r}_{24}{}^{2}}-\frac{\vec{r}_{12}%
{}^{2}}{\vec{r}_{04}{}^{2}\vec{r}_{14}{}^{2}\vec{r}_{24}{}^{2}}\right.
\]%
\begin{equation}
\left.  +\frac{\vec{r}_{12}{}^{2}}{\vec{r}_{01}{}^{2}\vec{r}_{04}{}^{2}\vec
{r}_{02}{}^{2}}+\frac{\vec{r}_{24}{}^{2}-\vec{r}_{14}{}^{2}}{\vec{r}_{04}%
{}^{2}\vec{r}_{14}{}^{2}\vec{r}_{02}{}^{2}}-\frac{\vec{r}_{12}{}^{2}}{\vec
{r}_{01}{}^{2}\vec{r}_{14}{}^{2}\vec{r}_{02}{}^{2}}\right]  \ln\left(
\frac{\vec{r}_{01}{}^{2}}{\vec{r}_{14}{}^{2}}\right)  .
\end{equation}%
\[
H_{4}=\frac{-1}{4\vec{r}_{04}{}^{4}}-\frac{1}{8}\left[  \frac{\vec{r}_{12}%
{}^{2}(\vec{r}_{14}{}^{2}-\vec{r}_{01}{}^{2})(\vec{r}_{02}{}^{2}+\vec{r}%
_{24}{}^{2})}{\vec{r}_{01}{}^{2}\vec{r}_{02}{}^{2}\vec{r}_{04}{}^{2}\vec
{r}_{14}{}^{2}\vec{r}_{24}{}^{2}}+\frac{\vec{r}_{12}{}^{2}}{\vec{r}_{01}{}%
^{2}\vec{r}_{14}{}^{2}\vec{r}_{24}{}^{2}}-\frac{\vec{r}_{12}{}^{4}{}}{\vec
{r}_{01}{}^{2}\vec{r}_{02}{}^{2}\vec{r}_{14}{}^{2}\vec{r}_{24}{}^{2}}\right.
\]%
\[
+\frac{\vec{r}_{24}{}^{2}+\vec{r}_{02}{}^{2}-\vec{r}_{14}{}^{2}}{\vec{r}%
_{02}{}^{2}\vec{r}_{04}{}^{2}\vec{r}_{14}{}^{2}}+\frac{\vec{r}_{01}{}^{2}%
-\vec{r}_{02}{}^{2}-\vec{r}_{24}{}^{2}}{\vec{r}_{01}{}^{2}\vec{r}_{04}{}%
^{2}\vec{r}_{24}{}^{2}}%
\]%
\[
+\frac{1}{\vec{r}_{01}{}^{2}-\vec{r}_{14}{}^{2}}\left(  \frac{\vec{r}_{12}%
{}^{2}-\vec{r}_{02}{}^{2}}{\vec{r}_{01}{}^{2}\vec{r}_{02}{}^{2}}-\frac{\vec
{r}_{12}{}^{2}+\vec{r}_{02}{}^{2}}{\vec{r}_{02}{}^{2}\vec{r}_{14}{}^{2}}%
-\frac{4\vec{r}_{14}{}^{2}}{\vec{r}_{04}{}^{4}{}}+\frac{8}{\vec{r}_{04}{}^{2}%
}\right)
\]%
\begin{equation}
\left.  +\frac{1}{\vec{r}_{01}{}^{2}\vec{r}_{24}{}^{2}-\vec{r}_{02}{}^{2}%
\vec{r}_{14}{}^{2}}\left(  \frac{2\vec{r}_{12}{}^{4}{}}{\vec{r}_{02}{}^{2}%
\vec{r}_{14}{}^{2}}+\frac{4\vec{r}_{02}{}^{2}\vec{r}_{14}{}^{2}}{\vec{r}%
_{04}{}^{4}{}}-\frac{8\vec{r}_{12}{}^{2}}{\vec{r}_{04}{}^{2}}\right)  \right]
\ln\left(  \frac{\vec{r}_{01}{}^{2}}{\vec{r}_{14}{}^{2}}\right)  .
\end{equation}
The fully connected ``triple'' contribution corresponding to the diagrams of
Fig. \ref{fig:typinlo} e,f can be taken from \cite{Balitsky:2013fea} or
\cite{Grabovsky:2013mba}\ and transformed to the form%
\[
\mathbf{G}_{\langle123\rangle}=H_{5}\left[  \left(  U_{0}U_{4}{}^{\dag}%
U_{3}\right)  \cdot\left(  U_{1}U_{0}{}^{\dag}U_{2}\right)  \cdot
U_{4}-\left(  U_{0}U_{4}{}^{\dag}U_{2}\right)  \cdot\left(  U_{1}U_{0}{}%
^{\dag}U_{3}\right)  \cdot U_{4}\right.
\]%
\[
\left.  +\left(  U_{2}U_{0}{}^{\dag}U_{1}\right)  \cdot\left(  U_{3}U_{4}%
{}^{\dag}U_{0}\right)  \cdot U_{4}-\left(  U_{2}U_{0}{}^{\dag}U_{4}\right)
\cdot\left(  U_{3}U_{4}{}^{\dag}U_{1}\right)  \cdot U_{0}+(4\leftrightarrow
0)\right]
\]%
\[
+H_{6}[\left(  U_{0}U_{4}{}^{\dag}U_{2}\right)  \cdot\left(  U_{1}U_{0}%
{}^{\dag}U_{3}\right)  \cdot U_{4}+\left(  U_{0}U_{4}{}^{\dag}U_{3}\right)
\cdot\left(  U_{1}U_{0}{}^{\dag}U_{2}\right)  \cdot U_{4}%
\]%
\[
+\left(  U_{2}U_{0}{}^{\dag}U_{1}\right)  \cdot\left(  U_{3}U_{4}{}^{\dag
}U_{0}\right)  \cdot U_{4}-\left(  U_{2}U_{0}{}^{\dag}U_{4}\right)
\cdot\left(  U_{3}U_{4}{}^{\dag}U_{1}\right)  \cdot U_{0}-(4\leftrightarrow
0)]
\]%
\begin{equation}
+(1\leftrightarrow2)+(1\leftrightarrow3). \label{<123>}%
\end{equation}%
\[
H_{5}=\frac{1}{8}\left[  \frac{\vec{r}_{13}{}^{2}\vec{r}_{02}{}^{2}}{\vec
{r}_{01}{}^{2}\vec{r}_{03}{}^{2}\vec{r}_{04}{}^{2}\vec{r}_{24}{}^{2}}%
-\frac{\vec{r}_{12}{}^{2}\vec{r}_{13}{}^{2}}{\vec{r}_{01}{}^{2}\vec{r}_{03}%
{}^{2}\vec{r}_{14}{}^{2}\vec{r}_{24}{}^{2}}+\frac{\vec{r}_{12}{}^{2}\vec
{r}_{34}{}^{2}}{\vec{r}_{03}{}^{2}\vec{r}_{04}{}^{2}\vec{r}_{14}{}^{2}\vec
{r}_{24}{}^{2}}-\frac{\vec{r}_{13}{}^{2}\vec{r}_{24}{}^{2}}{\vec{r}_{04}{}%
^{2}\vec{r}_{14}{}^{2}\vec{r}_{34}{}^{2}\vec{r}_{02}{}^{2}}\right.
\]%
\[
+\frac{\vec{r}_{12}{}^{2}\vec{r}_{13}{}^{2}}{\vec{r}_{01}{}^{2}\vec{r}_{14}%
{}^{2}\vec{r}_{34}{}^{2}\vec{r}_{02}{}^{2}}-\frac{\vec{r}_{03}{}^{2}\vec
{r}_{12}{}^{2}}{\vec{r}_{01}{}^{2}\vec{r}_{04}{}^{2}\vec{r}_{34}{}^{2}\vec
{r}_{02}{}^{2}}+\frac{\vec{r}_{14}{}^{2}-\vec{r}_{34}{}^{2}}{\vec{r}_{03}%
{}^{2}\vec{r}_{04}{}^{2}\vec{r}_{14}{}^{2}}-\frac{\vec{r}_{12}{}^{2}}{\vec
{r}_{04}{}^{2}\vec{r}_{14}{}^{2}\vec{r}_{24}{}^{2}}+\frac{\vec{r}_{03}{}%
^{2}-\vec{r}_{01}{}^{2}}{\vec{r}_{01}{}^{2}\vec{r}_{04}{}^{2}\vec{r}_{34}%
{}^{2}}%
\]%
\[
-\frac{\vec{r}_{13}{}^{2}}{\vec{r}_{01}{}^{2}\vec{r}_{14}{}^{2}\vec{r}_{34}%
{}^{2}}+\frac{\vec{r}_{13}{}^{2}}{\vec{r}_{04}{}^{2}\vec{r}_{14}{}^{2}\vec
{r}_{34}{}^{2}}+\frac{\vec{r}_{12}{}^{2}}{\vec{r}_{01}{}^{2}\vec{r}_{04}{}%
^{2}\vec{r}_{02}{}^{2}}+\frac{\vec{r}_{24}{}^{2}}{\vec{r}_{04}{}^{2}\vec
{r}_{14}{}^{2}\vec{r}_{02}{}^{2}}+\frac{\vec{r}_{13}{}^{2}}{\vec{r}_{01}{}%
^{2}\vec{r}_{03}{}^{2}\vec{r}_{14}{}^{2}}+\frac{\vec{r}_{12}{}^{2}}{\vec
{r}_{01}{}^{2}\vec{r}_{14}{}^{2}\vec{r}_{24}{}^{2}}%
\]%
\begin{equation}
\left.  +\frac{\vec{r}_{03}{}^{2}-\vec{r}_{23}{}^{2}}{\vec{r}_{03}{}^{2}%
\vec{r}_{04}{}^{2}\vec{r}_{24}{}^{2}}-\frac{\vec{r}_{12}{}^{2}}{\vec{r}_{01}%
{}^{2}\vec{r}_{14}{}^{2}\vec{r}_{02}{}^{2}}+\frac{\vec{r}_{23}{}^{2}-\vec
{r}_{34}{}^{2}}{\vec{r}_{04}{}^{2}\vec{r}_{34}{}^{2}\vec{r}_{02}{}^{2}}%
-\frac{\vec{r}_{02}{}^{2}}{\vec{r}_{01}{}^{2}\vec{r}_{04}{}^{2}\vec{r}_{24}%
{}^{2}}-\frac{\vec{r}_{13}{}^{2}}{\vec{r}_{01}{}^{2}\vec{r}_{03}{}^{2}\vec
{r}_{04}{}^{2}}\right]  \ln\left(  \frac{\vec{r}_{01}{}^{2}}{\vec{r}_{14}%
{}^{2}}\right)  .
\end{equation}%
\[
H_{6}=\frac{1}{8}\left[  \frac{\vec{r}_{12}{}^{2}\vec{r}_{13}{}^{2}}{\vec
{r}_{01}{}^{2}\vec{r}_{03}{}^{2}\vec{r}_{14}{}^{2}\vec{r}_{24}{}^{2}}%
-\frac{\vec{r}_{13}{}^{2}\vec{r}_{02}{}^{2}}{\vec{r}_{01}{}^{2}\vec{r}_{03}%
{}^{2}\vec{r}_{04}{}^{2}\vec{r}_{24}{}^{2}}-\frac{\vec{r}_{12}{}^{2}\vec
{r}_{34}{}^{2}}{\vec{r}_{03}{}^{2}\vec{r}_{04}{}^{2}\vec{r}_{14}{}^{2}\vec
{r}_{24}{}^{2}}\right.
\]%
\[
-\frac{\vec{r}_{03}{}^{2}\vec{r}_{12}{}^{2}}{\vec{r}_{01}{}^{2}\vec{r}_{04}%
{}^{2}\vec{r}_{34}{}^{2}\vec{r}_{02}{}^{2}}-\frac{\vec{r}_{13}{}^{2}\vec
{r}_{24}{}^{2}}{\vec{r}_{04}{}^{2}\vec{r}_{14}{}^{2}\vec{r}_{34}{}^{2}\vec
{r}_{02}{}^{2}}+\frac{\vec{r}_{12}{}^{2}\vec{r}_{13}{}^{2}}{\vec{r}_{01}{}%
^{2}\vec{r}_{14}{}^{2}\vec{r}_{34}{}^{2}\vec{r}_{02}{}^{2}}+\frac{\vec{r}%
_{03}{}^{2}-\vec{r}_{01}{}^{2}}{\vec{r}_{01}{}^{2}\vec{r}_{04}{}^{2}\vec
{r}_{34}{}^{2}}+\frac{\vec{r}_{13}{}^{2}-\vec{r}_{01}{}^{2}}{\vec{r}_{01}%
{}^{2}\vec{r}_{03}{}^{2}\vec{r}_{04}{}^{2}}%
\]%
\[
+\frac{\vec{r}_{23}{}^{2}-\vec{r}_{03}{}^{2}}{\vec{r}_{03}{}^{2}\vec{r}_{04}%
{}^{2}\vec{r}_{24}{}^{2}}+2\frac{\vec{r}_{04}{}^{2}-\vec{r}_{14}{}^{2}-\vec
{r}_{01}{}^{2}}{\vec{r}_{01}{}^{2}\vec{r}_{14}{}^{2}\vec{r}_{04}{}^{2}}%
+\frac{\vec{r}_{24}{}^{2}-\vec{r}_{14}{}^{2}}{\vec{r}_{04}{}^{2}\vec{r}_{14}%
{}^{2}\vec{r}_{02}{}^{2}}+\frac{\vec{r}_{02}{}^{2}}{\vec{r}_{01}{}^{2}\vec
{r}_{04}{}^{2}\vec{r}_{24}{}^{2}}+\frac{\vec{r}_{34}{}^{2}}{\vec{r}_{03}{}%
^{2}\vec{r}_{04}{}^{2}\vec{r}_{14}{}^{2}}%
\]%
\[
+\frac{\vec{r}_{12}{}^{2}}{\vec{r}_{04}{}^{2}\vec{r}_{14}{}^{2}\vec{r}_{24}%
{}^{2}}-\frac{\vec{r}_{13}{}^{2}}{\vec{r}_{01}{}^{2}\vec{r}_{14}{}^{2}\vec
{r}_{34}{}^{2}}+\frac{\vec{r}_{13}{}^{2}}{\vec{r}_{04}{}^{2}\vec{r}_{14}{}%
^{2}\vec{r}_{34}{}^{2}}-\frac{\vec{r}_{12}{}^{2}}{\vec{r}_{01}{}^{2}\vec
{r}_{14}{}^{2}\vec{r}_{24}{}^{2}}-\frac{\vec{r}_{13}{}^{2}}{\vec{r}_{01}{}%
^{2}\vec{r}_{03}{}^{2}\vec{r}_{14}{}^{2}}%
\]%
\begin{equation}
\left.  +\frac{\vec{r}_{12}{}^{2}}{\vec{r}_{01}{}^{2}\vec{r}_{04}{}^{2}\vec
{r}_{02}{}^{2}}-\frac{\vec{r}_{12}{}^{2}}{\vec{r}_{01}{}^{2}\vec{r}_{14}{}%
^{2}\vec{r}_{02}{}^{2}}+\frac{\vec{r}_{23}{}^{2}}{\vec{r}_{04}{}^{2}\vec
{r}_{34}{}^{2}\vec{r}_{02}{}^{2}}\right]  \ln\left(  \frac{\vec{r}_{01}{}^{2}%
}{\vec{r}_{14}{}^{2}}\right)  .
\end{equation}
The connection of our notations with the notations in \cite{Balitsky:2013fea}
is given in the appendix \ref{notation}.

\section{Construction of the kernel: gluon part}

Taking the contributions of the self-interaction on one Wilson line
(\ref{<1>23}), the connected contributions of 2 (\ref{<12>3}) and 3
(\ref{<123>}) Wilson lines from the previous section one can write for the
full contribution to the evolution of the 3QWL with 2-gluons intersecting the
shockwave%
\begin{equation}
\langle K_{NLO}\otimes B_{123}\rangle|_{2g}=\langle K_{NLO}\otimes U_{1}\cdot
U_{2}\cdot U_{3}\rangle|_{2g}=-{\frac{\alpha_{s}^{2}}{8\pi^{4}}}\!\int%
\!d\vec{r}_{0}d\vec{r}_{4}~\mathbf{G},
\end{equation}%
\begin{equation}
\mathbf{G=G}_{\langle1\rangle23}\mathbf{+G}_{1\langle2\rangle3}\mathbf{+G}%
_{12\langle3\rangle}\mathbf{+G}_{\langle12\rangle3}\mathbf{+G}_{1\langle
23\rangle}\mathbf{+G}_{\langle13\rangle2}\mathbf{+G}_{\langle123\rangle
}\mathbf{,}%
\end{equation}
Here $\langle\dots\rangle$ stands for the connected contribution, i.e.
$\mathbf{G}_{\langle1\rangle23}$ gives the contribution of the evolution of
line 1 (\ref{<1>23}), with lines 2 and 3 being spectators, $\mathbf{G}%
_{\langle12\rangle3}$ --- the connected contribution of the evolution of lines
1 and 2 (\ref{<12>3}), with line 3 being intact, and $\mathbf{G}%
_{\langle123\rangle}$ --- the fully connected contribution (\ref{<123>}). All
the rest can be obtained from them by $1\leftrightarrow2\leftrightarrow3$ transformation.

There are several useful $SU(3)$ identities, which help to reduce the number
of color structures. They are listed in the appendix \ref{identities}. First
we use (\ref{id1}) to get rid of the structure
\begin{equation}
\left(  U_{0}U_{4}{}^{\dag}U_{3}U_{0}{}^{\dag}U_{4}\right)  \cdot U_{1}\cdot
U_{2}%
\end{equation}
and the 2 ones it goes into after the $1\leftrightarrow2\leftrightarrow3$
transformations with their symmetric counterparts w.r.t. $0\leftrightarrow4$
exchange. Next we use (\ref{id2}) to eliminate 6 such contributions
antisymmetric w.r.t. $0\leftrightarrow4$ exchange as
\begin{equation}
\left(  U_{2}U_{0}{}^{\dag}U_{1}U_{4}{}^{\dag}U_{0}+U_{0}U_{4}{}^{\dag}%
U_{1}U_{0}{}^{\dag}U_{2}\right)  \cdot U_{3}\cdot U_{4}-(4\leftrightarrow0).
\end{equation}
After that we use (\ref{id3}) to express 6 structures like
\begin{equation}
\left(  U_{2}U_{4}{}^{\dag}U_{1}U_{0}{}^{\dag}U_{4}+U_{4}U_{0}{}^{\dag}%
U_{1}U_{4}{}^{\dag}U_{2}\right)  \cdot U_{0}\cdot U_{3}%
\end{equation}
and their symmetric counterparts w.r.t. $0\leftrightarrow4$ exchange through
other structures. Then via (\ref{id4}) we cancel 3 structures of the form
\begin{equation}
U_{2}\cdot U_{3}\cdot U_{4}\,\,tr\left(  U_{0}{}^{\dag}U_{1}\right)  tr\left(
U_{0}U_{4}{}^{\dag}\right)  -U_{2}\cdot U_{3}\cdot U_{0}\,\,tr\left(  U_{4}%
{}^{\dag}U_{1}\right)  tr\left(  U_{4}U_{0}{}^{\dag}\right)  .
\end{equation}
Finally, by means of (\ref{id5}) we discard the 3 nonconformal terms
proportional to
\begin{equation}
tr\left(  U_{0}U_{4}{}^{\dag}\right)  \left(  U_{1}U_{0}{}^{\dag}U_{2}%
+U_{2}U_{0}{}^{\dag}U_{1}\right)  \cdot U_{3}\cdot U_{4}-(4\leftrightarrow0)
\end{equation}
and the 2 structures they go into after the $1\leftrightarrow2\leftrightarrow
3$ transformations. Finally, we get%
\[
\mathbf{G=\{}(L_{12}+\tilde{L}_{12})\left(  U_{0}U_{4}{}^{\dag}U_{2}\right)
\cdot\left(  U_{1}U_{0}{}^{\dag}U_{4}\right)  \cdot U_{3}+L_{12}tr\left(
U_{0}U_{4}{}^{\dag}\right)  \left(  U_{1}U_{0}{}^{\dag}U_{2}\right)  \cdot
U_{3}\cdot U_{4}%
\]%
\[
+(M_{13}-M_{12}-M_{23}+M_{2})\left[  \left(  U_{0}U_{4}{}^{\dag}U_{3}\right)
\cdot\left(  U_{2}U_{0}{}^{\dag}U_{1}\right)  \cdot U_{4}+\left(  U_{1}U_{0}%
{}^{\dag}U_{2}\right)  \cdot\left(  U_{3}U_{4}{}^{\dag}U_{0}\right)  \cdot
U_{4}\right]
\]%
\begin{equation}
+(\text{all 5 permutations}\,1\leftrightarrow2\leftrightarrow
3)\}+(0\leftrightarrow4). \label{2gResult}%
\end{equation}%
\[
L_{12}=H_{3}+H_{4}-\frac{1}{2}G_{3}+\left(  1\leftrightarrow2\right)
\]%
\[
=\left[  \frac{1}{\vec{r}_{01}{}^{2}\vec{r}_{24}{}^{2}-\vec{r}_{02}{}^{2}%
\vec{r}_{14}{}^{2}}\left(  -\frac{\vec{r}_{12}{}^{4}}{8}\left(  \frac{1}%
{\vec{r}_{01}{}^{2}\vec{r}_{24}{}^{2}}+\frac{1}{\vec{r}_{02}{}^{2}\vec{r}%
_{14}{}^{2}}\right)  +\frac{\vec{r}_{12}{}^{2}}{\vec{r}_{04}{}^{2}}-\frac
{\vec{r}_{02}{}^{2}\vec{r}_{14}{}^{2}+\vec{r}_{01}{}^{2}\vec{r}_{24}{}^{2}%
}{4\vec{r}_{04}{}^{4}{}}\right)  \right.
\]%
\begin{equation}
\left.  +\frac{\vec{r}_{12}{}^{2}}{8\vec{r}_{04}{}^{2}}\left(  \frac{1}%
{\vec{r}_{02}{}^{2}\vec{r}_{14}{}^{2}}-\frac{1}{\vec{r}_{01}{}^{2}\vec{r}%
_{24}{}^{2}}\right)  \right]  \ln\left(  \frac{\vec{r}_{01}{}^{2}\vec{r}%
_{24}{}^{2}}{\vec{r}_{14}{}^{2}\vec{r}_{02}{}^{2}}\right)  +\frac{1}{2\vec
{r}_{04}{}^{4}{}}. \label{L12}%
\end{equation}%
\[
\tilde{L}_{12}=H_{1}+H_{2}-\frac{1}{2}G_{9}-\left(  1\leftrightarrow2\right)
\]%
\begin{equation}
=\frac{\vec{r}_{12}{}^{2}}{8}\left[  \frac{\vec{r}_{12}{}^{2}}{\vec{r}_{01}%
{}^{2}\vec{r}_{02}{}{}^{2}\vec{r}_{14}{}^{2}\vec{r}_{24}{}^{2}}-\frac{1}%
{\vec{r}_{01}{}^{2}\vec{r}_{04}{}^{2}\vec{r}_{24}{}^{2}}-\frac{1}{\vec{r}%
_{02}{}^{2}\vec{r}_{04}{}^{2}\vec{r}_{14}{}^{2}}\right]  \ln\left(  \frac
{\vec{r}_{01}{}^{2}\vec{r}_{24}{}^{2}}{\vec{r}_{14}{}^{2}\vec{r}_{02}{}^{2}%
}\right)  . \label{Ltilde12}%
\end{equation}%
\[
M_{12}=\frac{1}{2}\left\{  H_{1}+H_{2}-\frac{1}{2}G_{9}+\left(
1\leftrightarrow2\right)  \right\}
\]%
\begin{equation}
=\frac{\vec{r}_{12}{}^{2}}{16}\left[  \frac{\vec{r}_{12}{}^{2}}{\vec{r}_{01}%
{}^{2}\vec{r}_{02}{}{}^{2}\vec{r}_{14}{}^{2}\vec{r}_{24}{}^{2}}-\frac{1}%
{\vec{r}_{01}{}^{2}\vec{r}_{04}{}^{2}\vec{r}_{24}{}^{2}}-\frac{1}{\vec{r}%
_{02}{}^{2}\vec{r}_{04}{}^{2}\vec{r}_{14}{}^{2}}\right]  \ln\left(  \frac
{\vec{r}_{01}{}^{2}\vec{r}_{02}{}^{2}}{\vec{r}_{14}{}^{2}\vec{r}_{24}{}^{2}%
}\right)  .
\end{equation}%
\[
M_{2}=H_{2}(1\leftrightarrow2)+H_{2}(1\rightarrow2\rightarrow3\rightarrow
1)+H_{3}(1\rightarrow2\rightarrow3\rightarrow1)
\]%
\[
-H_{3}(1\leftrightarrow2)+H_{5}(1\leftrightarrow2)+H_{6}(1\leftrightarrow
2)-G_{9}(1\leftrightarrow2)
\]%
\begin{equation}
=\frac{1}{4}\left(  \frac{\vec{r}_{12}{}^{2}\vec{r}_{23}{}^{2}}{\vec{r}_{01}%
{}^{2}\vec{r}_{02}{}^{2}\vec{r}_{24}{}^{2}\vec{r}_{34}{}^{2}}-\frac{\vec
{r}_{14}{}^{2}\vec{r}_{23}{}^{2}}{\vec{r}_{01}{}^{2}\vec{r}_{04}{}^{2}\vec
{r}_{24}{}^{2}\vec{r}_{34}{}^{2}}-\frac{\vec{r}_{03}{}^{2}\vec{r}_{12}{}^{2}%
}{\vec{r}_{01}{}^{2}\vec{r}_{02}{}^{2}\vec{r}_{04}{}^{2}\vec{r}_{34}{}^{2}%
}+\frac{\vec{r}_{13}{}^{2}}{\vec{r}_{01}{}^{2}\vec{r}_{04}{}^{2}\vec{r}_{34}%
{}^{2}}\right)  \ln\left(  \frac{\vec{r}_{02}{}^{2}}{\vec{r}_{24}{}^{2}%
}\right)  . \label{M2}%
\end{equation}
These functions obey the identities%
\begin{equation}
M_{2}|_{\vec{r}_{1}\rightarrow\vec{r}_{3}}=\frac{\vec{r}_{23}{}^{2}}{4}\left(
\frac{\vec{r}_{23}{}^{2}}{\vec{r}_{03}{}^{2}\vec{r}_{02}{}^{2}\vec{r}_{24}%
{}^{2}\vec{r}_{34}{}^{2}}-\frac{1}{\vec{r}_{03}{}^{2}\vec{r}_{04}{}^{2}\vec
{r}_{24}{}{}^{2}}-\frac{1}{\vec{r}_{02}{}^{2}\vec{r}_{04}{}^{2}\vec{r}_{34}%
{}^{2}}\right)  \ln\left(  \frac{\vec{r}_{02}{}^{2}}{\vec{r}_{24}{}^{2}%
}\right)  .
\end{equation}%
\begin{equation}
M_{13}-M_{12}-M_{23}+M_{2}|_{\vec{r}_{1}\rightarrow\vec{r}_{3}}=\tilde{L}%
_{23}.
\end{equation}%
\begin{equation}
M_{13}-M_{12}-M_{23}+M_{2}|_{\vec{r}_{1}\rightarrow\vec{r}_{2}}=M_{13}%
-M_{12}-M_{23}+M_{2}|_{\vec{r}_{3}\rightarrow\vec{r}_{2}}=0.
\end{equation}
Using these identities and (\ref{3qWlidentity}) with $l=3$, we get the dipole
result%
\[
\mathbf{G}|_{\vec{r}_{1}\rightarrow\vec{r}_{3}}\mathbf{=}4(L_{32}+\tilde
{L}_{32})tr\left(  U_{0}{}^{\dag}U_{4}\right)  tr\left(  U_{3}{}^{\dag}%
U_{0}\right)  tr\left(  U_{4}{}^{\dag}U_{2}\right)
\]%
\begin{equation}
-4L_{32}tr\left(  U_{0}{}^{\dag}U_{2}U_{4}{}^{\dag}U_{0}U_{3}{}^{\dag}%
U_{4}\right)  +(0\leftrightarrow4).
\end{equation}
This expression is twice the corresponding part of the BK kernel for
$tr(U_{2}U_{3}^{\dag}).$

The only UV divergent term in (\ref{2gResult}) is the term proportional to
$L_{12}.$ This term has the same coordinate structure as the corresponding
term in the dipole kernel. Therefore we can do the same subtraction as in the
dipole case. Using (\ref{IDENTITY}), we get
\[
\left(  U_{0}U_{4}{}^{\dag}U_{2}\right)  \cdot\left(  U_{1}U_{0}{}^{\dag}%
U_{4}\right)  \cdot U_{3}+tr\left(  U_{0}U_{4}{}^{\dag}\right)  \left(
U_{1}U_{0}{}^{\dag}U_{2}\right)  \cdot U_{3}\cdot U_{4}+(2\leftrightarrow
1)|_{\vec{r}_{0}\rightarrow\vec{r}_{4}}%
\]%
\[
=3[tr\left(  U_{1}U_{4}{}^{\dag}\right)  U_{2}\cdot U_{3}\cdot U_{4}+tr\left(
U_{2}U_{4}{}^{\dag}\right)  U_{1}\cdot U_{3}\cdot U_{4}-tr\left(  U_{3}U_{4}%
{}^{\dag}\right)  U_{1}\cdot U_{2}\cdot U_{4}]-U_{1}\cdot U_{2}\cdot U_{3}%
\]%
\begin{equation}
=\frac{3}{2}[B_{144}B_{234}+B_{244}B_{134}-B_{344}B_{124}]-B_{123},
\end{equation}%
\begin{equation}
B_{123}=U_{1}\cdot U_{2}\cdot U_{3}=\varepsilon^{i^{\prime}j^{\prime}%
h^{\prime}}\varepsilon_{ijh}U_{1i^{\prime}}^{i}U_{2j^{\prime}}^{j}%
U_{3h^{\prime}}^{h}.
\end{equation}
Therefore we can separate the result into the UV finite and divergent parts%
\begin{equation}
\langle K_{NLO}\otimes B_{123}\rangle|_{2g}=-{\frac{\alpha_{s}^{2}}{8\pi^{4}}%
}\!\int\!d\vec{r}_{0}d\vec{r}_{4}~\mathbf{G}_{finite}-{\frac{\alpha_{s}^{2}%
}{8\pi^{3}}}\!\int\!d\vec{r}_{0}~\mathbf{G}_{UV},
\label{2gResultWithSubtruction}%
\end{equation}%
\[
\mathbf{G}_{finite}\mathbf{=\{}\tilde{L}_{12}\left(  U_{0}U_{4}{}^{\dag}%
U_{2}\right)  \cdot\left(  U_{1}U_{0}{}^{\dag}U_{4}\right)  \cdot U_{3}%
\]%
\[
\mathbf{+}L_{12}\left[  \left(  U_{0}U_{4}{}^{\dag}U_{2}\right)  \cdot\left(
U_{1}U_{0}{}^{\dag}U_{4}\right)  \cdot U_{3}+tr\left(  U_{0}U_{4}{}^{\dag
}\right)  \left(  U_{1}U_{0}{}^{\dag}U_{2}\right)  \cdot U_{3}\cdot U_{4}%
\frac{{}}{{}}\right.
\]%
\[
\left.  -\frac{3}{4}[B_{144}B_{234}+B_{244}B_{134}-B_{344}B_{124}]+\frac{1}%
{2}B_{123}\right]
\]%
\[
+(M_{13}-M_{12}-M_{23}+M_{2})\left[  \left(  U_{0}U_{4}{}^{\dag}U_{3}\right)
\cdot\left(  U_{2}U_{0}{}^{\dag}U_{1}\right)  \cdot U_{4}+\left(  U_{1}U_{0}%
{}^{\dag}U_{2}\right)  \cdot\left(  U_{3}U_{4}{}^{\dag}U_{0}\right)  \cdot
U_{4}\right]
\]%
\begin{equation}
+(\text{all 5 permutations}\,1\leftrightarrow2\leftrightarrow
3)\}+(0\leftrightarrow4). \label{Gfinite}%
\end{equation}
And $\mathbf{G}_{UV}$ is included into the term describing the contribution
with one gluon crossing the shockwave in \cite{Balitsky:2013fea}.

The contribution of the diagrams with 1 gluon intersecting the shockwave,
which are not proportional to the $\beta$-function one can take from (5.27) in
\cite{Grabovsky:2013mba}
\[
\langle\tilde{K}_{NLO}\otimes B_{123}\rangle|_{1g}=\frac{\alpha_{s}^{2}%
}{\left(  2\pi\right)  ^{3}}\int d\vec{r}_{0}\left[  \frac{\left(  \vec
{r}_{10}\vec{r}_{20}\right)  }{\vec{r}_{10}^{\,\,2}\vec{r}_{20}^{\,\,\,2}%
}-\frac{\left(  \vec{r}_{30}\vec{r}_{20}\right)  }{\vec{r}_{30}^{\,2}\vec
{r}_{20}^{\,\,\,2}}\right]  \ln\frac{\vec{r}_{30}^{\,\,2}}{\vec{r}%
_{31}^{\,\,2}}\ln\frac{\vec{r}_{10}^{\,\,2}}{\vec{r}_{31}^{\,\,2}}\left(
B_{100}B_{320}-B_{300}B_{210}\right)
\]%
\[
+\frac{\alpha_{s}^{2}}{\left(  2\pi\right)  ^{3}}\int d\vec{r}_{0}\left[
\frac{1}{\vec{r}_{10}^{\,\,2}}-\frac{\left(  \vec{r}_{30}\vec{r}_{10}\right)
}{\vec{r}_{30}^{\,2}\vec{r}_{10}^{\,\,\,2}}\right]  \ln\frac{\vec{r}%
_{30}^{\,\,2}}{\vec{r}_{31}^{\,\,2}}\ln\frac{\vec{r}_{10}^{\,\,2}}{\vec
{r}_{31}^{\,\,2}}\left(  B_{123}-\frac{1}{2}\left[  3B_{100}B_{320}%
+B_{300}B_{120}-B_{200}B_{130}\right]  \right)
\]%
\[
+\frac{\alpha_{s}^{2}}{\left(  2\pi\right)  ^{3}}\int d\vec{r}_{0}\left[
\frac{\left(  \vec{r}_{10}\vec{r}_{30}\right)  }{\vec{r}_{10}^{\,2}\vec
{r}_{30}^{\,\,\,2}}-\frac{1}{\vec{r}_{30}^{\,2}}\right]  \ln\frac{\vec{r}%
_{30}^{\,\,2}}{\vec{r}_{31}^{\,\,2}}\ln\frac{\vec{r}_{10}^{\,\,2}}{\vec
{r}_{31}^{\,\,2}}\left(  \frac{1}{2}\left[  3B_{300}B_{120}+B_{100}%
B_{320}-B_{200}B_{130}\right]  -B_{123}\right)
\]%
\begin{equation}
+(2\leftrightarrow1)+(2\leftrightarrow3).
\end{equation}
This term has the correct dipole limit (see (5.28) in \cite{Grabovsky:2013mba}).

The contribution proportional to $\beta$-function reads (from
\cite{Balitsky:2013fea})%
\[
\langle\tilde{K}_{NLO}\otimes B_{123}\rangle|_{1g}^{\beta}=\left[
-\frac{\alpha_{s}^{2}}{\left(  2\pi\right)  ^{3}}\frac{11}{2}\int d\vec{r}%
_{0}\left[  \ln\left(  \frac{\vec{r}_{01}^{\,\,2}}{\vec{r}_{02}^{\,\,2}%
}\right)  \left(  \frac{1}{\vec{r}_{02}^{\,\,2}}-\frac{1}{\vec{r}_{01}%
^{\,\,2}}\right)  -\frac{\vec{r}_{12}^{\,\,2}}{\vec{r}_{01}^{\,\,2}\vec
{r}_{02}^{\,\,2}}\ln\left(  \frac{\vec{r}_{12}^{\,\,2}}{\tilde{\mu}_{g}^{2}%
}\right)  \right.  \right.
\]%
\[
\left.  +\frac{1}{\vec{r}_{02}^{\,\,2}}\ln\left(  \frac{\vec{r}_{02}^{\,\,2}%
}{\tilde{\mu}_{g}^{2}}\right)  +\frac{1}{\vec{r}_{01}^{\,\,2}}\ln\left(
\frac{\vec{r}_{01}^{\,\,2}}{\tilde{\mu}_{g}^{2}}\right)  \right]
\]%
\[
\left.  \times\left(  U_{0}\cdot U_{3}\cdot(U_{2}U_{0}^{\dag}U_{1})+U_{0}\cdot
U_{3}\cdot(U_{1}U_{0}^{\dag}U_{2})+\frac{2}{3}U_{1}\cdot U_{2}\cdot
U_{3}\right)  +(1\leftrightarrow3)+(2\leftrightarrow3)\right]
\]%
\begin{equation}
+\left[  \frac{\alpha_{s}^{2}}{\left(  2\pi\right)  ^{3}}11\int\frac{d\vec
{r}_{0}}{\vec{r}_{01}^{\,\,2}}\ln\left(  \frac{\vec{r}_{01}^{\,\,2}}%
{\tilde{\mu}_{g}^{2}}\right)  \left(  U_{0}\cdot U_{2}\cdot U_{3}tr(U_{1}%
U_{0}^{\dag})-\frac{1}{3}U_{1}\cdot U_{2}\cdot U_{3}\right)
+(1\leftrightarrow3)+(1\leftrightarrow2)\right]  ,
\end{equation}%
\begin{equation}
\frac{11}{3}\ln\frac{1}{\tilde{\mu}_{g}^{2}}=\frac{11}{3}\ln\left(  \frac
{\mu^{2}}{4e^{2\psi\left(  1\right)  }}\right)  +\frac{67}{9}-\frac{\pi^{2}%
}{3}. \label{mu}%
\end{equation}
Or, after some algebra%
\[
\langle\tilde{K}_{NLO}\otimes B_{123}\rangle|_{1g}^{\beta}=-\frac{\alpha
_{s}^{2}}{\left(  2\pi\right)  ^{3}}\frac{11}{6}\int d\vec{r}_{0}\left[
\ln\left(  \frac{\vec{r}_{01}^{\,\,2}}{\vec{r}_{02}^{\,\,2}}\right)  \left(
\frac{1}{\vec{r}_{02}^{\,\,2}}-\frac{1}{\vec{r}_{01}^{\,\,2}}\right)
-\frac{\vec{r}_{12}^{\,\,2}}{\vec{r}_{01}^{\,\,2}\vec{r}_{02}^{\,\,2}}%
\ln\left(  \frac{\vec{r}_{12}^{\,\,2}}{\tilde{\mu}_{g}^{2}}\right)  \right]
\]%
\begin{equation}
\times\left(  \frac{3}{2}(B_{100}B_{230}+B_{200}B_{130}-B_{300}B_{210}%
)-B_{123}\right)  +(1\leftrightarrow3)+(2\leftrightarrow3). \label{beta-term}%
\end{equation}
It also has the correct dipole limit%
\[
\langle\tilde{K}_{NLO}\otimes B_{122}\rangle|_{1g}^{\beta}=-\frac{\alpha
_{s}^{2}}{\left(  2\pi\right)  ^{3}}\frac{11}{3}\int d\vec{r}_{0}\left[
\ln\left(  \frac{\vec{r}_{01}^{\,\,2}}{\vec{r}_{02}^{\,\,2}}\right)  \left(
\frac{1}{\vec{r}_{02}^{\,\,2}}-\frac{1}{\vec{r}_{01}^{\,\,2}}\right)
-\frac{\vec{r}_{12}^{\,\,2}}{\vec{r}_{01}^{\,\,2}\vec{r}_{02}^{\,\,2}}%
\ln\left(  \frac{\vec{r}_{12}^{\,\,2}}{\tilde{\mu}_{g}^{2}}\right)  \right]
\]%
\begin{equation}
\times\left(  \frac{3}{2}B_{100}B_{220}-B_{122}\right)  .
\end{equation}
and it matches the BFKL kernel \cite{Fadin:2009gh}. Therefore the real part of
the whole kernel reads%
\begin{equation}
\langle K_{NLO}\otimes B_{123}\rangle|_{\operatorname{real}}=-{\frac
{\alpha_{s}^{2}}{8\pi^{4}}}\!\int\!d\vec{r}_{0}d\vec{r}_{4}~\mathbf{G}%
_{finite}-{\frac{\alpha_{s}^{2}}{8\pi^{3}}}\!\int\!d\vec{r}_{0}~\mathbf{G}%
_{\operatorname{real}},
\end{equation}%
\[
\mathbf{G}_{\operatorname{real}}\mathbf{=-}\frac{1}{2}\left[  \frac{\left(
\vec{r}_{10}\vec{r}_{20}\right)  }{\vec{r}_{10}^{\,\,2}\vec{r}_{20}^{\,\,\,2}%
}-\frac{\left(  \vec{r}_{30}\vec{r}_{20}\right)  }{\vec{r}_{30}^{\,2}\vec
{r}_{20}^{\,\,\,2}}\right]  \ln\frac{\vec{r}_{30}^{\,\,2}}{\vec{r}%
_{31}^{\,\,2}}\ln\frac{\vec{r}_{10}^{\,\,2}}{\vec{r}_{31}^{\,\,2}}\left(
B_{100}B_{320}-B_{300}B_{210}\right)
\]%
\[
-\left[  \frac{1}{\vec{r}_{10}^{\,\,2}}-\frac{\left(  \vec{r}_{30}\vec{r}%
_{10}\right)  }{\vec{r}_{30}^{\,2}\vec{r}_{10}^{\,\,\,2}}\right]  \ln
\frac{\vec{r}_{30}^{\,\,2}}{\vec{r}_{31}^{\,\,2}}\ln\frac{\vec{r}_{10}%
^{\,\,2}}{\vec{r}_{31}^{\,\,2}}\left(  B_{123}-\frac{1}{2}\left[
3B_{100}B_{320}+B_{300}B_{120}-B_{200}B_{130}\right]  \right)
\]%
\[
+\frac{11}{12}\left[  \ln\left(  \frac{\vec{r}_{01}^{\,\,2}}{\vec{r}%
_{02}^{\,\,2}}\right)  \left(  \frac{1}{\vec{r}_{02}^{\,\,2}}-\frac{1}{\vec
{r}_{01}^{\,\,2}}\right)  -\frac{\vec{r}_{12}^{\,\,2}}{\vec{r}_{01}%
^{\,\,2}\vec{r}_{02}^{\,\,2}}\ln\left(  \frac{\vec{r}_{12}^{\,\,2}}{\tilde
{\mu}_{g}^{2}}\right)  \right]
\]%
\[
\times\left(  \frac{3}{2}(B_{100}B_{230}+B_{200}B_{130}-B_{300}B_{210}%
)-B_{123}\right)
\]%
\begin{equation}
+(\text{all 5 permutations}\,1\leftrightarrow2\leftrightarrow3), \label{Greal}%
\end{equation}
and $\mathbf{G}_{finite}$ is defined in (\ref{Gfinite}). If we put $\vec
{r}_{2}=\vec{r}_{3}$ here, we get the dipole result (see (100) in
\cite{Balitsky:2008zza})%
\[
\mathbf{G}_{\operatorname{real}}|_{\vec{r}_{2}=\vec{r}_{3}}\mathbf{=}\left\{
\frac{11}{3}\left[  \ln\left(  \frac{\vec{r}_{01}^{\,\,2}}{\vec{r}%
_{02}^{\,\,2}}\right)  \left(  \frac{1}{\vec{r}_{02}^{\,\,2}}-\frac{1}{\vec
{r}_{01}^{\,\,2}}\right)  -\frac{\vec{r}_{12}^{\,\,2}}{\vec{r}_{01}%
^{\,\,2}\vec{r}_{02}^{\,\,2}}\ln\left(  \frac{\vec{r}_{12}^{\,\,2}}{\tilde
{\mu}_{g}^{2}}\right)  \right]  +2\frac{\vec{r}_{12}^{\,\,2}}{\vec{r}%
_{20}^{\,2}\vec{r}_{10}^{\,\,\,2}}\ln\frac{\vec{r}_{20}^{\,\,2}}{\vec{r}%
_{21}^{\,\,2}}\ln\frac{\vec{r}_{10}^{\,\,2}}{\vec{r}_{21}^{\,\,2}}\right\}
\]%
\begin{equation}
\times\left(  \frac{3}{2}B_{100}B_{220}-B_{122}\right)  .
\end{equation}
Finally, from the condition that the kernel must vanish without the shockwave
(if all the $B=6)$ and that the virtual contribution is proportional to
$B_{123},$ we get the total kernel%
\begin{equation}
\langle K_{NLO}\otimes B_{123}\rangle=-{\frac{\alpha_{s}^{2}}{8\pi^{4}}}%
\!\int\!d\vec{r}_{0}d\vec{r}_{4}~\mathbf{G}_{finite}-{\frac{\alpha_{s}^{2}%
}{8\pi^{3}}}\!\int\!d\vec{r}_{0}~\mathbf{G}^{\prime},
\end{equation}%
\[
\mathbf{G}^{\prime}\mathbf{=}\frac{1}{2}\left[  \frac{\vec{r}_{13}^{\,\,2}%
}{\vec{r}_{10}^{\,\,2}\vec{r}_{30}^{\,\,\,2}}-\frac{\vec{r}_{32}^{\,2}}%
{\vec{r}_{30}^{\,2}\vec{r}_{20}^{\,\,\,2}}\right]  \ln\frac{\vec{r}%
_{20}^{\,\,2}}{\vec{r}_{21}^{\,\,2}}\ln\frac{\vec{r}_{10}^{\,\,2}}{\vec
{r}_{21}^{\,\,2}}\left(  B_{100}B_{320}-B_{200}B_{310}\right)
\]%
\[
-\frac{\vec{r}_{12}^{\,2}}{\vec{r}_{10}^{\,2}\vec{r}_{20}^{\,\,\,2}}\ln
\frac{\vec{r}_{10}^{\,\,2}}{\vec{r}_{12}^{\,\,2}}\ln\frac{\vec{r}_{20}%
^{\,\,2}}{\vec{r}_{12}^{\,\,2}}\left(  9B_{123}-\frac{1}{2}\left[  2\left(
B_{100}B_{320}+B_{200}B_{130}\right)  -B_{300}B_{120}\right]  \right)
\]%
\[
+\frac{11}{6}\left[  \ln\left(  \frac{\vec{r}_{01}^{\,\,2}}{\vec{r}%
_{02}^{\,\,2}}\right)  \left(  \frac{1}{\vec{r}_{02}^{\,\,2}}-\frac{1}{\vec
{r}_{01}^{\,\,2}}\right)  -\frac{\vec{r}_{12}^{\,\,2}}{\vec{r}_{01}%
^{\,\,2}\vec{r}_{02}^{\,\,2}}\ln\left(  \frac{\vec{r}_{12}^{\,\,2}}{\tilde
{\mu}_{g}^{2}}\right)  \right]
\]%
\begin{equation}
\times\left(  \frac{3}{2}(B_{100}B_{230}+B_{200}B_{130}-B_{300}B_{210}%
)-9B_{123}\right)  +(1\leftrightarrow3)+(2\leftrightarrow3). \label{Gall}%
\end{equation}
It differs from (\ref{Greal}) in the coefficients of $B_{123}$'s which turn
into 9's; $\mathbf{G}_{finite}$ is defined in (\ref{Gfinite}).

\section{Construction of the kernel: quark part}

One can take the quark contribution to the NLO evolution of 3QWL from
\cite{Balitsky:2013fea}. The contribution with 2 quarks intersecting the
shockwave without subtraction reads%
\begin{equation}
\langle K_{NLO}\otimes B_{123}\rangle|_{2g}^{q}=-{\frac{\alpha_{s}^{2}n_{f}%
}{8\pi^{4}}}\!\int\!d\vec{r}_{0}d\vec{r}_{4}~\mathbf{G}^{q},
\end{equation}%
\[
\mathbf{G}^{q}=\left[  \left(  (U_{1}U_{0}{}^{\dag}U_{4}+U_{4}U_{0}{}^{\dag
}U_{1})\cdot U_{2}\cdot U_{3}-3U_{2}\cdot U_{3}\cdot U_{4}tr(U_{0}{}^{\dag
}U_{1})-\frac{1}{3}U_{1}\cdot U_{2}\cdot U_{3}tr(U_{0}{}^{\dag}U_{4})\right)
\right.
\]%
\[
\left.  \times\frac{2}{3}\frac{1}{\vec{r}_{04}{}^{4}}\left\{  \frac{(\vec
{r}_{14}\vec{r}_{01})}{\vec{r}_{14}{}^{2}-\vec{r}_{01}{}^{2}}\ln\left(
\frac{\vec{r}_{14}{}^{2}}{\vec{r}_{01}{}^{2}}\right)  +1\right\}
+(1\leftrightarrow2)+(1\leftrightarrow3)\right]
\]%
\[
+\left[  \left(  \frac{1}{3}(U_{1}U_{0}{}^{\dag}U_{4}+U_{4}U_{0}{}^{\dag}%
U_{1})\cdot U_{2}\cdot U_{3}-\frac{1}{9}U_{1}\cdot U_{2}\cdot U_{3}tr(U_{0}%
{}^{\dag}U_{4})\right.  \right.
\]%
\[
\left.  \frac{{}}{{}}+(U_{1}U_{0}{}^{\dag}U_{2})\cdot U_{3}\cdot
U_{4}+(1\leftrightarrow2)\right)  \left(  \frac{1}{\vec{r}_{04}{}^{4}}\left\{
\frac{(\vec{r}_{14}\vec{r}_{01})}{\vec{r}_{14}{}^{2}-\vec{r}_{01}{}^{2}}%
\ln\left(  \frac{\vec{r}_{14}{}^{2}}{\vec{r}_{01}{}^{2}}\right)  +1\right\}
+\frac{L_{12}^{q}}{2}+(1\leftrightarrow2)\right)
\]%
\begin{equation}
\left.  \frac{{}}{{}}+(1\leftrightarrow3)+(2\leftrightarrow3)\right]  ,
\end{equation}
where%
\begin{equation}
L_{12}^{q}=\frac{1}{\vec{r}_{04}{}^{4}}\left\{  \frac{\vec{r}_{02}{}^{2}%
\vec{r}_{14}{}^{2}+\vec{r}_{01}{}^{2}\vec{r}_{24}{}^{2}-\vec{r}_{04}{}^{2}%
\vec{r}_{12}{}^{2}}{2(\vec{r}_{02}{}^{2}\vec{r}_{14}{}^{2}-\vec{r}_{01}{}%
^{2}\vec{r}_{24}{}^{2})}\ln\left(  \frac{\vec{r}_{02}{}^{2}\vec{r}_{14}{}^{2}%
}{\vec{r}_{01}{}^{2}\vec{r}_{24}{}^{2}}\right)  -1\right\}  . \label{Lq}%
\end{equation}
Using identity (\ref{id10}) one can see that this contribution is conformally
invariant, indeed%
\[
\mathbf{G}^{q}=\frac{1}{2}\left\{  \left(  \frac{1}{3}(U_{1}U_{0}{}^{\dag
}U_{4}+U_{4}U_{0}{}^{\dag}U_{1})\cdot U_{2}\cdot U_{3}-\frac{1}{9}U_{1}\cdot
U_{2}\cdot U_{3}tr(U_{0}{}^{\dag}U_{4})\right.  \right.
\]%
\begin{equation}
\left.  \left.  \frac{{}}{{}}+(U_{1}U_{0}{}^{\dag}U_{2})\cdot U_{3}\cdot
U_{4}+(1\leftrightarrow2)\right)  +(0\leftrightarrow4)\right\}  L_{12}%
^{q}+(1\leftrightarrow3)+(2\leftrightarrow3).
\end{equation}
In the dipole limit $\vec{r}_{3}\rightarrow\vec{r}_{2},$ one has%
\[
\mathbf{G}^{q}|_{\vec{r}_{3}\rightarrow\vec{r}_{2}}=\frac{2}{3}L_{12}%
^{q}\left\{  \frac{1}{3}tr(U_{0}{}^{\dag}U_{4})tr(U_{2}{}^{\dag}%
U_{1})+3tr(U_{0}{}^{\dag}U_{1})tr(U_{2}{}^{\dag}U_{4})\right.
\]%
\begin{equation}
\left.  -tr(U_{0}{}^{\dag}U_{1}U_{2}{}^{\dag}U_{4})-tr(U_{0}{}^{\dag}%
U_{4}U_{2}{}^{\dag}U_{1})+(0\leftrightarrow4)\right\}  ,
\end{equation}
which is twice the corresponding part of the BK kernel \cite{Balitsky:2006wa}.
Therefore one can do the same subtraction as in the BK case
\begin{equation}
\mathbf{G}^{q}=\mathbf{G}_{finite}^{q}+\mathbf{G}_{UV}^{q},
\end{equation}%
\[
\mathbf{G}_{finite}^{q}=\frac{1}{2}\left\{  \left(  \frac{1}{3}(U_{1}U_{0}%
{}^{\dag}U_{4}+U_{4}U_{0}{}^{\dag}U_{1})\cdot U_{2}\cdot U_{3}-\frac{1}%
{9}B_{123}tr(U_{0}{}^{\dag}U_{4})+(U_{1}U_{0}{}^{\dag}U_{2})\cdot U_{3}\cdot
U_{4}\right.  \right.
\]%
\[
\left.  \left.  \frac{{}}{{}}+\frac{1}{6}B_{123}-\frac{1}{4}(B_{013}%
B_{002}+B_{001}B_{023}-B_{012}B_{003})+(1\leftrightarrow2)\right)
+(0\leftrightarrow4)\right\}  L_{12}^{q}%
\]%
\begin{equation}
+(1\leftrightarrow3)+(2\leftrightarrow3), \label{GfiniteQ}%
\end{equation}
and $\mathbf{G}_{UV}^{q}$ is taken into account in the contribution with one
gluon crossing the shockwave in \cite{Balitsky:2013fea}. The latter
contribution one can restore from the gluon part via the substitutions%
\begin{equation}
\frac{11}{3}\rightarrow\beta=\left(  \frac{11}{3}-\frac{2}{3}\frac{n_{f}}%
{3}\right)  ,
\end{equation}%
\begin{equation}
\frac{11}{3}\ln\frac{1}{\tilde{\mu}_{g}^{2}}=\frac{11}{3}\ln\left(  \frac
{\mu^{2}}{4e^{2\psi\left(  1\right)  }}\right)  +\frac{67}{9}-\frac{\pi^{2}%
}{3}\rightarrow\beta\ln\frac{1}{\tilde{\mu}^{2}},
\end{equation}%
\begin{equation}
\beta\ln\frac{1}{\tilde{\mu}^{2}}=\left(  \frac{11}{3}-\frac{2}{3}\frac{n_{f}%
}{3}\right)  \ln\left(  \frac{\mu^{2}}{4e^{2\psi\left(  1\right)  }}\right)
+\frac{67}{9}-\frac{\pi^{2}}{3}-\frac{10}{9}\frac{n_{f}}{3}.
\label{mu-tilde-through-mu}%
\end{equation}
As a result the full kernel in QCD reads%
\begin{equation}
\langle K_{NLO}\otimes B_{123}\rangle=-{\frac{\alpha_{s}^{2}}{8\pi^{4}}}%
\!\int\!d\vec{r}_{0}d\vec{r}_{4}\,\left(  \mathbf{G}_{finite}+n_{f}%
\mathbf{G}_{finite}^{q}\right)  -{\frac{\alpha_{s}^{2}}{8\pi^{3}}}%
\!\int\!d\vec{r}_{0}~\mathbf{G}^{q\prime}, \label{KallWithquarks}%
\end{equation}%
\[
\mathbf{G}^{q\prime}\mathbf{=}\frac{1}{2}\left[  \frac{\vec{r}_{13}^{\,\,2}%
}{\vec{r}_{10}^{\,\,2}\vec{r}_{30}^{\,\,\,2}}-\frac{\vec{r}_{32}^{\,2}}%
{\vec{r}_{30}^{\,2}\vec{r}_{20}^{\,\,\,2}}\right]  \ln\frac{\vec{r}%
_{20}^{\,\,2}}{\vec{r}_{21}^{\,\,2}}\ln\frac{\vec{r}_{10}^{\,\,2}}{\vec
{r}_{21}^{\,\,2}}\left(  B_{100}B_{320}-B_{200}B_{310}\right)
\]%
\[
-\frac{\vec{r}_{12}^{\,2}}{\vec{r}_{10}^{\,2}\vec{r}_{20}^{\,\,\,2}}\ln
\frac{\vec{r}_{10}^{\,\,2}}{\vec{r}_{12}^{\,\,2}}\ln\frac{\vec{r}_{20}%
^{\,\,2}}{\vec{r}_{12}^{\,\,2}}\left(  9B_{123}-\frac{1}{2}\left[  2\left(
B_{100}B_{320}+B_{200}B_{130}\right)  -B_{300}B_{120}\right]  \right)
\]%
\[
+\frac{\beta}{2}\left[  \ln\left(  \frac{\vec{r}_{01}^{\,\,2}}{\vec{r}%
_{02}^{\,\,2}}\right)  \left(  \frac{1}{\vec{r}_{02}^{\,\,2}}-\frac{1}{\vec
{r}_{01}^{\,\,2}}\right)  -\frac{\vec{r}_{12}^{\,\,2}}{\vec{r}_{01}%
^{\,\,2}\vec{r}_{02}^{\,\,2}}\ln\left(  \frac{\vec{r}_{12}^{\,\,2}}{\tilde
{\mu}^{2}}\right)  \right]
\]%
\begin{equation}
\times\left(  \frac{3}{2}(B_{100}B_{230}+B_{200}B_{130}-B_{300}B_{210}%
)-9B_{123}\right)  +(1\leftrightarrow3)+(2\leftrightarrow3).
\end{equation}
Here $\mathbf{G}_{finite}$ is defined in (\ref{Gfinite}) and $\mathbf{G}%
_{finite}^{q}$ is defined in (\ref{GfiniteQ}).

\section{Evolution equation for composite 3QWL operator}

In this section we consider only the gluon part of the kernel since the quark
one is quasi-conformal. To construct composite conformal operators we will use
the prescription \cite{Balitsky:2009xg} (see also Ref.
\cite{NLOJIMWLKonformal})%
\begin{equation}
O^{conf}=O+\frac{1}{2}\frac{\partial O}{\partial\eta}\left\vert _{\frac
{\vec{r}_{mn}^{\,\,2}}{\vec{r}_{im}^{\,\,2}\vec{r}_{in}^{\,\,2}}%
\rightarrow\frac{\vec{r}_{mn}^{\,\,2}}{\vec{r}_{im}^{\,\,2}\vec{r}%
_{in}^{\,\,2}}\ln\left(  \frac{\vec{r}_{mn}^{\,\,2}a}{\vec{r}_{im}^{\,\,2}%
\vec{r}_{in}^{\,\,2}}\right)  }\right.  , \label{model}%
\end{equation}
where $a$ is an arbitrary constant. For the conformal 3QWL operator we have
the following ansatz%
\[
B_{123}^{conf}=B_{123}+\frac{\alpha_{s}3}{8\pi^{2}}\int d\vec{r}_{4}\left[
\frac{\vec{r}_{12}^{\,\,2}}{\vec{r}_{41}^{\,\,2}\vec{r}_{42}^{\,\,2}}%
\ln\left(  \frac{a\vec{r}_{12}^{\,\,2}}{\vec{r}_{41}^{\,\,2}\vec{r}%
_{42}^{\,\,2}}\right)  \right.
\]%
\begin{equation}
\left.  \times(-B_{123}+\frac{1}{6}(B_{144}B_{324}+B_{244}B_{314}%
-B_{344}B_{214}))+(1\leftrightarrow3)+(2\leftrightarrow3)\right]  .
\label{anzatz}%
\end{equation}
If we put $\vec{r}_{2}=\vec{r}_{3},$ then
\[
B_{122}^{conf}=B_{122}+\frac{\alpha_{s}3}{4\pi^{2}}\int d\vec{r}_{4}\frac
{\vec{r}_{12}^{\,\,2}}{\vec{r}_{41}^{\,\,2}\vec{r}_{42}^{\,\,2}}\ln\left(
\frac{a\vec{r}_{12}^{\,\,2}}{\vec{r}_{41}^{\,\,2}\vec{r}_{42}^{\,\,2}}\right)
(-B_{122}+\frac{1}{6}B_{144}B_{224}),
\]
or%
\begin{equation}
tr(U_{1}U_{2}^{\dag})^{conf}=tr(U_{1}U_{2}^{\dag})+\frac{\alpha_{s}}{4\pi^{2}%
}\int d\vec{r}_{4}\frac{\vec{r}_{12}^{\,\,2}}{\vec{r}_{41}^{\,\,2}\vec{r}%
_{42}^{\,\,2}}\ln\left(  \frac{a\vec{r}_{12}^{\,\,2}}{\vec{r}_{41}^{\,\,2}%
\vec{r}_{42}^{\,\,2}}\right)  (tr(U_{1}U_{4}^{\dag})tr(U_{4}U_{2}^{\dag
})-3tr(U_{1}U_{2}^{\dag})),
\end{equation}
which is exactly the composite dipole operator of \cite{Balitsky:2009xg}.
Using SU(3) identity (\ref{IDENTITY}) one can rewrite (\ref{anzatz}) as%
\[
B_{123}^{conf}=B_{123}+\frac{\alpha_{s}}{8\pi^{2}}\int d\vec{r}_{4}\left[
\frac{\vec{r}_{12}^{\,\,2}}{\vec{r}_{41}^{\,\,2}\vec{r}_{42}^{\,\,2}}%
\ln\left(  \frac{a\vec{r}_{12}^{\,\,2}}{\vec{r}_{41}^{\,\,2}\vec{r}%
_{42}^{\,\,2}}\right)  (\left(  U_{2}U_{4}^{\dag}U_{1}+U_{1}U_{4}^{\dag}%
U_{2}\right)  \cdot U_{4}\cdot U_{3}-2B_{123})\right.
\]%
\begin{equation}
\left.  \frac{{}}{{}}+(1\leftrightarrow3)+(2\leftrightarrow3)\right]  .
\end{equation}
Then as in \cite{Balitsky:2009xg}, for $(-B_{123}+\frac{1}{6}(B_{144}%
B_{324}+B_{244}B_{314}-B_{344}B_{214}))$ we have%
\[
(-3B_{123}+\frac{1}{2}(B_{144}B_{324}+B_{244}B_{314}-B_{344}B_{214}))^{conf}%
\]%
\[
=(-3B_{123}+\frac{1}{2}(B_{144}B_{324}+B_{244}B_{314}-B_{344}B_{214}))
\]%
\[
+\frac{\alpha_{s}}{8\pi^{2}}\int d\vec{r}_{0}\left(  A_{34}\frac{\vec{r}%
_{34}^{\,\,2}}{\vec{r}_{03}^{\,\,2}\vec{r}_{04}^{\,\,2}}\ln\left(  \frac
{\vec{r}_{34}^{\,\,2}a}{\vec{r}_{03}^{\,\,2}\vec{r}_{04}^{\,\,2}}\right)
+A_{13}\frac{\vec{r}_{13}^{\,2}}{\vec{r}_{03}^{\,\,2}\vec{r}_{01}^{\,\,2}}%
\ln\left(  \frac{\vec{r}_{13}^{\,2}a}{\vec{r}_{03}^{\,\,2}\vec{r}_{01}%
^{\,\,2}}\right)  +A_{23}\frac{\vec{r}_{23}^{\,\,2}}{\vec{r}_{03}^{\,\,2}%
\vec{r}_{02}^{\,\,2}}\ln\left(  \frac{\vec{r}_{23}^{\,\,2}a}{\vec{r}%
_{03}^{\,\,2}\vec{r}_{02}^{\,\,2}}\right)  \right.
\]%
\begin{equation}
\left.  +A_{14}\frac{\vec{r}_{14}^{\,\,2}}{\vec{r}_{01}^{\,\,2}\vec{r}%
_{04}^{\,\,2}}\ln\left(  \frac{\vec{r}_{14}^{\,\,2}a}{\vec{r}_{01}^{\,\,2}%
\vec{r}_{04}^{\,\,2}}\right)  +A_{24}\frac{\vec{r}_{24}^{\,\,2}}{\vec{r}%
_{02}^{\,\,2}\vec{z}_{04}^{\,\,2}}\ln\left(  \frac{\vec{r}_{24}^{\,\,2}a}%
{\vec{r}_{02}^{\,\,2}\vec{z}_{04}^{\,\,2}}\right)  +A_{12}\frac{\vec{r}%
_{12}^{\,\,2}}{\vec{r}_{01}^{\,\,2}\vec{r}_{02}^{\,\,2}}\ln\left(  \frac
{\vec{r}_{12}^{\,\,2}a}{\vec{r}_{01}^{\,\,2}\vec{r}_{02}^{\,\,2}}\right)
\right)  , \label{anzatz1}%
\end{equation}
where the functions $A$ are calculated in appendix \ref{4pointOperator}
(\ref{A34}--\ref{A23}) according to model (\ref{model}). Therefore the
evolution equation for $B_{123}^{conf}$ turns into%
\[
\frac{\partial B_{123}^{conf}}{\partial\eta}=\frac{\alpha_{s}3}{4\pi^{2}}\int
d\vec{r}_{4}\left[  \frac{\vec{r}_{12}^{\,\,2}}{\vec{r}_{41}^{\,\,2}\vec
{r}_{42}^{\,\,2}}(-B_{123}+\frac{1}{6}(B_{144}B_{324}+B_{244}B_{314}%
-B_{344}B_{214}))^{conf}\right.
\]%
\[
\left.  \frac{{}}{{}}+(1\leftrightarrow3)+(2\leftrightarrow3)\right]
-{\frac{\alpha_{s}^{2}}{8\pi^{4}}}\!\int\!d\vec{r}_{0}d\vec{r}_{4}%
~\mathbf{G}_{finite}-{\frac{\alpha_{s}^{2}}{8\pi^{3}}}\!\int\!d\vec{r}%
_{0}~\mathbf{G}^{\prime}%
\]%
\[
-\frac{\alpha_{s}}{4\pi^{2}}\frac{\alpha_{s}}{8\pi^{2}}\int d\vec{r}_{4}%
d\vec{r}_{0}\left[  \frac{\vec{r}_{12}^{\,\,2}}{\vec{r}_{41}^{\,\,2}\vec
{r}_{42}^{\,\,2}}\left(  A_{34}\frac{\vec{r}_{34}^{\,\,2}}{\vec{r}%
_{03}^{\,\,2}\vec{r}_{04}^{\,\,2}}\ln\left(  \frac{\vec{r}_{34}^{\,\,2}a}%
{\vec{r}_{03}^{\,\,2}\vec{r}_{04}^{\,\,2}}\right)  +A_{13}\frac{\vec{r}%
_{13}^{\,2}}{\vec{z}_{03}^{\,\,2}\vec{r}_{01}^{\,\,2}}\ln\left(  \frac{\vec
{r}_{13}^{\,2}a}{\vec{z}_{03}^{\,\,2}\vec{r}_{01}^{\,\,2}}\right)  \right.
\right.
\]%
\[
+A_{23}\frac{\vec{r}_{23}^{\,\,2}}{\vec{r}_{03}^{\,\,2}\vec{r}_{02}^{\,\,2}%
}\ln\left(  \frac{\vec{r}_{23}^{\,\,2}a}{\vec{r}_{03}^{\,\,2}\vec{r}%
_{02}^{\,\,2}}\right)  +A_{14}\frac{\vec{r}_{14}^{\,\,2}}{\vec{r}_{01}%
^{\,\,2}\vec{r}_{04}^{\,\,2}}\ln\left(  \frac{\vec{r}_{14}^{\,\,2}a}{\vec
{r}_{01}^{\,\,2}\vec{r}_{04}^{\,\,2}}\right)  +A_{24}\frac{\vec{r}%
_{24}^{\,\,2}}{\vec{r}_{02}^{\,\,2}\vec{z}_{04}^{\,\,2}}\ln\left(  \frac
{\vec{r}_{24}^{\,\,2}a}{\vec{r}_{02}^{\,\,2}\vec{z}_{04}^{\,\,2}}\right)
\]%
\[
\left.  \left.  +A_{12}\frac{\vec{r}_{12}^{\,\,2}}{\vec{r}_{01}^{\,\,2}\vec
{r}_{02}^{\,\,2}}\ln\left(  \frac{\vec{r}_{12}^{\,\,2}a}{\vec{r}_{01}%
^{\,\,2}\vec{r}_{02}^{\,\,2}}\right)  \right)  +(1\leftrightarrow
3)+(2\leftrightarrow3)\right]
\]%
\[
+\frac{\alpha_{s}}{8\pi^{2}}\frac{\alpha_{s}}{4\pi^{2}}\int d\vec{r}_{4}%
d\vec{r}_{0}\left[  \frac{\vec{r}_{12}^{\,\,2}}{\vec{r}_{41}^{\,\,2}\vec
{z}_{42}^{\,\,2}}\ln\left(  \frac{a\vec{r}_{12}^{\,\,2}}{\vec{r}_{41}%
^{\,\,2}\vec{r}_{42}^{\,\,2}}\right)  \right.
\]%
\[
\times\left(  A_{34}\frac{\vec{r}_{34}^{\,\,2}}{\vec{r}_{03}^{\,\,2}\vec
{z}_{04}^{\,\,2}}+A_{13}\frac{\vec{r}_{13}^{\,2}}{\vec{r}_{03}^{\,\,2}\vec
{z}_{01}^{\,\,2}}+A_{23}\frac{\vec{r}_{23}^{\,\,2}}{\vec{r}_{03}^{\,\,2}%
\vec{r}_{02}^{\,\,2}}+A_{14}\frac{\vec{r}_{14}^{\,\,2}}{\vec{r}_{01}%
^{\,\,2}\vec{r}_{04}^{\,\,2}}+A_{24}\frac{\vec{r}_{24}^{\,\,2}}{\vec{r}%
_{02}^{\,\,2}\vec{r}_{04}^{\,\,2}}+A_{12}\frac{\vec{r}_{12}^{\,\,2}}{\vec
{z}_{01}^{\,\,2}\vec{r}_{02}^{\,\,2}}\right)
\]%
\begin{equation}
\left.  \frac{{}}{{}}+(1\leftrightarrow3)+(2\leftrightarrow3)\right]  .
\end{equation}
After simplification one has%
\[
\frac{\partial B_{123}^{conf}}{\partial\eta}=\frac{\alpha_{s}3}{4\pi^{2}}\int
d\vec{r}_{4}\left[  \frac{\vec{r}_{12}^{\,\,2}}{\vec{r}_{41}^{\,\,2}\vec
{r}_{42}^{\,\,2}}(-B_{123}+\frac{1}{6}(B_{144}B_{324}+B_{244}B_{314}%
-B_{344}B_{214}))^{conf}\right.
\]%
\[
\left.  \frac{{}}{{}}+(1\leftrightarrow3)+(2\leftrightarrow3)\right]
-{\frac{\alpha_{s}^{2}}{8\pi^{4}}}\!\int\!d\vec{r}_{0}d\vec{r}_{4}%
~\mathbf{G}_{finite}-{\frac{\alpha_{s}^{2}}{8\pi^{3}}}\!\int\!d\vec{r}%
_{0}~\mathbf{G}^{\prime}%
\]%
\[
-\frac{\alpha_{s}^{2}}{32\pi^{4}}\int d\vec{r}_{4}d\vec{r}_{0}\left[
\frac{\vec{r}_{12}^{\,\,2}}{\vec{r}_{41}^{\,\,2}\vec{r}_{42}^{\,\,2}}\left(
A_{34}\frac{\vec{r}_{34}^{\,\,2}}{\vec{r}_{03}^{\,\,2}\vec{r}_{04}^{\,\,2}}%
\ln\left(  \frac{\vec{r}_{34}^{\,\,2}\vec{r}_{41}^{\,\,2}\vec{r}_{42}^{\,\,2}%
}{\vec{r}_{03}^{\,\,2}\vec{r}_{04}^{\,\,2}\vec{r}_{12}^{\,\,2}}\right)
+A_{13}\frac{\vec{r}_{13}^{\,2}}{\vec{r}_{03}^{\,\,2}\vec{r}_{01}^{\,\,2}}%
\ln\left(  \frac{\vec{r}_{13}^{\,2}\vec{r}_{41}^{\,\,2}\vec{r}_{42}^{\,\,2}%
}{\vec{r}_{03}^{\,\,2}\vec{r}_{01}^{\,\,2}\vec{r}_{12}^{\,\,2}}\right)
\right.  \right.
\]%
\[
+A_{23}\frac{\vec{r}_{23}^{\,\,2}}{\vec{r}_{03}^{\,\,2}\vec{r}_{02}^{\,\,2}%
}\ln\left(  \frac{\vec{r}_{23}^{\,\,2}\vec{r}_{41}^{\,\,2}\vec{r}_{42}%
^{\,\,2}}{\vec{r}_{03}^{\,\,2}\vec{r}_{02}^{\,\,2}\vec{r}_{12}^{\,\,2}%
}\right)  +A_{14}\frac{\vec{r}_{14}^{\,\,2}}{\vec{r}_{01}^{\,\,2}\vec{r}%
_{04}^{\,\,2}}\ln\left(  \frac{\vec{r}_{14}^{\,\,4}\vec{r}_{42}^{\,\,2}}%
{\vec{r}_{01}^{\,\,2}\vec{r}_{04}^{\,\,2}\vec{r}_{12}^{\,\,2}}\right)
+A_{24}\frac{\vec{r}_{24}^{\,\,2}}{\vec{r}_{02}^{\,\,2}\vec{r}_{04}^{\,\,2}%
}\ln\left(  \frac{\vec{r}_{24}^{\,\,4}\vec{r}_{41}^{\,\,2}}{\vec{r}%
_{02}^{\,\,2}\vec{r}_{04}^{\,\,2}\vec{r}_{12}^{\,\,2}}\right)
\]%
\begin{equation}
\left.  \left.  +A_{12}\frac{\vec{r}_{12}^{\,\,2}}{\vec{r}_{01}^{\,\,2}\vec
{z}_{02}^{\,\,2}}\ln\left(  \frac{\vec{r}_{41}^{\,\,2}\vec{r}_{42}^{\,\,2}%
}{\vec{r}_{01}^{\,\,2}\vec{r}_{02}^{\,\,2}}\right)  \right)
+(1\leftrightarrow3)+(2\leftrightarrow3)\right]  . \label{dBconf/deta}%
\end{equation}
Now we can symmetrize the last 3 lines of this expression w.r.t.
$0\leftrightarrow4$ transformation, i.e.%
\[
A_{ij}F(\vec{r}...)\rightarrow\left[  A_{ij}F(\vec{r}...)\right]  ^{sym}%
\]%
\begin{equation}
=\frac{\left[  A_{ij}+A_{ij}\left(  0\leftrightarrow4\right)  \right]  \left[
F+F\left(  0\leftrightarrow4\right)  \right]  +\left[  A_{ij}-A_{ij}\left(
0\leftrightarrow4\right)  \right]  \left[  F-F\left(  0\leftrightarrow
4\right)  \right]  }{4}.
\end{equation}
After that one can use (\ref{id5}) to show that all the nonconformal terms
have the $SU(3)$ coefficients independent either of $\vec{r}_{4}$ or of
$\vec{r}_{0}.$

So first we add the symmetrized last 3 lines of the previous expression to the
nonconformal part of $\mathbf{G}_{finite}$\ (\ref{Gfinite}). Taking into
account (\ref{IDENTITY}), (\ref{id5}), and (\ref{id8}), we have%
\[
-{\frac{\alpha_{s}^{2}}{8\pi^{4}}}\!\int\!d\vec{r}_{0}d\vec{r}_{4}%
\mathbf{\tilde{G}}=-{\frac{\alpha_{s}^{2}}{8\pi^{4}}}\!\int\!d\vec{r}_{0}%
d\vec{r}_{4}\left[  \left\{  (M_{13}-M_{12}-M_{23}+M_{2})[(U_{0}U_{4}{}^{\dag
}U_{3})\cdot(U_{2}U_{0}{}^{\dag}U_{1})\cdot U_{4}\right.  \right.
\]%
\[
\left.  \left.  +(U_{1}U_{0}{}^{\dag}U_{2})\cdot(U_{3}U_{4}{}^{\dag}%
U_{0})\cdot U_{4}]+(\text{all 5 permutations}\,1\leftrightarrow
2\leftrightarrow3)\right\}  +(0\leftrightarrow4)\right]
\]%
\[
-\frac{\alpha_{s}^{2}}{32\pi^{4}}\int d\vec{r}_{4}d\vec{r}_{0}\left[
\frac{\vec{r}_{12}^{\,\,2}}{\vec{r}_{41}^{\,\,2}\vec{r}_{42}^{\,\,2}}\left(
A_{34}\frac{\vec{r}_{34}^{\,\,2}}{\vec{r}_{03}^{\,\,2}\vec{r}_{04}^{\,\,2}}%
\ln\left(  \frac{\vec{r}_{34}^{\,\,2}\vec{r}_{41}^{\,\,2}\vec{r}_{42}^{\,\,2}%
}{\vec{r}_{03}^{\,\,2}\vec{r}_{04}^{\,\,2}\vec{r}_{12}^{\,\,2}}\right)
\right.  \right.
\]%
\[
+A_{23}\frac{\vec{r}_{23}^{\,\,2}}{\vec{r}_{03}^{\,\,2}\vec{r}_{02}^{\,\,2}%
}\ln\left(  \frac{\vec{r}_{23}^{\,\,2}\vec{r}_{41}^{\,\,2}\vec{r}_{42}%
^{\,\,2}}{\vec{r}_{03}^{\,\,2}\vec{r}_{02}^{\,\,2}\vec{r}_{12}^{\,\,2}%
}\right)  +A_{14}\frac{\vec{r}_{14}^{\,\,2}}{\vec{r}_{01}^{\,\,2}\vec{r}%
_{04}^{\,\,2}}\ln\left(  \frac{\vec{r}_{14}^{\,\,4}\vec{r}_{42}^{\,\,2}}%
{\vec{r}_{01}^{\,\,2}\vec{r}_{04}^{\,\,2}\vec{r}_{12}^{\,\,2}}\right)
+A_{24}\frac{\vec{r}_{24}^{\,\,2}}{\vec{r}_{02}^{\,\,2}\vec{r}_{04}^{\,\,2}%
}\ln\left(  \frac{\vec{r}_{24}^{\,\,4}\vec{r}_{41}^{\,\,2}}{\vec{r}%
_{02}^{\,\,2}\vec{r}_{04}^{\,\,2}\vec{r}_{12}^{\,\,2}}\right)
\]%
\[
\left.  \left.  +A_{13}\frac{\vec{r}_{13}^{\,2}}{\vec{r}_{03}^{\,\,2}\vec
{r}_{01}^{\,\,2}}\ln\left(  \frac{\vec{r}_{13}^{\,2}\vec{r}_{41}^{\,\,2}%
\vec{r}_{42}^{\,\,2}}{\vec{r}_{03}^{\,\,2}\vec{r}_{01}^{\,\,2}\vec{r}%
_{12}^{\,\,2}}\right)  +A_{12}\frac{\vec{r}_{12}^{\,\,2}}{\vec{r}_{01}%
^{\,\,2}\vec{z}_{02}^{\,\,2}}\ln\left(  \frac{\vec{r}_{41}^{\,\,2}\vec{r}%
_{42}^{\,\,2}}{\vec{r}_{01}^{\,\,2}\vec{r}_{02}^{\,\,2}}\right)  \right)
+(1\leftrightarrow3)+(2\leftrightarrow3)\right]  ^{sym}%
\]%
\[
=-{\frac{\alpha_{s}^{2}}{8\pi^{4}}}\!\int\!d\vec{r}_{0}d\vec{r}_{4}\left\{
\left(  \frac{\vec{r}_{12}{}^{4}{}}{8\vec{r}_{01}{}^{2}\vec{r}_{02}{}^{2}%
\vec{r}_{14}{}^{2}\vec{r}_{24}{}^{2}}\ln\left(  \frac{\vec{r}_{14}{}^{2}%
\vec{r}_{24}{}^{2}}{\vec{r}_{01}{}^{2}\vec{r}_{02}{}^{2}}\right)  \left(
\frac{1}{2}B_{003}B_{012}-2B_{001}B_{023}\right)  \right.  \right.
\]%
\[
+\frac{\vec{r}_{12}{}^{2}}{8\vec{r}_{01}{}^{2}\vec{r}_{02}{}^{2}}\left(
\frac{\vec{r}_{13}{}^{2}}{\vec{r}_{14}{}^{2}\vec{r}_{34}{}^{2}}\ln\left(
\frac{\vec{r}_{12}{}^{2}\vec{r}_{14}{}^{2}\vec{r}_{34}{}^{2}}{\vec{r}_{01}%
{}^{2}\vec{r}_{02}{}^{2}\vec{r}_{13}{}^{2}}\right)  +\frac{\vec{r}_{03}{}^{2}%
}{\vec{r}_{04}{}^{2}\vec{r}_{34}{}^{2}}\ln\left(  \frac{\vec{r}_{01}{}^{2}%
\vec{r}_{02}{}^{2}\vec{r}_{03}{}^{2}}{\vec{r}_{04}{}^{2}\vec{r}_{12}{}^{2}%
\vec{r}_{34}{}^{2}}\right)  \right)
\]%
\[
\times\left(  B_{003}B_{012}-B_{001}B_{023}\right)
\]%
\[
+\frac{\vec{r}_{12}{}^{2}}{8\vec{r}_{02}{}^{2}\vec{r}_{04}{}^{2}\vec{r}_{14}%
{}^{2}}\left[  \ln\left(  \frac{\vec{r}_{01}{}^{4}{}\vec{r}_{02}{}^{2}}%
{\vec{r}_{04}{}^{2}\vec{r}_{12}{}^{2}\vec{r}_{14}{}^{2}}\right)  \left(
2B_{003}B_{012}-2B_{002}B_{013}-3B_{001}B_{023}+4B_{123}\right)  \right.
\]%
\[
+\ln\left(  \frac{\vec{r}_{04}{}^{2}\vec{r}_{12}{}^{2}}{\vec{r}_{01}{}^{2}%
\vec{r}_{24}{}^{2}}\right)  \left(  -tr\left(  U_{0}{}^{\dag}U_{4}\right)
\left(  U_{1}U_{4}{}^{\dag}U_{2}+U_{2}U_{4}{}^{\dag}U_{1}\right)  \cdot
U_{0}\cdot U_{3}\right.
\]%
\[
\left.  +2\left(  U_{1}\cdot U_{2}\cdot U_{3}-\left(  U_{0}U_{4}{}^{\dag}%
U_{1}\right)  \cdot\left(  U_{2}U_{0}{}^{\dag}U_{4}\right)  \cdot
U_{3}\right)  \right)
\]%
\[
+\ln\left(  \frac{\vec{r}_{01}{}^{4}{}\vec{r}_{02}{}^{2}\vec{r}_{34}{}^{4}{}%
}{\vec{r}_{03}{}^{4}{}\vec{r}_{04}{}^{2}\vec{r}_{12}{}^{2}\vec{r}_{14}{}^{2}%
}\right)  \left(  tr\left(  U_{4}{}^{\dag}U_{1}\right)  \left(  \left(
U_{2}U_{0}{}^{\dag}U_{4}\right)  \cdot U_{0}\cdot U_{3}+\left(  U_{4}U_{0}%
{}^{\dag}U_{2}\right)  \cdot U_{0}\cdot U_{3}\right)  \frac{{}}{{}}\right.
\]%
\[
\left.  \left.  +\frac{{}}{{}}\left(  U_{0}U_{4}{}^{\dag}U_{1}U_{0}{}^{\dag
}U_{2}\right)  \cdot U_{3}\cdot U_{4}+\left(  U_{2}U_{0}{}^{\dag}U_{1}U_{4}%
{}^{\dag}U_{0}\right)  \cdot U_{3}\cdot U_{4}\right)  \right]
\]%
\[
+\left(  \frac{\vec{r}_{12}{}^{2}\vec{r}_{13}{}^{2}}{16\vec{r}_{01}{}^{2}%
\vec{r}_{02}{}^{2}\vec{r}_{14}{}^{2}\vec{r}_{34}{}^{2}}\ln\left(  \frac
{\vec{r}_{01}{}^{2}\vec{r}_{12}{}^{2}\vec{r}_{34}{}^{2}}{\vec{r}_{02}{}%
^{2}\vec{r}_{13}{}^{2}\vec{r}_{14}{}^{2}}\right)  +\frac{\vec{r}_{03}{}%
^{2}\vec{r}_{12}{}^{2}}{8\vec{r}_{01}{}^{2}\vec{r}_{02}{}^{2}\vec{r}_{04}%
{}^{2}\vec{r}_{34}{}^{2}}\ln\left(  \frac{\vec{r}_{02}{}^{2}\vec{r}_{03}{}%
^{2}\vec{r}_{14}{}^{4}{}}{\vec{r}_{01}{}^{2}\vec{r}_{04}{}^{2}\vec{r}_{12}%
{}^{2}\vec{r}_{34}{}^{2}}\right)  \right)
\]%
\[
\times\left(  \left(  U_{0}U_{4}{}^{\dag}U_{3}\right)  \cdot\left(  U_{1}%
U_{0}{}^{\dag}U_{2}\right)  \cdot U_{4}+\left(  U_{2}U_{0}{}^{\dag}%
U_{1}\right)  \cdot\left(  U_{3}U_{4}{}^{\dag}U_{0}\right)  \cdot
U_{4}\right.
\]%
\[
\left.  -\left(  U_{2}U_{0}{}^{\dag}U_{4}\right)  \cdot\left(  U_{3}U_{4}%
{}^{\dag}U_{1}\right)  \cdot U_{0}-\left(  U_{1}U_{4}{}^{\dag}U_{3}\right)
\cdot\left(  U_{4}U_{0}{}^{\dag}U_{2}\right)  \cdot U_{0}\right)
\]%
\begin{equation}
\left.  \left.  +\frac{{}}{{}}(\text{all 5 permutations}\,1\leftrightarrow
2\leftrightarrow3)\right)  +(0\leftrightarrow4)\right\}  . \label{Gtilde}%
\end{equation}
Indeed, in this expression all the nonconformal terms have the $SU(3)$
coefficients independent either of $\vec{r}_{4}$ or of $\vec{r}_{0}.$
Therefore one can integrate them w.r.t. $\vec{r}_{4}$ or $\vec{r}_{0}.$
However, it is easier to transform (\ref{Gall})\ using integral (116) from
\cite{Balitsky:2009xg}. We use it in the symmetric form%
\[
\frac{\vec{r}_{12}^{\,2}}{\vec{r}_{10}^{\,2}\vec{r}_{20}^{\,\,\,2}}\ln
\frac{\vec{r}_{10}^{\,\,2}}{\vec{r}_{12}^{\,\,2}}\ln\frac{\vec{r}_{20}%
^{\,\,2}}{\vec{r}_{12}^{\,\,2}}=2\pi\zeta\left(  3\right)  \left(
\delta\left(  \vec{r}_{10}\right)  +\delta\left(  \vec{r}_{20}\right)
\right)
\]%
\begin{equation}
+\frac{\vec{r}_{12}^{\,2}}{\vec{r}_{10}^{\,2}\vec{r}_{20}^{\,\,\,2}}\int%
\frac{d\vec{r}_{4}}{2\pi}\left(  \frac{\vec{r}_{20}^{\,2}}{\vec{r}_{04}%
^{\,2}\vec{r}_{24}^{\,\,\,2}}+\frac{\vec{r}_{10}^{\,2}}{\vec{r}_{04}^{\,2}%
\vec{r}_{14}^{\,\,\,2}}-\frac{\vec{r}_{12}^{\,2}}{\vec{r}_{24}^{\,2}\vec
{r}_{14}^{\,\,\,2}}\right)  \ln\left(  \frac{\vec{r}_{10}^{\,2}\vec{r}%
_{20}^{\,2}}{\vec{r}_{14}^{\,2}\vec{r}_{24}^{\,2}}\right)  .
\end{equation}
Then,%
\[
\mathbf{G}^{\prime}\mathbf{=}\frac{1}{2}\left[  \frac{\vec{r}_{13}^{\,\,2}%
\vec{r}_{20}^{\,\,\,2}}{\vec{r}_{30}^{\,\,\,2}\vec{r}_{12}^{\,2}}-\frac
{\vec{r}_{32}^{\,2}\vec{r}_{10}^{\,2}}{\vec{r}_{30}^{\,2}\vec{r}_{12}^{\,2}%
}\right]  \frac{\vec{r}_{12}^{\,2}}{\vec{r}_{10}^{\,2}\vec{r}_{20}^{\,\,\,2}%
}\int\frac{d\vec{r}_{4}}{2\pi}\left(  \frac{\vec{r}_{20}^{\,2}}{\vec{r}%
_{04}^{\,2}\vec{r}_{24}^{\,\,\,2}}+\frac{\vec{r}_{10}^{\,2}}{\vec{r}%
_{04}^{\,2}\vec{r}_{14}^{\,\,\,2}}-\frac{\vec{r}_{12}^{\,2}}{\vec{r}%
_{24}^{\,2}\vec{r}_{14}^{\,\,\,2}}\right)  \ln\left(  \frac{\vec{r}_{10}%
^{\,2}\vec{r}_{20}^{\,2}}{\vec{r}_{14}^{\,2}\vec{r}_{24}^{\,2}}\right)
\]%
\[
\times\left(  B_{100}B_{320}-B_{200}B_{310}\right)
\]%
\[
\mathbf{+}\left[  \frac{\vec{r}_{13}^{\,\,2}\vec{r}_{20}^{\,\,\,2}}{\vec
{r}_{30}^{\,\,\,2}\vec{r}_{12}^{\,2}}-\frac{\vec{r}_{32}^{\,2}\vec{r}%
_{10}^{\,2}}{\vec{r}_{30}^{\,2}\vec{r}_{12}^{\,2}}\right]  \zeta\left(
3\right)  \pi\left(  \delta\left(  \vec{r}_{10}\right)  +\delta\left(  \vec
{r}_{20}\right)  \right)  \left(  B_{100}B_{320}-B_{200}B_{310}\right)
\]%
\[
-\frac{\vec{r}_{12}^{\,2}}{\vec{r}_{10}^{\,2}\vec{r}_{20}^{\,\,\,2}}\int%
\frac{d\vec{r}_{4}}{2\pi}\left(  \frac{\vec{r}_{20}^{\,2}}{\vec{r}_{04}%
^{\,2}\vec{r}_{24}^{\,\,\,2}}+\frac{\vec{r}_{10}^{\,2}}{\vec{r}_{04}^{\,2}%
\vec{r}_{14}^{\,\,\,2}}-\frac{\vec{r}_{12}^{\,2}}{\vec{r}_{24}^{\,2}\vec
{r}_{14}^{\,\,\,2}}\right)  \ln\left(  \frac{\vec{r}_{10}^{\,2}\vec{r}%
_{20}^{\,2}}{\vec{r}_{14}^{\,2}\vec{r}_{24}^{\,2}}\right)
\]%
\[
\times\left(  9B_{123}-\frac{1}{2}\left[  2\left(  B_{100}B_{320}%
+B_{200}B_{130}\right)  -B_{300}B_{120}\right]  \right)
\]%
\[
-2\pi\zeta\left(  3\right)  \left(  \delta\left(  \vec{r}_{10}\right)
+\delta\left(  \vec{r}_{20}\right)  \right)  \left(  9B_{123}-\frac{1}%
{2}\left[  2\left(  B_{100}B_{320}+B_{200}B_{130}\right)  -B_{300}%
B_{120}\right]  \right)
\]%
\[
+\frac{11}{6}\left[  \ln\left(  \frac{\vec{r}_{01}^{\,\,2}}{\vec{r}%
_{02}^{\,\,2}}\right)  \left(  \frac{1}{\vec{r}_{02}^{\,\,2}}-\frac{1}{\vec
{r}_{01}^{\,\,2}}\right)  -\frac{\vec{r}_{12}^{\,\,2}}{\vec{r}_{01}%
^{\,\,2}\vec{r}_{02}^{\,\,2}}\ln\left(  \frac{\vec{r}_{12}^{\,\,2}}{\tilde
{\mu}^{2}}\right)  \right]
\]%
\begin{equation}
\times\left(  \frac{3}{2}(B_{100}B_{230}+B_{200}B_{130}-B_{300}B_{210}%
)-9B_{123}\right)  +(1\leftrightarrow3)+(2\leftrightarrow3).
\end{equation}
One can write it as
\[
\mathbf{G}^{\prime}\mathbf{=}\frac{1}{2}\left[  \frac{\vec{r}_{13}^{\,\,2}%
\vec{r}_{20}^{\,\,\,2}}{\vec{r}_{30}^{\,\,\,2}\vec{r}_{12}^{\,2}}-\frac
{\vec{r}_{32}^{\,2}\vec{r}_{10}^{\,2}}{\vec{r}_{30}^{\,2}\vec{r}_{12}^{\,2}%
}\right]  \frac{\vec{r}_{12}^{\,2}}{\vec{r}_{10}^{\,2}\vec{r}_{20}^{\,\,\,2}%
}\int\frac{d\vec{r}_{4}}{2\pi}\left(  \frac{\vec{r}_{20}^{\,2}}{\vec{r}%
_{04}^{\,2}\vec{r}_{24}^{\,\,\,2}}+\frac{\vec{r}_{10}^{\,2}}{\vec{r}%
_{04}^{\,2}\vec{r}_{14}^{\,\,\,2}}-\frac{\vec{r}_{12}^{\,2}}{\vec{r}%
_{24}^{\,2}\vec{r}_{14}^{\,\,\,2}}\right)  \ln\left(  \frac{\vec{r}_{10}%
^{\,2}\vec{r}_{20}^{\,2}}{\vec{r}_{14}^{\,2}\vec{r}_{24}^{\,2}}\right)
\]%
\[
\times\left(  B_{100}B_{320}-B_{200}B_{310}\right)  -\frac{\vec{r}_{12}^{\,2}%
}{\vec{r}_{10}^{\,2}\vec{r}_{20}^{\,\,\,2}}\int\frac{d\vec{r}_{4}}{2\pi
}\left(  \frac{\vec{r}_{20}^{\,2}}{\vec{r}_{04}^{\,2}\vec{r}_{24}^{\,\,\,2}%
}+\frac{\vec{r}_{10}^{\,2}}{\vec{r}_{04}^{\,2}\vec{r}_{14}^{\,\,\,2}}%
-\frac{\vec{r}_{12}^{\,2}}{\vec{r}_{24}^{\,2}\vec{r}_{14}^{\,\,\,2}}\right)
\ln\left(  \frac{\vec{r}_{10}^{\,2}\vec{r}_{20}^{\,2}}{\vec{r}_{14}^{\,2}%
\vec{r}_{24}^{\,2}}\right)
\]%
\[
\times\left(  9B_{123}-\frac{1}{2}\left[  2\left(  B_{100}B_{320}%
+B_{200}B_{130}\right)  -B_{300}B_{120}\right]  \right)
\]%
\[
+\frac{11}{6}\left[  \ln\left(  \frac{\vec{r}_{01}^{\,\,2}}{\vec{r}%
_{02}^{\,\,2}}\right)  \left(  \frac{1}{\vec{r}_{02}^{\,\,2}}-\frac{1}{\vec
{r}_{01}^{\,\,2}}\right)  -\frac{\vec{r}_{12}^{\,\,2}}{\vec{r}_{01}%
^{\,\,2}\vec{r}_{02}^{\,\,2}}\ln\left(  \frac{\vec{r}_{12}^{\,\,2}}{\tilde
{\mu}^{2}}\right)  \right]
\]%
\begin{equation}
\times\left(  \frac{3}{2}(B_{100}B_{230}+B_{200}B_{130}-B_{300}B_{210}%
)-9B_{123}\right)  +(1\leftrightarrow3)+(2\leftrightarrow3).
\end{equation}
Next, we symmetrize the previous expression w.r.t. $0\leftrightarrow4$
exchange and combine it with (\ref{Gfinite}), (\ref{dBconf/deta}), and
(\ref{Gtilde}) to obtain the NLO kernel for the composite 3QWL operator
$B_{123}^{conf}$
\[
\langle K_{NLO}\otimes B_{123}^{conf}\rangle=-{\frac{\alpha_{s}^{2}}{8\pi^{3}%
}}\!\int\!d\vec{r}_{0}\mathbf{G}^{\prime}-{\frac{\alpha_{s}^{2}}{8\pi^{4}}%
}\!\int\!d\vec{r}_{0}d\vec{r}_{4}\mathbf{\tilde{G}}%
\]%
\[
-{\frac{\alpha_{s}^{2}}{8\pi^{4}}}\!\int\!d\vec{r}_{0}d\vec{r}_{4}~\left(
\left\{  \tilde{L}_{12}\left(  U_{0}U_{4}{}^{\dag}U_{2}\right)  \cdot\left(
U_{1}U_{0}{}^{\dag}U_{4}\right)  \cdot U_{3}\frac{{}}{{}}\right.  \right.
\]%
\[
\mathbf{+}L_{12}\left[  \left(  U_{0}U_{4}{}^{\dag}U_{2}\right)  \cdot\left(
U_{1}U_{0}{}^{\dag}U_{4}\right)  \cdot U_{3}+tr\left(  U_{0}U_{4}{}^{\dag
}\right)  \left(  U_{1}U_{0}{}^{\dag}U_{2}\right)  \cdot U_{3}\cdot U_{4}%
\frac{{}}{{}}\right.
\]%
\[
\left.  -\frac{3}{4}[B_{144}B_{234}+B_{244}B_{134}-B_{344}B_{124}]+\frac{1}%
{2}B_{123}\right]
\]%
\begin{equation}
\left.  \left.  +\frac{{}}{{}}(\text{all 5 permutations}\,1\leftrightarrow
2\leftrightarrow3)\right\}  +(0\leftrightarrow4)\right)  .
\end{equation}
Using (\ref{id5}) to get rid of the terms like
\begin{equation}
\left(  U_{0}U_{4}{}^{\dag}U_{1}U_{0}{}^{\dag}U_{2}\right)  \cdot U_{3}\cdot
U_{4}+\left(  U_{2}U_{0}{}^{\dag}U_{1}U_{4}{}^{\dag}U_{0}\right)  \cdot
U_{3}\cdot U_{4}%
\end{equation}
it can be transformed to%
\[
\langle K_{NLO}\otimes B_{123}^{conf}\rangle=-{\frac{\alpha_{s}^{2}}{8\pi^{4}%
}}\!\int\!d\vec{r}_{0}d\vec{r}_{4}~\left(  \left\{  \tilde{L}_{12}^{C}\left(
U_{0}U_{4}{}^{\dag}U_{2}\right)  \cdot\left(  U_{1}U_{0}{}^{\dag}U_{4}\right)
\cdot U_{3}\frac{{}}{{}}\right.  \right.
\]%
\[
\mathbf{+}L_{12}^{C}\left[  \left(  U_{0}U_{4}{}^{\dag}U_{2}\right)
\cdot\left(  U_{1}U_{0}{}^{\dag}U_{4}\right)  \cdot U_{3}+tr\left(  U_{0}%
U_{4}{}^{\dag}\right)  \left(  U_{1}U_{0}{}^{\dag}U_{2}\right)  \cdot
U_{3}\cdot U_{4}\frac{{}}{{}}\right.
\]%
\[
\left.  -\frac{3}{4}[B_{144}B_{234}+B_{244}B_{134}-B_{344}B_{124}]+\frac{1}%
{2}B_{123}\right]
\]%
\[
+M_{12}^{C}\left[  \left(  U_{0}U_{4}{}^{\dag}U_{3}\right)  \cdot\left(
U_{2}U_{0}{}^{\dag}U_{1}\right)  \cdot U_{4}+\left(  U_{1}U_{0}{}^{\dag}%
U_{2}\right)  \cdot\left(  U_{3}U_{4}{}^{\dag}U_{0}\right)  \cdot
U_{4}\right]
\]%
\[
\left.  \left.  +Z_{12}B_{003}B_{012}+\frac{{}}{{}}(\text{all 5 permutations}%
\,1\leftrightarrow2\leftrightarrow3)\right\}  +(0\leftrightarrow4)\right)
\]%
\[
-{\frac{\alpha_{s}^{2}}{8\pi^{3}}}\!\int\!d\vec{r}_{0}\left(  \frac{11}%
{6}\left[  \ln\left(  \frac{\vec{r}_{01}^{\,\,2}}{\vec{r}_{02}^{\,\,2}%
}\right)  \left(  \frac{1}{\vec{r}_{02}^{\,\,2}}-\frac{1}{\vec{r}_{01}%
^{\,\,2}}\right)  -\frac{\vec{r}_{12}^{\,\,2}}{\vec{r}_{01}^{\,\,2}\vec
{r}_{02}^{\,\,2}}\ln\left(  \frac{\vec{r}_{12}^{\,\,2}}{\tilde{\mu}^{2}%
}\right)  \right]  \right.
\]%
\begin{equation}
\left.  \times\left(  \frac{3}{2}(B_{100}B_{230}+B_{200}B_{130}-B_{300}%
B_{210})-9B_{123}\right)  +(1\leftrightarrow3)+(2\leftrightarrow3)\right)  .
\end{equation}
Here%
\begin{equation}
L_{12}^{C}=L_{12}+\frac{\vec{r}_{12}{}^{2}}{4\vec{r}_{01}{}^{2}\vec{r}_{04}%
{}^{2}\vec{r}_{24}{}^{2}}\ln\left(  \frac{\vec{r}_{02}{}^{2}\vec{r}_{14}{}%
^{2}}{\vec{r}_{04}{}^{2}\vec{r}_{12}{}^{2}}\right)  +\frac{\vec{r}_{12}{}^{2}%
}{4\vec{r}_{02}{}^{2}\vec{r}_{04}{}^{2}\vec{r}_{14}{}^{2}}\ln\left(
\frac{\vec{r}_{01}{}^{2}\vec{r}_{24}{}^{2}}{\vec{r}_{04}{}^{2}\vec{r}_{12}%
{}^{2}}\right)  , \label{Lc}%
\end{equation}%
\begin{equation}
\tilde{L}_{12}^{C}=\tilde{L}_{12}+\frac{\vec{r}_{12}{}^{2}}{4\vec{r}_{01}%
{}^{2}\vec{r}_{04}{}^{2}\vec{r}_{24}{}^{2}}\ln\left(  \frac{\vec{r}_{02}{}%
^{2}\vec{r}_{14}{}^{2}}{\vec{r}_{04}{}^{2}\vec{r}_{12}{}^{2}}\right)
-\frac{\vec{r}_{12}{}^{2}}{4\vec{r}_{02}{}^{2}\vec{r}_{04}{}^{2}\vec{r}_{14}%
{}^{2}}\ln\left(  \frac{\vec{r}_{01}{}^{2}\vec{r}_{24}{}^{2}}{\vec{r}_{04}%
{}^{2}\vec{r}_{12}{}^{2}}\right)  ,
\end{equation}%
\[
M_{12}^{C}=\frac{\vec{r}_{12}{}^{2}}{16\vec{r}_{02}{}^{2}\vec{r}_{04}{}%
^{2}\vec{r}_{14}{}^{2}}\ln\left(  \frac{\vec{r}_{01}{}^{2}\vec{r}_{02}{}%
^{2}\vec{r}_{34}{}^{4}{}}{\vec{r}_{03}{}^{4}{}\vec{r}_{14}{}^{2}\vec{r}_{24}%
{}^{2}}\right)  +\frac{\vec{r}_{12}{}^{2}}{16\vec{r}_{01}{}^{2}\vec{r}_{04}%
{}^{2}\vec{r}_{24}{}^{2}}\ln\left(  \frac{\vec{r}_{03}{}^{4}{}\vec{r}_{04}%
{}^{4}{}\vec{r}_{12}{}^{4}{}\vec{r}_{24}{}^{2}}{\vec{r}_{01}{}^{2}\vec{r}%
_{02}{}^{6}{}\vec{r}_{14}{}^{2}\vec{r}_{34}{}^{4}{}}\right)
\]%
\[
+\frac{\vec{r}_{23}{}^{2}}{16\vec{r}_{02}{}^{2}\vec{r}_{04}{}^{2}\vec{r}%
_{34}{}^{2}}\ln\left(  \frac{\vec{r}_{01}{}^{4}{}\vec{r}_{03}{}^{2}\vec
{r}_{24}{}^{6}{}\vec{r}_{34}{}^{2}}{\vec{r}_{02}{}^{2}\vec{r}_{04}{}^{4}{}%
\vec{r}_{14}{}^{4}{}\vec{r}_{23}{}^{4}{}}\right)  +\frac{\vec{r}_{23}{}^{2}%
}{16\vec{r}_{03}{}^{2}\vec{r}_{04}{}^{2}\vec{r}_{24}{}^{2}}\ln\left(
\frac{\vec{r}_{02}{}^{2}\vec{r}_{03}{}^{2}\vec{r}_{14}{}^{4}{}}{\vec{r}_{01}%
{}^{4}{}\vec{r}_{24}{}^{2}\vec{r}_{34}{}^{2}}\right)
\]%
\[
+\frac{\vec{r}_{13}{}^{2}}{16\vec{r}_{03}{}^{2}\vec{r}_{04}{}^{2}\vec{r}%
_{14}{}^{2}}\ln\left(  \frac{\vec{r}_{02}{}^{4}{}\vec{r}_{14}{}^{2}\vec
{r}_{34}{}^{2}}{\vec{r}_{01}{}^{2}\vec{r}_{03}{}^{2}\vec{r}_{24}{}^{4}{}%
}\right)  +\frac{\vec{r}_{13}{}^{2}}{16\vec{r}_{01}{}^{2}\vec{r}_{04}{}%
^{2}\vec{r}_{34}{}^{2}}\ln\left(  \frac{\vec{r}_{02}{}^{4}{}\vec{r}_{14}{}%
^{2}\vec{r}_{34}{}^{2}}{\vec{r}_{01}{}^{2}\vec{r}_{03}{}^{2}\vec{r}_{24}{}%
^{4}{}}\right)
\]%
\[
+\frac{\vec{r}_{03}{}^{2}\vec{r}_{12}{}^{2}}{8\vec{r}_{01}{}^{2}\vec{r}_{02}%
{}^{2}\vec{r}_{04}{}^{2}\vec{r}_{34}{}^{2}}\ln\left(  \frac{\vec{r}_{01}{}%
^{2}\vec{r}_{03}{}^{2}\vec{r}_{24}{}^{4}{}}{\vec{r}_{02}{}^{2}\vec{r}_{04}%
{}^{2}\vec{r}_{12}{}^{2}\vec{r}_{34}{}^{2}}\right)  +\frac{\vec{r}_{23}{}%
^{2}\vec{r}_{12}{}^{2}}{8\vec{r}_{01}{}^{2}\vec{r}_{02}{}^{2}\vec{r}_{24}%
{}^{2}\vec{r}_{34}{}^{2}}\ln\left(  \frac{\vec{r}_{02}{}^{2}\vec{r}_{12}{}%
^{2}\vec{r}_{34}{}^{2}}{\vec{r}_{01}{}^{2}\vec{r}_{23}{}^{2}\vec{r}_{24}{}%
^{2}}\right)
\]%
\begin{equation}
+\frac{\vec{r}_{14}{}^{2}\vec{r}_{23}{}^{2}}{8\vec{r}_{01}{}^{2}\vec{r}_{04}%
{}^{2}\vec{r}_{24}{}^{2}\vec{r}_{34}{}^{2}}\ln\left(  \frac{\vec{r}_{01}{}%
^{2}\vec{r}_{04}{}^{2}\vec{r}_{23}{}^{2}\vec{r}_{24}{}^{2}}{\vec{r}_{02}{}%
^{4}{}\vec{r}_{14}{}^{2}\vec{r}_{34}{}^{2}}\right)  , \label{Mc}%
\end{equation}%
\[
Z_{12}=\frac{\vec{r}_{12}{}^{2}}{8\vec{r}_{01}{}^{2}\vec{r}_{02}{}^{2}}\left[
\left(  \frac{\vec{r}_{03}{}^{2}}{\vec{r}_{04}{}^{2}\vec{r}_{34}{}^{2}}%
-\frac{\vec{r}_{02}{}^{2}}{\vec{r}_{04}{}^{2}\vec{r}_{24}{}^{2}}\right)
\ln\left(  \frac{\vec{r}_{02}{}^{2}\vec{r}_{14}{}^{2}}{\vec{r}_{04}{}^{2}%
\vec{r}_{12}{}^{2}}\right)  \right.
\]%
\begin{equation}
\left.  +\frac{\vec{r}_{01}{}^{2}}{\vec{r}_{04}{}^{2}\vec{r}_{14}{}^{2}}%
\ln\left(  \frac{\vec{r}_{02}{}^{2}\vec{r}_{34}{}^{2}}{\vec{r}_{03}{}^{2}%
\vec{r}_{24}{}^{2}}\right)  +\frac{\vec{r}_{13}{}^{2}}{\vec{r}_{14}{}^{2}%
\vec{r}_{34}{}^{2}}\ln\left(  \frac{\vec{r}_{03}{}^{2}\vec{r}_{12}{}^{2}}%
{\vec{r}_{02}{}^{2}\vec{r}_{13}{}^{2}}\right)  \right]  -(1\leftrightarrow3),
\end{equation}
and $L_{12}$ and $\tilde{L}_{12}$ are the elements of the nonconformal kernel
defined in (\ref{L12}) and (\ref{Ltilde12}). Checking that $L_{12}^{C},$
$\tilde{L}_{12}^{C},$ $M_{12}^{C},$ and $Z_{12}$ have integrable singularities
at $\vec{r}_{4}=\vec{r}_{0}$ and that $L_{12}^{C},$ $\tilde{L}_{12}^{C},$ and
$Z_{12}$ have integrable singularities at $\vec{r}_{4}=\vec{r}_{1,2,3}$ is
straightforward. To prove that all the terms with $M^{C}$ have safe behavior
at $\vec{r}_{4}=\vec{r}_{1,2,3}$ one has to use $SU(3)$ identity (\ref{id9}).

Now one can see that the NLO kernel for the evolution equation for the
composite 3QWL operator $B_{123}^{conf}$ (\ref{anzatz}) is quasi-conformal if
one expresses the LO kernel in terms of composite operator (\ref{anzatz1}).

The term with $Z$ can be integrated w.r.t. $\vec{r}_{4}$. The integral is
calculated in the appendix \ref{integrals}
\begin{equation}
\int\frac{d\vec{r}_{4}}{\pi}Z_{12}=\frac{\vec{r}_{32}{}^{2}}{8\vec{r}_{03}%
{}^{2}\vec{r}_{02}{}^{2}}\ln^{2}\left(  \frac{\vec{r}_{32}{}^{2}\vec{r}_{10}%
{}^{2}}{\vec{r}_{13}{}^{2}\vec{r}_{20}{}^{2}}\right)  -\frac{\vec{r}_{12}%
{}^{2}}{8\vec{r}_{01}{}^{2}\vec{r}_{02}{}^{2}}\ln^{2}\left(  \frac{\vec
{r}_{12}{}^{2}\vec{r}_{30}{}^{2}}{\vec{r}_{13}{}^{2}\vec{r}_{20}{}^{2}%
}\right)  . \label{Z}%
\end{equation}
Finally, the kernel reads%
\[
\langle K_{NLO}\otimes B_{123}^{conf}\rangle=-{\frac{\alpha_{s}^{2}}{8\pi^{4}%
}}\!\int\!d\vec{r}_{0}d\vec{r}_{4}~\left(  \left\{  \tilde{L}_{12}^{C}\left(
U_{0}U_{4}{}^{\dag}U_{2}\right)  \cdot\left(  U_{1}U_{0}{}^{\dag}U_{4}\right)
\cdot U_{3}\frac{{}}{{}}\right.  \right.
\]%
\[
\mathbf{+}L_{12}^{C}\left[  \left(  U_{0}U_{4}{}^{\dag}U_{2}\right)
\cdot\left(  U_{1}U_{0}{}^{\dag}U_{4}\right)  \cdot U_{3}+tr\left(  U_{0}%
U_{4}{}^{\dag}\right)  \left(  U_{1}U_{0}{}^{\dag}U_{2}\right)  \cdot
U_{3}\cdot U_{4}\frac{{}}{{}}\right.
\]%
\[
\left.  -\frac{3}{4}[B_{144}B_{234}+B_{244}B_{134}-B_{344}B_{124}]+\frac{1}%
{2}B_{123}\right]
\]%
\[
+M_{12}^{C}\left[  \left(  U_{0}U_{4}{}^{\dag}U_{3}\right)  \cdot\left(
U_{2}U_{0}{}^{\dag}U_{1}\right)  \cdot U_{4}+\left(  U_{1}U_{0}{}^{\dag}%
U_{2}\right)  \cdot\left(  U_{3}U_{4}{}^{\dag}U_{0}\right)  \cdot
U_{4}\right]
\]%
\[
\left.  \left.  +\frac{{}}{{}}(\text{all 5 permutations}\,1\leftrightarrow
2\leftrightarrow3)\right\}  +(0\leftrightarrow4)\right)
\]%
\[
-{\frac{\alpha_{s}^{2}}{8\pi^{3}}}\!\int\!d\vec{r}_{0}\left(  \frac{11}%
{6}\left[  \ln\left(  \frac{\vec{r}_{01}^{\,\,2}}{\vec{r}_{02}^{\,\,2}%
}\right)  \left(  \frac{1}{\vec{r}_{02}^{\,\,2}}-\frac{1}{\vec{r}_{01}%
^{\,\,2}}\right)  -\frac{\vec{r}_{12}^{\,\,2}}{\vec{r}_{01}^{\,\,2}\vec
{r}_{02}^{\,\,2}}\ln\left(  \frac{\vec{r}_{12}^{\,\,2}}{\tilde{\mu}^{2}%
}\right)  \right]  \right.
\]%
\[
\left.  \times\left(  \frac{3}{2}(B_{100}B_{230}+B_{200}B_{130}-B_{300}%
B_{210})-9B_{123}\right)  +(1\leftrightarrow3)+(2\leftrightarrow3)\right)
\]%
\[
-{\frac{\alpha_{s}^{2}}{32\pi^{3}}}\!\int\!d\vec{r}_{0}\left(  B_{003}%
B_{012}\left[  \frac{\vec{r}_{32}{}^{2}}{\vec{r}_{03}{}^{2}\vec{r}_{02}{}^{2}%
}\ln^{2}\left(  \frac{\vec{r}_{32}{}^{2}\vec{r}_{10}{}^{2}}{\vec{r}_{13}{}%
^{2}\vec{r}_{20}{}^{2}}\right)  -\frac{\vec{r}_{12}{}^{2}}{\vec{r}_{01}{}%
^{2}\vec{r}_{02}{}^{2}}\ln^{2}\left(  \frac{\vec{r}_{12}{}^{2}\vec{r}_{30}%
{}^{2}}{\vec{r}_{13}{}^{2}\vec{r}_{20}{}^{2}}\right)  \right]  \right.
\]%
\begin{equation}
\left.  \frac{{}}{{}}+(\text{all 5 permutations}\,1\leftrightarrow
2\leftrightarrow3)\right)  . \label{QCkernel}%
\end{equation}
In the quark-diquark limit $\vec{r}_{3}\rightarrow\vec{r}_{2}$ one has
\[
\left\{  M_{12}^{C}\left[  \left(  U_{0}U_{4}{}^{\dag}U_{3}\right)
\cdot\left(  U_{2}U_{0}{}^{\dag}U_{1}\right)  \cdot U_{4}+\left(  U_{1}U_{0}%
{}^{\dag}U_{2}\right)  \cdot\left(  U_{3}U_{4}{}^{\dag}U_{0}\right)  \cdot
U_{4}\right]  \frac{{}}{{}}\right.
\]%
\[
\left.  +\frac{{}}{{}}(\text{all 5 permutations}\,1\leftrightarrow
2\leftrightarrow3)\right\}  +(0\leftrightarrow4)
\]%
\[
\rightarrow2\tilde{L}_{12}^{C}\left[  tr\left(  U_{0}{}^{\dagger}U_{4}\right)
\left(  tr\left(  U_{2}{}^{\dagger}U_{0}U_{4}{}^{\dagger}U_{1}\right)
+tr\left(  U_{2}{}^{\dagger}U_{1}U_{4}{}^{\dagger}U_{0}\right)  \right)
\right.
\]%
\begin{equation}
\left.  +2tr\left(  U_{0}{}^{\dagger}U_{1}\right)  tr\left(  U_{2}{}^{\dagger
}U_{4}\right)  tr\left(  U_{4}{}^{\dagger}U_{0}\right)  -(0\leftrightarrow
4)\right]  ,
\end{equation}%
\[
\left\{  \tilde{L}_{12}^{C}\left(  U_{0}U_{4}{}^{\dag}U_{2}\right)
\cdot\left(  U_{1}U_{0}{}^{\dag}U_{4}\right)  \cdot U_{3}+(\text{all 5
permutations}\,1\leftrightarrow2\leftrightarrow3)\right\}  +(0\leftrightarrow
4)
\]%
\begin{equation}
\rightarrow2\tilde{L}_{12}^{C}\left[  tr\left(  U_{4}{}^{\dagger}U_{0}\right)
\left(  tr\left(  U_{0}{}^{\dagger}U_{1}U_{2}{}^{\dagger}U_{4}\right)
+tr\left(  U_{0}{}^{\dagger}U_{4}U_{2}{}^{\dagger}U_{1}\right)  \right)
-(0\leftrightarrow4)\right]  ,
\end{equation}%
\[
L_{12}^{C}\left[  \left(  U_{0}U_{4}{}^{\dag}U_{2}\right)  \cdot\left(
U_{1}U_{0}{}^{\dag}U_{4}\right)  \cdot U_{3}+tr\left(  U_{0}U_{4}{}^{\dag
}\right)  \left(  U_{1}U_{0}{}^{\dag}U_{2}\right)  \cdot U_{3}\cdot
U_{4}+\frac{1}{2}B_{123}\right.
\]%
\[
\left.  -\frac{3}{4}[B_{144}B_{234}+B_{244}B_{134}-B_{344}B_{124}]+(\text{all
5 permutations}\,1\leftrightarrow2\leftrightarrow3)\right]  +(0\leftrightarrow
4)
\]%
\[
\rightarrow4L_{12}^{C}\left[  tr\left(  U_{2}{}^{\dagger}U_{1}\right)
-3tr\left(  U_{0}{}^{\dagger}U_{1}\right)  tr\left(  U_{2}{}^{\dagger}%
U_{0}\right)  +tr\left(  U_{0}{}^{\dagger}U_{1}\right)  tr\left(  U_{2}%
{}^{\dagger}U_{4}\right)  tr\left(  U_{4}{}^{\dagger}U_{0}\right)  \right.
\]%
\begin{equation}
\left.  -tr\left(  U_{0}{}^{\dagger}U_{1}U_{4}{}^{\dagger}U_{0}U_{2}%
{}^{\dagger}U_{4}\right)  +(0\leftrightarrow4)\right]  .
\end{equation}
Therefore,%
\[
\langle K_{NLO}\otimes B_{122}^{conf}\rangle=-{\frac{\alpha_{s}^{2}}{2\pi^{4}%
}}\!\int\!d\vec{r}_{0}d\vec{r}_{4}~\left(  \left\{  \left(  \tilde{L}_{12}%
^{C}+L_{12}^{C}\right)  tr\left(  U_{0}{}^{\dagger}U_{1}\right)  tr\left(
U_{2}{}^{\dagger}U_{4}\right)  tr\left(  U_{4}{}^{\dagger}U_{0}\right)
\frac{{}}{{}}\right.  \right.
\]%
\[
\left.  \left.  \frac{{}}{{}}+L_{12}^{C}\left[  tr\left(  U_{2}{}^{\dagger
}U_{1}\right)  -3tr\left(  U_{0}{}^{\dagger}U_{1}\right)  tr\left(  U_{2}%
{}^{\dagger}U_{0}\right)  -tr\left(  U_{0}{}^{\dagger}U_{1}U_{4}{}^{\dagger
}U_{0}U_{2}{}^{\dagger}U_{4}\right)  \right]  \right\}  +(0\leftrightarrow
4)\right)
\]%
\[
-{\frac{3\alpha_{s}^{2}}{2\pi^{3}}}\!\int\!d\vec{r}_{0}\frac{11}{6}\left[
\ln\left(  \frac{\vec{r}_{01}^{\,\,2}}{\vec{r}_{02}^{\,\,2}}\right)  \left(
\frac{1}{\vec{r}_{02}^{\,\,2}}-\frac{1}{\vec{r}_{01}^{\,\,2}}\right)
-\frac{\vec{r}_{12}^{\,\,2}}{\vec{r}_{01}^{\,\,2}\vec{r}_{02}^{\,\,2}}%
\ln\left(  \frac{\vec{r}_{12}^{\,\,2}}{\tilde{\mu}^{2}}\right)  \right]
\]%
\begin{equation}
\times\left(  tr\left(  U_{0}{}^{\dagger}U_{1}\right)  tr\left(  U_{2}%
{}^{\dagger}U_{0}\right)  -3tr\left(  U_{2}{}^{\dagger}U_{1}\right)  \right)
.
\end{equation}
This is twice the gluon part of the BK kernel (see (67) in
\cite{Balitsky:2009xg}).

\section{Linearization}

In the 3-gluon approximation
\begin{equation}
B_{003}B_{012}\overset{\mathrm{3g}}{=}6B_{003}+6B_{012}-36.
\end{equation}
We use the following identity to linearize the color structures in
(\ref{QCkernel}).%
\[
(U_{0}U_{4}^{\dag}U_{2})\cdot(U_{1}U_{0}^{\dag}U_{4})\cdot U_{3}%
+(1\leftrightarrow2,0\leftrightarrow4)=
\]%
\[
=(U_{0}U_{4}^{\dag}-E)(U_{2}-U_{4})\cdot(U_{1}-U_{0})U_{0}^{\dag}U_{4}\cdot
U_{3}+U_{4}U_{0}^{\dag}(U_{1}-U_{0})\cdot(U_{2}-U_{4})(U_{4}^{\dag}%
U_{0}-E)\cdot U_{3}%
\]%
\[
+U_{0}\cdot(U_{1}U_{0}^{\dag}U_{4}+U_{4}U_{0}^{\dag}U_{1})\cdot U_{3}%
+(U_{0}U_{4}^{\dag}U_{2}+U_{2}U_{4}^{\dag}U_{0})\cdot U_{4}\cdot U_{3}%
\]%
\[
+(U_{2}-U_{4})\cdot(U_{1}-U_{0})(U_{0}^{\dag}U_{4}-E)\cdot U_{3}+(U_{4}%
U_{0}^{\dag}-E)(U_{1}-U_{0})\cdot(U_{2}-U_{4})\cdot U_{3}%
\]%
\begin{equation}
+2(U_{2}-U_{4})\cdot(U_{1}-U_{0})\cdot U_{3}-2U_{0}\cdot U_{4}\cdot U_{3}.
\end{equation}
Here $E$ is the identity matrix. In the 3-gluon approximation it reads%
\[
(U_{0}U_{4}^{\dag}U_{2})\cdot(U_{1}U_{0}^{\dag}U_{4})\cdot U_{3}%
+(1\leftrightarrow2,0\leftrightarrow4)
\]%
\[
\overset{\mathrm{3g}}{=}(U_{0}-U_{4})(U_{2}-U_{4})\cdot(U_{1}-U_{0})\cdot
E+(U_{1}-U_{0})\cdot(U_{2}-U_{4})(U_{0}-U_{4})\cdot E
\]%
\[
+U_{0}\cdot(U_{1}U_{0}^{\dag}U_{4}+U_{4}U_{0}^{\dag}U_{1})\cdot U_{3}%
+(U_{0}U_{4}^{\dag}U_{2}+U_{2}U_{4}^{\dag}U_{0})\cdot U_{4}\cdot U_{3}%
\]%
\[
+(U_{2}-U_{4})\cdot(U_{1}-U_{0})(U_{4}-U_{0})\cdot E+(U_{4}-U_{0})(U_{1}%
-U_{0})\cdot(U_{2}-U_{4})\cdot E
\]%
\begin{equation}
+2(U_{2}-U_{4})\cdot(U_{1}-U_{0})\cdot U_{3}-2U_{0}\cdot U_{4}\cdot U_{3}.
\end{equation}
Using identity (\ref{IDENTITY}) and the fact that in the 3-gluon approximation%
\[
((U_{0}-U_{4})(U_{2}-U_{4})+(U_{2}-U_{4})(U_{0}-U_{4}))\cdot(U_{1}-U_{0})\cdot
E
\]%
\begin{equation}
\overset{\mathrm{3g}}{=}-(U_{2}-U_{4})\cdot(U_{0}-U_{4})\cdot(U_{1}-U_{0}),
\label{3gLinearization}%
\end{equation}
we get
\[
(U_{0}U_{4}^{\dag}U_{2})\cdot(U_{1}U_{0}^{\dag}U_{4})\cdot U_{3}%
+(1\leftrightarrow2,0\leftrightarrow4)
\]%
\[
\overset{\mathrm{3g}}{=}-B_{134}+\frac{1}{2}(B_{100}B_{340}+B_{400}%
B_{130}-B_{300}B_{140})
\]%
\[
-B_{023}+\frac{1}{2}(B_{044}B_{234}+B_{244}B_{034}-B_{344}B_{024})
\]%
\[
+2(U_{2}-U_{4})\cdot(U_{1}-U_{0})\cdot U_{3}-2U_{0}\cdot U_{4}\cdot U_{3}%
\]%
\[
=B_{123}-3B_{134}+\frac{1}{2}(B_{100}B_{340}+B_{400}B_{130}-B_{300}%
B_{140})+(1\leftrightarrow2,0\leftrightarrow4)
\]%
\begin{equation}
\overset{\mathrm{3g}}{=}B_{123}+3(B_{100}+B_{340}+B_{400}+B_{130}%
-B_{300}-B_{140}-B_{134}-6)+(1\leftrightarrow2,0\leftrightarrow4).
\end{equation}
As a result the coefficient of $\tilde{L}_{12}^{C}$ in (\ref{QCkernel}) reads%
\[
(\left(  U_{0}U_{4}{}^{\dag}U_{2}\right)  \cdot\left(  U_{1}U_{0}{}^{\dag
}U_{4}\right)  \cdot U_{3}+(1\leftrightarrow2,0\leftrightarrow
4))-(0\leftrightarrow4)
\]%
\begin{equation}
\overset{\mathrm{3g}}{=}(3B_{001}+6B_{130}-(1\leftrightarrow
2))-(0\leftrightarrow4).
\end{equation}
Using integrals (114) and (125) from \cite{Balitsky:2009xg},%
\begin{equation}
\int d\vec{r}_{4}\tilde{L}_{12}=\frac{\pi^{2}}{2}\zeta\left(  3\right)
\left(  \delta\left(  \vec{r}_{10}\right)  -\delta\left(  \vec{r}_{20}\right)
\right)  ,
\end{equation}
and%
\[
\int d\vec{r}_{4}\left[  \frac{\vec{r}_{12}{}^{2}}{4\vec{r}_{01}{}^{2}\vec
{r}_{04}{}^{2}\vec{r}_{24}{}^{2}}\ln\left(  \frac{\vec{r}_{02}{}^{2}\vec
{r}_{14}{}^{2}}{\vec{r}_{04}{}^{2}\vec{r}_{12}{}^{2}}\right)  -\frac{\vec
{r}_{12}{}^{2}}{4\vec{r}_{02}{}^{2}\vec{r}_{04}{}^{2}\vec{r}_{14}{}^{2}}%
\ln\left(  \frac{\vec{r}_{01}{}^{2}\vec{r}_{24}{}^{2}}{\vec{r}_{04}{}^{2}%
\vec{r}_{12}{}^{2}}\right)  \right]
\]%
\begin{equation}
=\pi^{2}\zeta\left(  3\right)  \left(  \delta\left(  \vec{r}_{10}\right)
-\delta\left(  \vec{r}_{20}\right)  \right)  ,
\end{equation}
one has%
\begin{equation}
\int d\vec{r}_{4}\tilde{L}_{12}^{C}=\frac{3}{2}\pi^{2}\zeta\left(  3\right)
\left(  \delta\left(  \vec{r}_{10}\right)  -\delta\left(  \vec{r}_{20}\right)
\right)  , \label{int_L12C}%
\end{equation}
and%
\[
-{\frac{\alpha_{s}^{2}}{8\pi^{4}}}\!\int\!d\vec{r}_{0}d\vec{r}_{4}~\left(
\left\{  \tilde{L}_{12}^{C}\left(  U_{0}U_{4}{}^{\dag}U_{2}\right)
\cdot\left(  U_{1}U_{0}{}^{\dag}U_{4}\right)  \cdot U_{3}\frac{{}}{{}}\right.
\right.
\]%
\[
\left.  \left.  +\frac{{}}{{}}(\text{all 5 permutations}\,1\leftrightarrow
2\leftrightarrow3)\right\}  +(0\leftrightarrow4)\right)
\]%
\[
\overset{\mathrm{3g}}{=}-{\frac{\alpha_{s}^{2}}{8\pi^{4}}}\!\int\!d\vec{r}%
_{0}(3B_{001}+6B_{130}-(1\leftrightarrow2))3\pi^{2}\zeta\left(  3\right)
\left(  \delta\left(  \vec{r}_{10}\right)  -\delta\left(  \vec{r}_{20}\right)
\right)  +(\,1\leftrightarrow3)+(2\leftrightarrow3)
\]%
\begin{equation}
=-{\frac{9\alpha_{s}^{2}}{8\pi^{2}}}\!\zeta\left(  3\right)  (36+B_{131}%
+B_{133}+B_{121}+B_{212}+B_{232}+B_{233}-12B_{231})).
\end{equation}
The second structure reads%
\[
\left(  U_{1}U_{0}{}^{\dag}U_{2}+U_{2}U_{0}{}^{\dag}U_{1}\right)  \cdot
U_{3}\cdot U_{4}%
\]%
\[
=(U_{1}-U_{0})U_{0}{}^{\dag}(U_{2}-U_{0}))\cdot U_{3}\cdot(U_{4}-U_{0}%
)+(U_{2}-U_{0})U_{0}{}^{\dag}(U_{1}-U_{0}))\cdot U_{3}\cdot(U_{4}-U_{0})
\]%
\begin{equation}
+2(U_{2}+U_{1}-U_{0})\cdot U_{3}\cdot(U_{4}-U_{0})+\left(  U_{1}U_{0}{}^{\dag
}U_{2}+U_{2}U_{0}{}^{\dag}U_{1}\right)  \cdot U_{3}\cdot U_{0}.
\end{equation}
Again, applying identity (\ref{IDENTITY}) and equality (\ref{3gLinearization})
one gets in the 3-gluon approximation%
\[
\left(  U_{1}U_{0}{}^{\dag}U_{2}+U_{2}U_{0}{}^{\dag}U_{1}\right)  \cdot
U_{3}\cdot U_{4}\overset{\mathrm{3g}}{=}-(U_{1}-U_{0})\cdot(U_{2}-U_{0}%
)\cdot(U_{4}-U_{0})
\]%
\begin{equation}
+2(U_{2}+U_{1}-U_{0})\cdot U_{3}\cdot(U_{4}-U_{0})-B_{123}+\frac{1}{2}\left(
B_{100}B_{230}+B_{200}B_{130}-B_{300}B_{120}\right)  . \label{(102)(201)}%
\end{equation}
Finally, the coefficient of $L_{12}^{C}$ in (\ref{QCkernel}) reads%
\[
\left[  \left(  U_{0}U_{4}{}^{\dag}U_{2}\right)  \cdot\left(  U_{1}U_{0}%
{}^{\dag}U_{4}\right)  \cdot U_{3}+tr\left(  U_{0}U_{4}{}^{\dag}\right)
\left(  U_{1}U_{0}{}^{\dag}U_{2}\right)  \cdot U_{3}\cdot U_{4}\frac{{}}{{}%
}\right.
\]%
\[
\left.  -\frac{3}{4}[B_{144}B_{234}+B_{244}B_{134}-B_{344}B_{124}]+\frac{1}%
{2}B_{123}+(1\leftrightarrow2)\right]  +(0\leftrightarrow4)
\]%
\begin{equation}
\overset{\mathrm{3g}}{=}9(B_{044}+B_{004}-12).
\end{equation}
Therefore,%
\[
-{\frac{\alpha_{s}^{2}}{8\pi^{4}}}\!\int\!d\vec{r}_{0}d\vec{r}_{4}~\left(
\left\{  L_{12}^{C}\left[  \left(  U_{0}U_{4}{}^{\dag}U_{2}\right)
\cdot\left(  U_{1}U_{0}{}^{\dag}U_{4}\right)  \cdot U_{3}+tr\left(  U_{0}%
U_{4}{}^{\dag}\right)  \left(  U_{1}U_{0}{}^{\dag}U_{2}\right)  \cdot
U_{3}\cdot U_{4}\frac{{}}{{}}\right.  \right.  \right.
\]%
\[
\left.  -\frac{3}{4}[B_{144}B_{234}+B_{244}B_{134}-B_{344}B_{124}]+\frac{1}%
{2}B_{123}\right]
\]%
\[
\left.  \left.  +\frac{{}}{{}}(\text{all 5 permutations}\,1\leftrightarrow
2\leftrightarrow3)\right\}  +(0\leftrightarrow4)\right)
\]%
\begin{equation}
\overset{\mathrm{3g}}{=}-{\frac{9\alpha_{s}^{2}}{8\pi^{4}}}\!\int\!d\vec
{r}_{0}d\vec{r}_{4}(L_{12}^{C}+L_{13}^{C}+L_{23}^{C})(B_{044}+B_{004}-12)~.
\end{equation}
The third structure reads%
\[
(U_{2}U_{0}^{\dag}U_{1})\cdot U_{4}\cdot(U_{0}U_{4}^{\dag}U_{3})+(U_{1}%
U_{0}^{\dag}U_{2})\cdot U_{4}\cdot(U_{3}U_{4}^{\dag}U_{0})
\]%
\[
=U_{1}\cdot U_{4}\cdot(U_{3}U_{4}^{\dag}U_{0}+U_{0}U_{4}^{\dag}U_{3}%
)+(U_{1}U_{0}^{\dag}U_{2}+U_{2}U_{0}^{\dag}U_{1})\cdot U_{4}\cdot U_{0}%
\]%
\[
+(U_{2}-U_{0})\cdot U_{4}\cdot((U_{0}-U_{4})U_{4}^{\dag}(U_{3}-U_{4}%
)+(U_{3}-U_{4})U_{4}^{\dag}(U_{0}-U_{4}))
\]%
\[
+((U_{1}-U_{0})U_{0}^{\dag}(U_{2}-U_{0}))\cdot U_{4}\cdot((U_{3}-U_{4}%
)U_{4}^{\dag}U_{0})
\]%
\[
+((U_{2}-U_{0})U_{0}^{\dag}(U_{1}-U_{0}))\cdot U_{4}\cdot(U_{0}U_{4}^{\dag
}(U_{3}-U_{4}))
\]%
\begin{equation}
+2(U_{2}-U_{0})\cdot U_{4}\cdot(U_{3}-U_{4})-2U_{1}\cdot U_{4}\cdot U_{0}.
\end{equation}
Using (\ref{IDENTITY})\ and (\ref{3gLinearization}),
\[
(U_{2}U_{0}^{\dag}U_{1})\cdot U_{4}\cdot(U_{0}U_{4}^{\dag}U_{3})+(U_{1}%
U_{0}^{\dag}U_{2})\cdot U_{4}\cdot(U_{3}U_{4}^{\dag}U_{0})
\]%
\[
\overset{\mathrm{3g}}{=}-B_{013}+\frac{1}{2}(B_{344}B_{014}+B_{044}%
B_{134}-B_{144}B_{034})-B_{124}+\frac{1}{2}(B_{100}B_{240}+B_{200}%
B_{140}-B_{004}B_{012})
\]%
\[
-(U_{2}-U_{0})\cdot(U_{1}-3U_{4})\cdot(U_{3}-U_{4})-2U_{1}\cdot U_{4}\cdot
U_{0}%
\]%
\[
\overset{\mathrm{3g}}{=}3(B_{010}-B_{441}+B_{020}-B_{442}-B_{040}%
+2B_{440}-B_{120}+B_{140}%
\]%
\begin{equation}
+B_{341}+B_{240}-2B_{340}+B_{342}+B_{443})-B_{231}-36.
\end{equation}
As a result%
\[
-{\frac{\alpha_{s}^{2}}{8\pi^{4}}}\!\int\!d\vec{r}_{0}d\vec{r}_{4}~\left(
\left\{  M_{12}^{C}\left[  \left(  U_{0}U_{4}{}^{\dag}U_{3}\right)
\cdot\left(  U_{2}U_{0}{}^{\dag}U_{1}\right)  \cdot U_{4}+\left(  U_{1}U_{0}%
{}^{\dag}U_{2}\right)  \cdot\left(  U_{3}U_{4}{}^{\dag}U_{0}\right)  \cdot
U_{4}\right]  \frac{{}}{{}}\right.  \right.
\]%
\[
\left.  \left.  +\frac{{}}{{}}(\text{all 5 permutations}\,1\leftrightarrow
2\leftrightarrow3)\right\}  +(0\leftrightarrow4)\right)
\]%
\[
\overset{\mathrm{3g}}{=}-{\frac{3\alpha_{s}^{2}}{32\pi^{4}}}\!\int\!d\vec
{r}_{0}d\vec{r}_{4}~\left(  \frac{3}{2}F_{0}(B_{040}-B_{044})+\left\{
\frac{3}{2}F_{140}B_{140}+F_{100}B_{100}+F_{230}B_{230}\frac{{}}{{}}\right.
\right.
\]%
\begin{equation}
\left.  \left.  \frac{{}}{{}}+(0\leftrightarrow4)\right\}  +(\text{all 5
permutations}\,1\leftrightarrow2\leftrightarrow3)\right)  .
\end{equation}
Here%
\[
F_{0}=\frac{\vec{r}_{12}{}^{2}}{2\vec{r}_{14}{}^{2}\vec{r}_{24}{}^{2}}\left(
\frac{\vec{r}_{24}{}^{2}}{\vec{r}_{02}{}^{2}\vec{r}_{04}{}^{2}}\ln\left(
\frac{\vec{r}_{01}{}^{2}\vec{r}_{02}{}^{2}\vec{r}_{34}{}^{4}{}}{{}\vec{r}%
_{14}{}^{2}\vec{r}_{24}{}^{2}\vec{r}_{03}{}^{4}}\right)  -\frac{\vec{r}_{13}%
{}^{2}}{\vec{r}_{01}{}^{2}\vec{r}_{03}{}^{2}}\ln\left(  \frac{\vec{r}_{01}%
{}^{2}\vec{r}_{13}{}^{2}\vec{r}_{24}{}^{2}}{\vec{r}_{03}{}^{2}\vec{r}_{12}%
{}^{2}\vec{r}_{14}{}^{2}}\right)  \right.
\]%
\begin{equation}
\left.  +\frac{2\vec{r}_{34}{}^{2}}{\vec{r}_{03}{}^{2}\vec{r}_{04}{}^{2}}%
\ln\left(  \frac{\vec{r}_{01}{}^{2}\vec{r}_{02}{}^{2}\vec{r}_{34}{}^{2}}%
{\vec{r}_{03}{}^{2}\vec{r}_{04}{}^{2}\vec{r}_{12}{}^{2}}\right)  \right)
-(0\leftrightarrow4). \label{F0}%
\end{equation}%
\[
F_{140}=\frac{\vec{r}_{12}{}^{2}}{\vec{r}_{02}{}^{2}\vec{r}_{04}{}^{2}\vec
{r}_{14}{}^{2}}\ln\left(  \frac{\vec{r}_{02}{}^{2}\vec{r}_{04}{}^{2}\vec
{r}_{12}{}^{2}\vec{r}_{34}{}^{4}{}}{\vec{r}_{03}{}^{4}{}\vec{r}_{14}{}^{2}%
\vec{r}_{24}{}^{4}{}}\right)
\]%
\[
-\frac{\vec{r}_{01}{}^{2}\vec{r}_{23}{}^{2}}{\vec{r}_{02}{}^{2}\vec{r}_{03}%
{}^{2}\vec{r}_{04}{}^{2}\vec{r}_{14}{}^{2}}\ln\left(  \frac{\vec{r}_{01}{}%
^{2}\vec{r}_{24}{}^{2}\vec{r}_{34}{}^{2}}{\vec{r}_{04}{}^{2}\vec{r}_{14}{}%
^{2}\vec{r}_{23}{}^{2}}\right)  -\frac{\vec{r}_{23}{}^{2}\vec{r}_{12}{}^{2}%
}{\vec{r}_{02}{}^{2}\vec{r}_{03}{}^{2}\vec{r}_{14}{}^{2}\vec{r}_{24}{}^{2}}%
\ln\left(  \frac{\vec{r}_{02}{}^{2}\vec{r}_{14}{}^{2}\vec{r}_{23}{}^{2}}%
{\vec{r}_{03}{}^{2}\vec{r}_{12}{}^{2}\vec{r}_{24}{}^{2}}\right)
\]%
\begin{equation}
+\frac{\vec{r}_{23}{}^{2}}{\vec{r}_{03}{}^{2}\vec{r}_{04}{}^{2}\vec{r}_{24}%
{}^{2}}\ln\left(  \frac{\vec{r}_{02}{}^{2}\vec{r}_{34}{}^{2}}{\vec{r}_{04}%
{}^{2}\vec{r}_{23}{}^{2}}\right)  +\frac{\vec{r}_{02}{}^{2}\vec{r}_{13}{}^{2}%
}{\vec{r}_{01}{}^{2}\vec{r}_{03}{}^{2}\vec{r}_{04}{}^{2}\vec{r}_{24}{}^{2}}%
\ln\left(  \frac{\vec{r}_{01}{}^{2}\vec{r}_{02}{}^{2}\vec{r}_{34}{}^{4}{}%
}{\vec{r}_{03}{}^{2}\vec{r}_{04}{}^{2}\vec{r}_{13}{}^{2}\vec{r}_{24}{}^{2}%
}\right)  . \label{F140}%
\end{equation}%
\[
F_{100}=\frac{\vec{r}_{23}{}^{2}}{2\vec{r}_{03}{}^{2}\vec{r}_{04}{}^{2}\vec
{r}_{24}{}^{2}}\ln\left(  \frac{\vec{r}_{01}{}^{8}{}\vec{r}_{04}{}^{2}\vec
{r}_{23}{}^{2}\vec{r}_{24}{}^{4}{}\vec{r}_{34}{}^{2}}{\vec{r}_{02}{}^{6}{}%
\vec{r}_{03}{}^{4}{}\vec{r}_{14}{}^{8}{}}\right)  -\frac{\vec{r}_{02}{}%
^{2}\vec{r}_{13}{}^{2}}{\vec{r}_{01}{}^{2}\vec{r}_{03}{}^{2}\vec{r}_{04}{}%
^{2}\vec{r}_{24}{}^{2}}\ln\left(  \frac{\vec{r}_{01}{}^{2}\vec{r}_{04}{}%
^{2}\vec{r}_{13}{}^{2}\vec{r}_{24}{}^{2}}{\vec{r}_{02}{}^{2}\vec{r}_{03}{}%
^{2}\vec{r}_{14}{}^{4}{}}\right)
\]%
\[
-\frac{\vec{r}_{34}{}^{2}\vec{r}_{12}{}^{2}}{2\vec{r}_{03}{}^{2}\vec{r}_{04}%
{}^{2}\vec{r}_{14}{}^{2}\vec{r}_{24}{}^{2}}\ln\left(  \frac{\vec{r}_{01}{}%
^{8}{}\vec{r}_{02}{}^{4}{}\vec{r}_{24}{}^{2}\vec{r}_{34}{}^{6}{}}{\vec{r}%
_{03}{}^{6}{}\vec{r}_{04}{}^{6}{}\vec{r}_{12}{}^{6}{}\vec{r}_{14}{}^{2}%
}\right)  -\frac{\vec{r}_{12}{}^{2}}{2\vec{r}_{02}{}^{2}\vec{r}_{04}{}^{2}%
\vec{r}_{14}{}^{2}}\ln\left(  \frac{\vec{r}_{01}{}^{4}{}\vec{r}_{02}{}^{2}%
\vec{r}_{34}{}{}^{4}}{\vec{r}_{03}{}^{4}{}\vec{r}_{04}{}^{2}\vec{r}_{12}{}%
^{2}\vec{r}_{14}{}^{2}}\right)
\]%
\[
+\frac{\vec{r}_{23}{}^{2}\vec{r}_{12}{}^{2}}{2\vec{r}_{02}{}^{2}\vec{r}_{03}%
{}^{2}\vec{r}_{14}{}^{2}\vec{r}_{24}{}^{2}}\ln\left(  \frac{\vec{r}_{02}{}%
^{2}\vec{r}_{14}{}^{2}\vec{r}_{23}{}^{2}}{\vec{r}_{03}{}^{2}\vec{r}_{12}{}%
^{2}\vec{r}_{24}{}^{2}}\right)  +\frac{\vec{r}_{12}{}^{2}}{\vec{r}_{01}{}%
^{2}\vec{r}_{04}{}^{2}\vec{r}_{24}{}^{2}}\ln\left(  \frac{\vec{r}_{02}{}%
^{2}\vec{r}_{14}{}^{2}}{\vec{r}_{04}{}^{2}\vec{r}_{12}{}^{2}}\right)
\]%
\begin{equation}
+\frac{\vec{r}_{13}{}^{2}\vec{r}_{12}{}^{2}}{\vec{r}_{01}{}^{2}\vec{r}_{03}%
{}^{2}\vec{r}_{14}{}^{2}\vec{r}_{24}{}^{2}}\ln\left(  \frac{\vec{r}_{01}{}%
^{2}\vec{r}_{13}{}^{2}\vec{r}_{24}{}^{2}}{\vec{r}_{03}{}^{2}\vec{r}_{12}{}%
^{2}\vec{r}_{14}{}^{2}}\right)  +\frac{\vec{r}_{01}{}^{2}\vec{r}_{23}{}^{2}%
}{2\vec{r}_{02}{}^{2}\vec{r}_{03}{}^{2}\vec{r}_{04}{}^{2}\vec{r}_{14}{}^{2}%
}\ln\left(  \frac{\vec{r}_{01}{}^{2}\vec{r}_{24}{}^{2}\vec{r}_{34}{}^{2}}%
{\vec{r}_{04}{}^{2}\vec{r}_{14}{}^{2}\vec{r}_{23}{}^{2}}\right)  .
\label{F100}%
\end{equation}%
\[
F_{230}=\frac{\vec{r}_{02}{}^{2}\vec{r}_{13}{}^{2}}{2\vec{r}_{01}{}^{2}\vec
{r}_{03}{}^{2}\vec{r}_{04}{}^{2}\vec{r}_{24}{}^{2}}\ln\left(  \frac{\vec
{r}_{01}{}^{2}\vec{r}_{04}{}^{2}\vec{r}_{13}{}^{2}\vec{r}_{24}{}^{2}}{\vec
{r}_{02}{}^{2}\vec{r}_{03}{}^{2}\vec{r}_{14}{}^{4}{}}\right)  -\frac{\vec
{r}_{23}{}^{2}}{2\vec{r}_{03}{}^{2}\vec{r}_{04}{}^{2}\vec{r}_{24}{}^{2}}%
\ln\left(  \frac{\vec{r}_{01}{}^{4}{}\vec{r}_{04}{}^{4}{}\vec{r}_{23}{}^{4}%
{}\vec{r}_{24}{}^{2}}{\vec{r}_{02}{}^{6}{}\vec{r}_{03}{}^{2}\vec{r}_{14}{}%
^{4}{}\vec{r}_{34}{}^{2}}\right)
\]%
\[
+\frac{\vec{r}_{34}{}^{2}\vec{r}_{12}{}^{2}}{2\vec{r}_{03}{}^{2}\vec{r}_{04}%
{}^{2}\vec{r}_{14}{}^{2}\vec{r}_{24}{}^{2}}\ln\left(  \frac{\vec{r}_{01}{}%
^{4}{}\vec{r}_{02}{}^{8}{}\vec{r}_{14}{}^{2}\vec{r}_{34}{}^{6}{}}{\vec{r}%
_{03}{}^{6}{}\vec{r}_{04}{}^{6}{}\vec{r}_{12}{}^{6}{}\vec{r}_{24}{}^{2}%
}\right)  +\frac{\vec{r}_{12}{}^{2}}{2\vec{r}_{02}{}^{2}\vec{r}_{04}{}^{2}%
\vec{r}_{14}{}^{2}}\ln\left(  \frac{\vec{r}_{01}{}^{2}\vec{r}_{02}{}^{4}{}%
\vec{r}_{04}{}^{2}\vec{r}_{12}{}^{2}\vec{r}_{34}{}^{8}{}}{\vec{r}_{03}{}^{8}%
{}\vec{r}_{14}{}^{4}{}\vec{r}_{24}{}^{6}{}}\right)
\]%
\[
-\frac{\vec{r}_{23}{}^{2}\vec{r}_{12}{}^{2}}{\vec{r}_{02}{}^{2}\vec{r}_{03}%
{}^{2}\vec{r}_{14}{}^{2}\vec{r}_{24}{}^{2}}\ln\left(  \frac{\vec{r}_{02}{}%
^{2}\vec{r}_{14}{}^{2}\vec{r}_{23}{}^{2}}{\vec{r}_{03}{}^{2}\vec{r}_{12}{}%
^{2}\vec{r}_{24}{}^{2}}\right)  -\frac{\vec{r}_{12}{}^{2}}{2\vec{r}_{01}{}%
^{2}\vec{r}_{04}{}^{2}\vec{r}_{24}{}^{2}}\ln\left(  \frac{\vec{r}_{02}{}%
^{2}\vec{r}_{14}{}^{2}}{\vec{r}_{04}{}^{2}\vec{r}_{12}{}^{2}}\right)
\]%
\begin{equation}
-\frac{\vec{r}_{13}{}^{2}\vec{r}_{12}{}^{2}}{2\vec{r}_{01}{}^{2}\vec{r}_{03}%
{}^{2}\vec{r}_{14}{}^{2}\vec{r}_{24}{}^{2}}\ln\left(  \frac{\vec{r}_{01}{}%
^{2}\vec{r}_{13}{}^{2}\vec{r}_{24}{}^{2}}{\vec{r}_{03}{}^{2}\vec{r}_{12}{}%
^{2}\vec{r}_{14}{}^{2}}\right)  -\frac{\vec{r}_{01}{}^{2}\vec{r}_{23}{}^{2}%
}{\vec{r}_{02}{}^{2}\vec{r}_{03}{}^{2}\vec{r}_{04}{}^{2}\vec{r}_{14}{}^{2}}%
\ln\left(  \frac{\vec{r}_{01}{}^{2}\vec{r}_{24}{}^{2}\vec{r}_{34}{}^{2}}%
{\vec{r}_{04}{}^{2}\vec{r}_{14}{}^{2}\vec{r}_{23}{}^{2}}\right)  .
\label{F230}%
\end{equation}
One can integrate $F_{100}$ and $F_{230}$ w.r.t. $\vec{r}_{4}.$ The integrals
are given in appendix \ref{integrals} (\ref{F100integrated}) and
(\ref{F230integrated}).

The color structure in the quark part of the kernel can be linearized via
(\ref{(102)(201)})
\[
\frac{1}{2}\left\{  \left(  \frac{1}{3}(U_{1}U_{0}{}^{\dag}U_{4}+U_{4}U_{0}%
{}^{\dag}U_{1})\cdot U_{2}\cdot U_{3}-\frac{1}{9}B_{123}tr(U_{0}{}^{\dag}%
U_{4})+(U_{1}U_{0}{}^{\dag}U_{2})\cdot U_{3}\cdot U_{4}\right.  \right.
\]%
\[
\left.  \left.  \frac{{}}{{}}+\frac{1}{6}B_{123}-\frac{1}{4}(B_{013}%
B_{002}+B_{001}B_{023}-B_{012}B_{003})+(1\leftrightarrow2)\right)
+(0\leftrightarrow4)\right\}
\]%
\[
\overset{\mathrm{3g}}{=}\frac{1}{6}\left(  12-B_{004}-B_{044}+2\left(
2B_{014}-B_{001}-B_{144}\right)  \right.
\]%
\begin{equation}
+2\left(  2B_{024}-B_{002}-B_{244}\right)  -4\left(  2B_{034}-B_{344}%
-B_{003}\right)  ).
\end{equation}
Therefore%
\[
-{\frac{\alpha_{s}^{2}n_{f}}{8\pi^{4}}}\!\int\!d\vec{r}_{0}d\vec{r}%
_{4}~\mathbf{G}^{q}\overset{\mathrm{3g}}{=}-{\frac{\alpha_{s}^{2}n_{f}}%
{48\pi^{4}}}\!\int\!d\vec{r}_{0}d\vec{r}_{4}~\left\{  \left(  12-B_{004}%
-B_{044}\right)  (L_{12}^{q}+L_{13}^{q}+L_{23}^{q})\right.
\]%
\[
+2\left(  2B_{014}-B_{001}-B_{144}\right)  (L_{12}^{q}+L_{13}^{q}-2L_{32}%
^{q})+2\left(  2B_{024}-B_{002}-B_{244}\right)  (L_{12}^{q}+L_{23}^{q}%
-2L_{31}^{q})
\]%
\begin{equation}
\left.  +2\left(  2B_{034}-B_{344}-B_{003}\right)  (L_{32}^{q}+L_{13}%
^{q}-2L_{12}^{q})\right\}  .
\end{equation}
Putting things together we have for the linearized kernel
\[
\langle K_{NLO}\otimes B_{123}^{conf}\rangle\overset{\mathrm{3g}}{=}%
{\frac{27\alpha_{s}^{2}}{4\pi^{2}}}\zeta(3)(3-\delta_{23}-\delta_{13}%
-\delta_{21})(B_{123}-6)
\]%
\[
-{\frac{9\alpha_{s}^{2}}{8\pi^{4}}}\!\int\!d\vec{r}_{0}d\vec{r}_{4}\left(
L_{12}^{C}+L_{13}^{C}+L_{23}^{C}-\frac{n_{f}}{54}(L_{12}^{q}+L_{13}^{q}%
+L_{23}^{q})\right)  (B_{044}+B_{004}-12)
\]%
\[
-{\frac{\alpha_{s}^{2}n_{f}}{24\pi^{4}}}\!\int\!d\vec{r}_{0}d\vec{r}%
_{4}~\left\{  \left(  2B_{014}-B_{001}-B_{144}\right)  (L_{12}^{q}+L_{13}%
^{q}-2L_{32}^{q})+(1\leftrightarrow3)+(1\leftrightarrow2)\right\}
\]%
\[
-{\frac{9\alpha_{s}^{2}}{64\pi^{4}}}\!\int\!d\vec{r}_{0}d\vec{r}_{4}~\left(
F_{0}(B_{040}-B_{044})+\left\{  F_{140}+(0\leftrightarrow4)\right\}
B_{140}+(\text{all 5 perm.}\,1\leftrightarrow2\leftrightarrow3)\right)
\]%
\[
-{\frac{9\alpha_{s}^{2}}{64\pi^{3}}}\!\int\!d\vec{r}_{0}\left(  \tilde
{F}_{100}B_{100}+\tilde{F}_{230}B_{230}+(1\leftrightarrow3)+(1\leftrightarrow
2)\right)
\]%
\[
-{\frac{9\alpha_{s}^{2}}{16\pi^{3}}}\!\int\!d\vec{r}_{0}\left(  \beta\left[
\ln\left(  \frac{\vec{r}_{01}^{\,\,2}}{\vec{r}_{02}^{\,\,2}}\right)  \left(
\frac{1}{\vec{r}_{02}^{\,\,2}}-\frac{1}{\vec{r}_{01}^{\,\,2}}\right)
-\frac{\vec{r}_{12}^{\,\,2}}{\vec{r}_{01}^{\,\,2}\vec{r}_{02}^{\,\,2}}%
\ln\left(  \frac{\vec{r}_{12}^{\,\,2}}{\tilde{\mu}^{2}}\right)  \right]
\right.
\]%
\begin{equation}
\left.  \times\left(  B_{100}+B_{230}+B_{200}+B_{130}-B_{300}-B_{210}%
-B_{123}-6\right)  +(1\leftrightarrow3)+(2\leftrightarrow3)\right)  .
\label{QCkernel3g}%
\end{equation}
Here $\delta_{ij}=1,$ if $\vec{r}_{i}=\vec{r}_{j}$ and $\delta_{ij}=0$
otherwise; $F_{0}$ and $F_{140}$ are defined in (\ref{F0}) and (\ref{F140});
$L_{12}^{q}$ is defined in (\ref{Lq}) and $L_{12}^{C}$ is defined in
(\ref{Lc}) \
\[
L_{12}^{C}=\frac{\vec{r}_{12}{}^{2}}{4\vec{r}_{01}{}^{2}\vec{r}_{04}{}^{2}%
\vec{r}_{24}{}^{2}}\ln\left(  \frac{\vec{r}_{02}{}^{2}\vec{r}_{14}{}^{2}}%
{\vec{r}_{04}{}^{2}\vec{r}_{12}{}^{2}}\right)  +\frac{\vec{r}_{12}{}^{2}%
}{4\vec{r}_{02}{}^{2}\vec{r}_{04}{}^{2}\vec{r}_{14}{}^{2}}\ln\left(
\frac{\vec{r}_{01}{}^{2}\vec{r}_{24}{}^{2}}{\vec{r}_{04}{}^{2}\vec{r}_{12}%
{}^{2}}\right)
\]%
\[
+\left[  \frac{1}{\vec{r}_{01}{}^{2}\vec{r}_{24}{}^{2}-\vec{r}_{02}{}^{2}%
\vec{r}_{14}{}^{2}}\left(  -\frac{\vec{r}_{12}{}^{4}}{8}\left(  \frac{1}%
{\vec{r}_{01}{}^{2}\vec{r}_{24}{}^{2}}+\frac{1}{\vec{r}_{02}{}^{2}\vec{r}%
_{14}{}^{2}}\right)  +\frac{\vec{r}_{12}{}^{2}}{\vec{r}_{04}{}^{2}}-\frac
{\vec{r}_{02}{}^{2}\vec{r}_{14}{}^{2}+\vec{r}_{01}{}^{2}\vec{r}_{24}{}^{2}%
}{4\vec{r}_{04}{}^{4}{}}\right)  \right.
\]%
\begin{equation}
\left.  +\frac{\vec{r}_{12}{}^{2}}{8\vec{r}_{04}{}^{2}}\left(  \frac{1}%
{\vec{r}_{02}{}^{2}\vec{r}_{14}{}^{2}}-\frac{1}{\vec{r}_{01}{}^{2}\vec{r}%
_{24}{}^{2}}\right)  \right]  \ln\left(  \frac{\vec{r}_{01}{}^{2}\vec{r}%
_{24}{}^{2}}{\vec{r}_{14}{}^{2}\vec{r}_{02}{}^{2}}\right)  +\frac{1}{2\vec
{r}_{04}{}^{4}{}} \label{Lc_apparent}%
\end{equation}
and%
\[
\tilde{F}_{100}=\left(  \frac{\vec{r}_{12}{}^{2}}{\vec{r}_{01}{}^{2}\vec
{r}_{02}{}^{2}}-\frac{\vec{r}_{13}{}^{2}}{\vec{r}_{01}{}^{2}\vec{r}_{03}{}%
^{2}}-\frac{2\vec{r}_{23}{}^{2}}{\vec{r}_{02}{}^{2}\vec{r}_{03}{}^{2}}\right)
\ln^{2}\left(  \frac{\vec{r}_{02}{}^{2}\vec{r}_{13}{}^{2}}{\vec{r}_{01}{}%
^{2}\vec{r}_{23}{}^{2}}\right)  +\frac{\vec{r}_{23}{}^{2}}{2\vec{r}_{02}{}%
^{2}\vec{r}_{03}{}^{2}}\ln^{2}\left(  \frac{\vec{r}_{03}{}^{2}\vec{r}_{12}%
{}^{2}}{\vec{r}_{02}{}^{2}\vec{r}_{13}{}^{2}}\right)
\]%
\begin{equation}
+\tilde{S}_{123}I\left(  \frac{\vec{r}_{12}{}^{2}}{\vec{r}_{01}{}^{2}\vec
{r}_{02}{}^{2}},\frac{\vec{r}_{13}{}^{2}}{\vec{r}_{01}{}^{2}\vec{r}_{03}{}%
^{2}},\frac{\vec{r}_{23}{}^{2}}{\vec{r}_{02}{}^{2}\vec{r}_{03}{}^{2}}\right)
+(2\leftrightarrow3), \label{F100tilde}%
\end{equation}%
\[
\tilde{F}_{230}=\left(  \frac{2\vec{r}_{12}{}^{2}}{\vec{r}_{01}{}^{2}\vec
{r}_{02}{}^{2}}-\frac{\vec{r}_{23}{}^{2}}{2\vec{r}_{02}{}^{2}\vec{r}_{03}%
{}^{2}}\right)  \ln^{2}\left(  \frac{\vec{r}_{03}{}^{2}\vec{r}_{12}{}^{2}%
}{\vec{r}_{02}{}^{2}\vec{r}_{13}{}^{2}}\right)  +\left(  \frac{\vec{r}_{13}%
{}^{2}}{\vec{r}_{01}{}^{2}\vec{r}_{03}{}^{2}}-\frac{\vec{r}_{12}{}^{2}}%
{\vec{r}_{01}{}^{2}\vec{r}_{02}{}^{2}}\right)  \ln^{2}\left(  \frac{\vec
{r}_{02}{}^{2}\vec{r}_{13}{}^{2}}{\vec{r}_{01}{}^{2}\vec{r}_{23}{}^{2}%
}\right)
\]%
\begin{equation}
-\tilde{S}_{123}I\left(  \frac{\vec{r}_{12}{}^{2}}{\vec{r}_{01}{}^{2}\vec
{r}_{02}{}^{2}},\frac{\vec{r}_{13}{}^{2}}{\vec{r}_{01}{}^{2}\vec{r}_{03}{}%
^{2}},\frac{\vec{r}_{23}{}^{2}}{\vec{r}_{02}{}^{2}\vec{r}_{03}{}^{2}}\right)
+(2\leftrightarrow3). \label{F230tilde}%
\end{equation}
The functions $\tilde{S}_{123}$ and $I$ are defined in appendix
\ref{integrals} (\ref{Sinv}) and (\ref{integral I}). If we consider the dipole
limit $\vec{r}_{3}=\vec{r}_{2}$ and take into account that in this limit%
\begin{equation}
\tilde{F}_{230}|_{\vec{r}_{3}=\vec{r}_{2}}=\tilde{F}_{130}|_{\vec{r}_{3}%
=\vec{r}_{2}}=\tilde{F}_{210}|_{\vec{r}_{3}=\vec{r}_{2}}=\tilde{F}%
_{100}|_{\vec{r}_{3}=\vec{r}_{2}}=\tilde{F}_{200}|_{\vec{r}_{3}=\vec{r}_{2}%
}=\tilde{F}_{300}|_{\vec{r}_{3}=\vec{r}_{2}}=0,
\end{equation}%
\begin{equation}
F_{0}+(\text{all 5 permutations}\,1\leftrightarrow2\leftrightarrow3)|_{\vec
{r}_{3}=\vec{r}_{2}}=-16\tilde{L}_{12}^{C},
\end{equation}%
\[
(F_{140}+(0\leftrightarrow4))+(2\leftrightarrow3)|_{\vec{r}_{3}=\vec{r}_{2}%
}=0,
\]%
\begin{equation}
(F_{240}+(0\leftrightarrow4))+(1\leftrightarrow3)|_{\vec{r}_{3}=\vec{r}_{2}%
}=(F_{340}+(0\leftrightarrow4))+(2\leftrightarrow1)|_{\vec{r}_{3}=\vec{r}_{2}%
}=0,
\end{equation}
we have the linearized BK kernel in the 3-gluon approximation, whose C-even
part is the BFKL kernel \cite{Fadin:2009gh}%
\[
\langle K_{NLO}\otimes B_{122}^{conf}\rangle\overset{\mathrm{3g}}{=}%
-{\frac{9\alpha_{s}^{2}}{4\pi^{4}}}\!\int\!d\vec{r}_{0}d\vec{r}_{4}\left(
L_{12}^{C}-\frac{n_{f}}{54}L_{12}^{q}\right)  (B_{044}+B_{004}-12)
\]%
\[
-{\frac{\alpha_{s}^{2}n_{f}}{12\pi^{4}}}\!\int\!d\vec{r}_{0}d\vec{r}%
_{4}\left\{  \left(  2B_{014}-B_{001}-B_{144}\right)  -\left(  2B_{024}%
-B_{002}-B_{244}\right)  \right\}  L_{12}^{q}%
\]%
\[
+{\frac{27\alpha_{s}^{2}}{2\pi^{2}}}\zeta\left(  3\right)  (B_{122}%
-6)-{\frac{9\alpha_{s}^{2}}{4\pi^{4}}}\!\int\!d\vec{r}_{0}d\vec{r}_{4}%
~\tilde{L}_{12}^{C}(B_{044}-B_{040})
\]%
\begin{equation}
-{\frac{9\alpha_{s}^{2}}{8\pi^{3}}}\!\beta\int\!d\vec{r}_{0}\left[  \ln\left(
\frac{\vec{r}_{01}^{\,\,2}}{\vec{r}_{02}^{\,\,2}}\right)  \left(  \frac
{1}{\vec{r}_{02}^{\,\,2}}-\frac{1}{\vec{r}_{01}^{\,\,2}}\right)  -\frac
{\vec{r}_{12}^{\,\,2}}{\vec{r}_{01}^{\,\,2}\vec{r}_{02}^{\,\,2}}\ln\left(
\frac{\vec{r}_{12}^{\,\,2}}{\tilde{\mu}^{2}}\right)  \right]  \left(
B_{100}+B_{220}-B_{122}-6\right)  . \label{BK3g}%
\end{equation}
Let us compare this kernel for $B_{122}=2tr(U_{1}U_{2}^{\dag})$ with the
linearized BK kernel in the 2-gluon approximation from \cite{Balitsky:2009xg}.
One can see that their C-even parts coincide as they are fixed by the
BFKL\ kernel \cite{Fadin:2009gh}. However the 2-gluon approximation is not
enough to catch the correct C-odd part of the kernel. But the 3-gluon
approximation (\ref{BK3g}) allows one to write it. At once we see that even
for the color dipole the C-odd part of the kernel in the 3-gluon approximation
can not be expressed through dipoles only. One necessarily needs to introduce
the 3QWL operators as is clear from the second line of this expression. One
can check it by direct calculation via (\ref{(102)(201)}), indeed%
\[
12\left\{  tr(U_{2}^{\dag}t^{a}U_{1}t^{b})tr(U_{4}^{\dag}t^{a}U_{0}%
t^{b})-(0\rightarrow4)\right\}  +(0\leftrightarrow4)
\]%
\[
=\left\{  \frac{1}{3}tr(U_{0}{}U_{4}^{\dag})tr(U_{2}{}^{\dag}U_{1}%
)+3tr(U_{4}{}^{\dag}U_{1})tr(U_{2}{}^{\dag}U_{0})\right.
\]%
\[
\left.  \frac{{}}{{}}-tr(U_{0}{}U_{4}^{\dag}U_{1}U_{2}{}^{\dag})-tr(U_{0}%
{}U_{2}{}^{\dag}U_{1}U_{4}^{\dag})-(0\rightarrow4)\right\}  +(0\leftrightarrow
4)
\]%
\[
=\left\{  \frac{1}{12}B_{044}{}B_{122}{}+\frac{3}{4}B_{144}{}B_{022}{}%
-\frac{1}{2}tr(U_{1}U_{2}{}^{\dag}U_{0}{}+U_{0}{}U_{2}{}^{\dag}U_{1})\cdot
U_{4}\cdot U_{4}-(0\rightarrow4)\right\}  +(0\leftrightarrow4)
\]%
\begin{equation}
\overset{\mathrm{3g}}{=}\frac{1}{2}\left\{  12-B_{044}-B_{004}+2(2B_{014}%
-B_{001}-B_{144})-2(2B_{024}-B_{002}-B_{244})\right\}  .
\end{equation}

As in \cite{Gerasimov:2012bj}\ to separate the C-even and C-odd contributions
we introduce C-even (pomeron) and C-odd (odderon) Green functions%
\begin{equation}
B_{123}^{+}=B_{123}+B_{\bar{1}\bar{2}\bar{3}}-12, \label{pomeronFG}%
\end{equation}
and%
\begin{equation}
B_{123}^{-}=B_{123}-B_{\bar{1}\bar{2}\bar{3}}, \label{odderonFG}%
\end{equation}
where $B_{\bar{1}\bar{2}\bar{3}}$ is the 3-antiquark Wilson loop operator%
\begin{equation}
B_{\bar{1}\bar{2}\bar{3}}=U_{1}^{\dag}\cdot U_{2}^{\dag}\cdot U_{3}^{\dag}.
\end{equation}
The NLO kernel for the C-even Green function in the 3-gluon approximation
reads%
\[
\langle K_{NLO}\otimes B_{123}^{+conf}\rangle\overset{\mathrm{3g}}{=}%
-{\frac{9\alpha_{s}^{2}}{4\pi^{4}}}\!\int\!d\vec{r}_{0}d\vec{r}_{4}(L_{12}%
^{C}+L_{13}^{C}+L_{23}^{C}-\frac{n_{f}}{54}(L_{12}^{q}+L_{13}^{q}+L_{23}%
^{q}))B_{044}^{+}%
\]%
\[
+{\frac{27\alpha_{s}^{2}}{4\pi^{2}}}\zeta(3)(3-\delta_{23}-\delta_{13}%
-\delta_{21})B_{123}^{+}%
\]%
\[
-{\frac{9\alpha_{s}^{2}}{64\pi^{4}}}\!\int\!d\vec{r}_{0}d\vec{r}_{4}~\left(
\left\{  F_{140}+(0\leftrightarrow4)\right\}  B_{140}^{+}+(\text{all 5
permutations}\,1\leftrightarrow2\leftrightarrow3)\right)
\]%
\[
-{\frac{9\alpha_{s}^{2}}{64\pi^{3}}}\!\int\!d\vec{r}_{0}\left(  \tilde
{F}_{100}B_{100}^{+}+\tilde{F}_{230}B_{230}^{+}+(1\leftrightarrow
3)+(1\leftrightarrow2)\right)
\]%
\[
-{\frac{9\alpha_{s}^{2}}{16\pi^{3}}}\!\int\!d\vec{r}_{0}\left(  \beta\left[
\ln\left(  \frac{\vec{r}_{01}^{\,\,2}}{\vec{r}_{02}^{\,\,2}}\right)  \left(
\frac{1}{\vec{r}_{02}^{\,\,2}}-\frac{1}{\vec{r}_{01}^{\,\,2}}\right)
-\frac{\vec{r}_{12}^{\,\,2}}{\vec{r}_{01}^{\,\,2}\vec{r}_{02}^{\,\,2}}%
\ln\left(  \frac{\vec{r}_{12}^{\,\,2}}{\tilde{\mu}^{2}}\right)  \right]
\right.
\]%
\begin{equation}
\left.  \times\left(  B_{100}^{+}+B_{230}^{+}+B_{200}^{+}+B_{130}^{+}%
-B_{300}^{+}-B_{210}^{+}-B_{123}^{+}\right)  +(1\leftrightarrow
3)+(2\leftrightarrow3)\right)  . \label{CevenKnlo3g}%
\end{equation}
In the 3-gluon approximation \cite{Gerasimov:2012bj}
\begin{equation}
B_{123}^{+}\overset{\mathrm{3g}}{=}\frac{1}{2}(B_{133}^{+}+B_{211}^{+}%
+B_{322}^{+}),
\end{equation}
which kills all the terms in the third line in (\ref{QCkernel3g}) in the
C-even case. This equality also leads to the fact that for model (\ref{model})%
\begin{equation}
B_{123}^{+conf}\overset{\mathrm{3g}}{=}\frac{1}{2}(B_{133}^{+conf}%
+B_{211}^{+conf}+B_{322}^{+conf})
\end{equation}
we have%
\begin{equation}
\langle K_{NLO}\otimes B_{123}^{+conf}\rangle\overset{\mathrm{3g}}{=}\frac
{1}{2}\langle K_{NLO}\otimes(B_{133}^{+conf}+B_{211}^{+conf}+B_{322}%
^{+conf})\rangle.
\end{equation}
This equality imposes the following constraints%
\begin{equation}
0=~\left\{  F_{140}+(0\leftrightarrow4)\right\}  +(\text{all 5 permutations}%
\,1\leftrightarrow2\leftrightarrow3), \label{constraint1}%
\end{equation}%
\begin{equation}
0=\int\!d\vec{r}_{0}\tilde{F}_{230}, \label{constraint2}%
\end{equation}%
\begin{equation}
0=\!\int\frac{d\vec{r}_{4}}{\pi}~\left(  \left\{  F_{140}+(0\leftrightarrow
4)\right\}  +(2\leftrightarrow3)\right)  +\tilde{F}_{100}+\frac{1}{2}\tilde
{F}_{230}|_{1\leftrightarrow3}+\frac{1}{2}\tilde{F}_{230}|_{1\leftrightarrow
2}. \label{constraint3}%
\end{equation}
Constraint (\ref{constraint1}) follows from definition of $F_{140}$
(\ref{F140}) directly. Constraint (\ref{constraint2}) holds since thanks to
conformal invariance
\begin{equation}
\int\!d\vec{r}_{0}\tilde{F}_{230}=\int\!d\vec{r}_{0}\tilde{F}_{230}|_{2=3}=0.
\end{equation}
Using (\ref{F100tilde}) and (\ref{F230tilde}) one can rewrite constraint
(\ref{constraint3}) as%
\[
\!\int\frac{d\vec{r}_{4}}{\pi}~\left(  \left\{  F_{140}+(0\leftrightarrow
4)\right\}  +(2\leftrightarrow3)\right)  =\frac{\vec{r}_{23}{}^{2}}{2\vec
{r}_{02}{}^{2}\vec{r}_{03}{}^{2}}\left(  \ln^{2}\left(  \frac{\vec{r}_{03}%
{}^{2}\vec{r}_{12}{}^{2}}{\vec{r}_{01}{}^{2}\vec{r}_{23}{}^{2}}\right)
+\ln^{2}\left(  \frac{\vec{r}_{02}{}^{2}\vec{r}_{13}{}^{2}}{\vec{r}_{01}{}%
^{2}\vec{r}_{23}{}^{2}}\right)  \right)
\]%
\begin{equation}
-\frac{1}{2}\left(  \frac{\vec{r}_{12}{}^{2}}{\vec{r}_{01}{}^{2}\vec{r}_{02}%
{}^{2}}+\frac{\vec{r}_{13}{}^{2}}{\vec{r}_{01}{}^{2}\vec{r}_{03}{}^{2}%
}\right)  \ln^{2}\left(  \frac{\vec{r}_{02}{}^{2}\vec{r}_{13}{}^{2}}{\vec
{r}_{03}{}^{2}\vec{r}_{12}{}^{2}}\right)  . \label{constraint3-1}%
\end{equation}
The calculation of the integral and proof of this identity is given in
appendix \ref{integrals}.

The NLO kernel for the C-odd Green function in the 3-gluon approximation reads%
\[
\langle K_{NLO}\otimes B_{123}^{-conf}\rangle\overset{\mathrm{3g}}{=}%
{\frac{27\alpha_{s}^{2}}{4\pi^{2}}}\zeta(3)(3-\delta_{23}-\delta_{13}%
-\delta_{21})B_{123}^{-}%
\]%
\[
-{\frac{\alpha_{s}^{2}n_{f}}{24\pi^{4}}}\!\int\!d\vec{r}_{0}d\vec{r}%
_{4}~\left\{  \left(  2B_{014}^{-}-B_{001}^{-}-B_{144}^{-}\right)  (L_{12}%
^{q}+L_{13}^{q}-2L_{32}^{q})+(1\leftrightarrow3)+(1\leftrightarrow2)\right\}
\]%
\[
-{\frac{9\alpha_{s}^{2}}{64\pi^{4}}}\!\int\!d\vec{r}_{0}d\vec{r}_{4}~\left(
2F_{0}B_{040}^{-}+\left\{  F_{140}+(0\leftrightarrow4)\right\}  B_{140}%
^{-}+(\text{all 5 permutations}\,1\leftrightarrow2\leftrightarrow3)\right)
\]%
\[
-{\frac{9\alpha_{s}^{2}}{64\pi^{3}}}\!\int\!d\vec{r}_{0}\left(  \tilde
{F}_{100}B_{100}^{-}+\tilde{F}_{230}B_{230}^{-}+(1\leftrightarrow
3)+(1\leftrightarrow2)\right)
\]%
\[
-{\frac{9\alpha_{s}^{2}}{16\pi^{3}}}\!\int\!d\vec{r}_{0}\left(  \beta\left[
\ln\left(  \frac{\vec{r}_{01}^{\,\,2}}{\vec{r}_{02}^{\,\,2}}\right)  \left(
\frac{1}{\vec{r}_{02}^{\,\,2}}-\frac{1}{\vec{r}_{01}^{\,\,2}}\right)
-\frac{\vec{r}_{12}^{\,\,2}}{\vec{r}_{01}^{\,\,2}\vec{r}_{02}^{\,\,2}}%
\ln\left(  \frac{\vec{r}_{12}^{\,\,2}}{\tilde{\mu}^{2}}\right)  \right]
\right.
\]%
\begin{equation}
\left.  \times\left(  B_{100}^{-}+B_{230}^{-}+B_{200}^{-}+B_{130}^{-}%
-B_{300}^{-}-B_{210}^{-}-B_{123}^{-}\right)  +(1\leftrightarrow
3)+(2\leftrightarrow3)\right)  . \label{CoddKnlo3g}%
\end{equation}

\section{Results}

In this section we list the main results of the paper. Taking the LO equation
(\ref{LO})  and using (\ref{Gall}) we can write the NLO evolution equation for
3QWL operator as
\[
\frac{\partial B_{123}}{\partial\eta}=\frac{\alpha_{s}(\mu^{2})}{8\pi^{2}}\int
d\vec{r}_{0}\left[  (B_{100}B_{320}+B_{200}B_{310}-B_{300}B_{210}%
-6B_{123})\frac{{}}{{}}\right.
\]%
\[
~\times\left\{  \frac{\vec{r}_{12}^{\,\,2}}{\vec{r}_{01}^{\,\,2}\vec{r}%
_{02}^{\,\,2}}-{\frac{3\alpha_{s}}{4\pi}}\!\beta\left[  \ln\left(  \frac
{\vec{r}_{01}^{\,\,2}}{\vec{r}_{02}^{\,\,2}}\right)  \left(  \frac{1}{\vec
{r}_{02}^{\,\,2}}-\frac{1}{\vec{r}_{01}^{\,\,2}}\right)  -\frac{\vec{r}%
_{12}^{\,\,2}}{\vec{r}_{01}^{\,\,2}\vec{r}_{02}^{\,\,2}}\ln\left(  \frac
{\vec{r}_{12}^{\,\,2}}{\tilde{\mu}^{2}}\right)  \right]  \right\}
\]%
\[
-{\frac{\alpha_{s}}{\pi}}\!~\ln\frac{\vec{r}_{20}^{\,\,2}}{\vec{r}%
_{21}^{\,\,2}}\ln\frac{\vec{r}_{10}^{\,\,2}}{\vec{r}_{21}^{\,\,2}}\left\{
\frac{1}{2}\left[  \frac{\vec{r}_{13}^{\,\,2}}{\vec{r}_{10}^{\,\,2}\vec
{r}_{30}^{\,\,\,2}}-\frac{\vec{r}_{32}^{\,2}}{\vec{r}_{30}^{\,2}\vec{r}%
_{20}^{\,\,\,2}}\right]  \left(  B_{100}B_{320}-B_{200}B_{310}\right)
\right.
\]%
\[
\left.  \left.  -\frac{\vec{r}_{12}^{\,2}}{\vec{r}_{10}^{\,2}\vec{r}%
_{20}^{\,\,\,2}}\left(  9B_{123}-\frac{1}{2}\left[  2\left(  B_{100}%
B_{320}+B_{200}B_{130}\right)  -B_{300}B_{120}\right]  \right)  \right\}
+(1\leftrightarrow3)+(2\leftrightarrow3)\right]
\]%
\[
-{\frac{\alpha_{s}^{2}n_{f}}{16\pi^{4}}}\!\int\!d\vec{r}_{0}d\vec{r}%
_{4}\left[  \left\{  \left(  \frac{1}{3}(U_{1}U_{0}{}^{\dag}U_{4}+U_{4}U_{0}%
{}^{\dag}U_{1})\cdot U_{2}\cdot U_{3}-\frac{1}{9}B_{123}tr(U_{0}{}^{\dag}%
U_{4})\right.  \right.  \right.
\]%
\[
+(U_{1}U_{0}{}^{\dag}U_{2})\cdot U_{3}\cdot U_{4}+\frac{1}{6}B_{123}-\frac
{1}{4}(B_{013}B_{002}+B_{001}B_{023}-B_{012}B_{003})
\]%
\[
\left.  \left.  \left.  \frac{{}}{{}}+(1\leftrightarrow2)\right)
+(0\leftrightarrow4)\right\}  L_{12}^{q}+(1\leftrightarrow3)+(2\leftrightarrow
3)\right]
\]%
\[
-{\frac{\alpha_{s}^{2}}{8\pi^{4}}}\!\int\!d\vec{r}_{0}d\vec{r}_{4}~\left[
\mathbf{\{}\tilde{L}_{12}\left(  U_{0}U_{4}{}^{\dag}U_{2}\right)  \cdot\left(
U_{1}U_{0}{}^{\dag}U_{4}\right)  \cdot U_{3}\frac{{}}{{}}\right.
\]%
\[
\mathbf{+}L_{12}\left[  \left(  U_{0}U_{4}{}^{\dag}U_{2}\right)  \cdot\left(
U_{1}U_{0}{}^{\dag}U_{4}\right)  \cdot U_{3}+tr\left(  U_{0}U_{4}{}^{\dag
}\right)  \left(  U_{1}U_{0}{}^{\dag}U_{2}\right)  \cdot U_{3}\cdot U_{4}%
\frac{{}}{{}}\right.
\]%
\[
\left.  -\frac{3}{4}[B_{144}B_{234}+B_{244}B_{134}-B_{344}B_{124}]+\frac{1}%
{2}B_{123}\right]
\]%
\[
+(M_{13}-M_{12}-M_{23}+M_{2})\left[  \left(  U_{0}U_{4}{}^{\dag}U_{3}\right)
\cdot\left(  U_{2}U_{0}{}^{\dag}U_{1}\right)  \cdot U_{4}+\left(  U_{1}U_{0}%
{}^{\dag}U_{2}\right)  \cdot\left(  U_{3}U_{4}{}^{\dag}U_{0}\right)  \cdot
U_{4}\right]
\]%
\begin{equation}
\left.  +(\text{all 5 permutations}\,1\leftrightarrow2\leftrightarrow
3)\}+\frac{{}}{{}}(0\leftrightarrow4)\right]  .
\end{equation}
Here the functions $L_{12},\tilde{L}_{12},M_{12},M_{2}$ are defined in
(\ref{L12}-\ref{M2}), $L_{12}^{q}$ is defined in (\ref{Lq}),\ the
$\overline{MS}$ renormalization scale $\mu^{2}$ is related to scale
$\tilde{\mu}^{2}$ through (\ref{mu-tilde-through-mu}),%
\begin{equation}
\beta=\left(  \frac{11}{3}-\frac{2}{3}\frac{n_{f}}{3}\right)  .
\end{equation}
As we mentioned, all the expressions in this paper are written in the
$\overline{MS}$ renormalization scheme.

The evolution equation for the composite 3QWL operator $B_{123}^{conf}%
$\ (\ref{anzatz})%
\[
B_{123}^{conf}=B_{123}+\frac{\alpha_{s}3}{8\pi^{2}}\int d\vec{r}_{4}\left[
\frac{\vec{r}_{12}^{\,\,2}}{\vec{r}_{41}^{\,\,2}\vec{r}_{42}^{\,\,2}}%
\ln\left(  \frac{a\vec{r}_{12}^{\,\,2}}{\vec{r}_{41}^{\,\,2}\vec{r}%
_{42}^{\,\,2}}\right)  \right.
\]%
\begin{equation}
\left.  \times(-B_{123}+\frac{1}{6}(B_{144}B_{324}+B_{244}B_{314}%
-B_{344}B_{214}))+(1\leftrightarrow3)+(2\leftrightarrow3)\right]
\end{equation}
follows from (\ref{QCkernel})%
\[
\frac{\partial B_{123}^{conf}}{\partial\eta}=\frac{\alpha_{s}\left(  \mu
^{2}\right)  }{8\pi^{2}}\int d\vec{r}_{0}\left[  ((B_{100}B_{320}%
+B_{200}B_{310}-B_{300}B_{210})-6B_{123})^{conf}\frac{{}}{{}}\right.
\]%
\[
\left.  \times\left(  \frac{\vec{r}_{12}^{\,\,2}}{\vec{r}_{01}^{\,\,2}\vec
{r}_{02}^{\,\,2}}-\!\frac{3\alpha_{s}}{4\pi}\beta\left[  \ln\left(  \frac
{\vec{r}_{01}^{\,\,2}}{\vec{r}_{02}^{\,\,2}}\right)  \left(  \frac{1}{\vec
{r}_{02}^{\,\,2}}-\frac{1}{\vec{r}_{01}^{\,\,2}}\right)  -\frac{\vec{r}%
_{12}^{\,\,2}}{\vec{r}_{01}^{\,\,2}\vec{r}_{02}^{\,\,2}}\ln\left(  \frac
{\vec{r}_{12}^{\,\,2}}{\tilde{\mu}^{2}}\right)  \right]  \right)
+(1\leftrightarrow3)+(2\leftrightarrow3)\right]
\]%
\[
-{\frac{\alpha_{s}^{2}}{32\pi^{3}}}\!\int\!d\vec{r}_{0}\left(  B_{003}%
B_{012}\left[  \frac{\vec{r}_{32}{}^{2}}{\vec{r}_{03}{}^{2}\vec{r}_{02}{}^{2}%
}\ln^{2}\left(  \frac{\vec{r}_{32}{}^{2}\vec{r}_{10}{}^{2}}{\vec{r}_{13}{}%
^{2}\vec{r}_{20}{}^{2}}\right)  -\frac{\vec{r}_{12}{}^{2}}{\vec{r}_{01}{}%
^{2}\vec{r}_{02}{}^{2}}\ln^{2}\left(  \frac{\vec{r}_{12}{}^{2}\vec{r}_{30}%
{}^{2}}{\vec{r}_{13}{}^{2}\vec{r}_{20}{}^{2}}\right)  \right]  \right.
\]%
\[
\left.  \frac{{}}{{}}+(\text{all 5 permutations}\,1\leftrightarrow
2\leftrightarrow3)\right)
\]%
\[
-{\frac{\alpha_{s}^{2}n_{f}}{16\pi^{4}}}\!\int\!d\vec{r}_{0}d\vec{r}%
_{4}\left[  \left\{  \left(  \frac{1}{3}(U_{1}U_{0}{}^{\dag}U_{4}+U_{4}U_{0}%
{}^{\dag}U_{1})\cdot U_{2}\cdot U_{3}-\frac{1}{9}B_{123}tr(U_{0}{}^{\dag}%
U_{4})\right.  \right.  \right.
\]%
\[
+(U_{1}U_{0}{}^{\dag}U_{2})\cdot U_{3}\cdot U_{4}+\frac{1}{6}B_{123}-\frac
{1}{4}(B_{013}B_{002}+B_{001}B_{023}-B_{012}B_{003})
\]%
\[
\left.  \left.  \left.  \frac{{}}{{}}+(1\leftrightarrow2)\right)
+(0\leftrightarrow4)\right\}  L_{12}^{q}+(1\leftrightarrow3)+(2\leftrightarrow
3)\right]
\]%
\[
-{\frac{\alpha_{s}^{2}}{8\pi^{4}}}\!\int\!d\vec{r}_{0}d\vec{r}_{4}~\left(
\left\{  \tilde{L}_{12}^{C}\left(  U_{0}U_{4}{}^{\dag}U_{2}\right)
\cdot\left(  U_{1}U_{0}{}^{\dag}U_{4}\right)  \cdot U_{3}\frac{{}}{{}}\right.
\right.
\]%
\[
\mathbf{+}L_{12}^{C}\left[  \left(  U_{0}U_{4}{}^{\dag}U_{2}\right)
\cdot\left(  U_{1}U_{0}{}^{\dag}U_{4}\right)  \cdot U_{3}+tr\left(  U_{0}%
U_{4}{}^{\dag}\right)  \left(  U_{1}U_{0}{}^{\dag}U_{2}\right)  \cdot
U_{3}\cdot U_{4}\frac{{}}{{}}\right.
\]%
\[
\left.  -\frac{3}{4}[B_{144}B_{234}+B_{244}B_{134}-B_{344}B_{124}]+\frac{1}%
{2}B_{123}\right]
\]%
\[
+M_{12}^{C}\left[  \left(  U_{0}U_{4}{}^{\dag}U_{3}\right)  \cdot\left(
U_{2}U_{0}{}^{\dag}U_{1}\right)  \cdot U_{4}+\left(  U_{1}U_{0}{}^{\dag}%
U_{2}\right)  \cdot\left(  U_{3}U_{4}{}^{\dag}U_{0}\right)  \cdot
U_{4}\right]
\]%
\begin{equation}
\left.  \left.  +\frac{{}}{{}}(\text{all 5 permutations}\,1\leftrightarrow
2\leftrightarrow3)\right\}  +(0\leftrightarrow4)\right)  .
\end{equation}
Here the composite operator $([B_{100}B_{320}+B_{200}B_{310}-B_{300}%
B_{210}]-6B_{123})^{conf}$ is defined in (\ref{anzatz1}) according to the
prescription (\ref{model}) and the functions $L_{12}^{C},\tilde{L}_{12}%
^{C},M_{12}^{C}$ are defined in (\ref{Lc}-\ref{Mc}).

The equation for the composite 3QWL operator $B_{123}^{conf}$\ (\ref{anzatz})
linearized in the 3-gluon approximation is the result of (\ref{QCkernel3g})
and (\ref{compositeOperator3g})%
\[
\frac{\partial B_{123}^{conf}}{\partial\eta}\overset{\mathrm{3g}}{=}%
\frac{3\alpha_{s}\left(  \mu^{2}\right)  }{4\pi^{2}}\int d\vec{r}_{0}\left[
(B_{100}^{conf}+B_{320}^{conf}+B_{200}^{conf}+B_{310}^{conf}-B_{300}%
^{conf}-B_{210}^{conf}-B_{123}^{conf}-6)\right.
\]%
\[
\!\left.  \times\left(  \frac{\vec{r}_{12}^{\,\,2}}{\vec{r}_{01}^{\,\,2}%
\vec{r}_{02}^{\,\,2}}-{\frac{3\alpha_{s}}{4\pi}}\beta\left[  \ln\left(
\frac{\vec{r}_{01}^{\,\,2}}{\vec{r}_{02}^{\,\,2}}\right)  \left(  \frac
{1}{\vec{r}_{02}^{\,\,2}}-\frac{1}{\vec{r}_{01}^{\,\,2}}\right)  -\frac
{\vec{r}_{12}^{\,\,2}}{\vec{r}_{01}^{\,\,2}\vec{r}_{02}^{\,\,2}}\ln\left(
\frac{\vec{r}_{12}^{\,\,2}}{\tilde{\mu}^{2}}\right)  \right]  \right)
+(1\leftrightarrow3)+(2\leftrightarrow3)\right]
\]%
\[
-{\frac{9\alpha_{s}^{2}}{64\pi^{3}}}\!\int\!d\vec{r}_{0}\left(  \tilde
{F}_{100}B_{100}+\tilde{F}_{230}B_{230}+(1\leftrightarrow3)+(1\leftrightarrow
2)\right)
\]%
\[
+{\frac{27\alpha_{s}^{2}}{4\pi^{2}}}\zeta(3)(3-\delta_{23}-\delta_{13}%
-\delta_{21})(B_{123}-6)
\]%
\[
-{\frac{9\alpha_{s}^{2}}{8\pi^{4}}}\!\int\!d\vec{r}_{0}d\vec{r}_{4}\left(
L_{12}^{C}+L_{13}^{C}+L_{23}^{C}-\frac{n_{f}}{54}(L_{12}^{q}+L_{13}^{q}%
+L_{23}^{q})\right)  (B_{044}+B_{004}-12)
\]%
\[
-{\frac{\alpha_{s}^{2}n_{f}}{24\pi^{4}}}\!\int\!d\vec{r}_{0}d\vec{r}%
_{4}~\left\{  \left(  2B_{014}-B_{001}-B_{144}\right)  (L_{12}^{q}+L_{13}%
^{q}-2L_{32}^{q})+(1\leftrightarrow3)+(1\leftrightarrow2)\right\}
\]%
\begin{equation}
-{\frac{9\alpha_{s}^{2}}{64\pi^{4}}}\!\int\!d\vec{r}_{0}d\vec{r}_{4}~\left(
\left\{  F_{0}B_{040}+F_{140}B_{140}+(0\leftrightarrow4)\right\}  +(\text{all
5 permutations}\,1\leftrightarrow2\leftrightarrow3)\right)  .
\end{equation}
Here $\delta_{ij}=1,$ if $\vec{r}_{i}=\vec{r}_{j}$ and $\delta_{ij}=0$
otherwise; the functions $F_{0}$ and $F_{140}$ are defined in (\ref{F0}) and
(\ref{F140}); $\tilde{F}_{100}$ and $\tilde{F}_{230}$ are defined in
(\ref{F100tilde}-\ref{F230tilde}).

The linearized equation for C-even composite 3QWL Green function is the
consequence of (\ref{CevenKnlo3g}) and (\ref{compositeOperator3g})%
\[
\frac{\partial B_{123}^{+conf}}{\partial\eta}\overset{\mathrm{3g}}{=}%
\frac{3\alpha_{s}\left(  \mu^{2}\right)  }{4\pi^{2}}\int d\vec{r}_{0}\left[
\left(  B_{100}^{+conf}+B_{320}^{+conf}+B_{200}^{+conf}+B_{310}^{+conf}%
\right.  \right.
\]%
\[
\left.  -B_{300}^{+conf}-B_{210}^{+conf}-B_{123}^{+conf}\right)
\]%
\[
\!\left.  \times\left(  \frac{\vec{r}_{12}^{\,\,2}}{\vec{r}_{01}^{\,\,2}%
\vec{r}_{02}^{\,\,2}}-{\frac{3\alpha_{s}}{4\pi}}\beta\left[  \ln\left(
\frac{\vec{r}_{01}^{\,\,2}}{\vec{r}_{02}^{\,\,2}}\right)  \left(  \frac
{1}{\vec{r}_{02}^{\,\,2}}-\frac{1}{\vec{r}_{01}^{\,\,2}}\right)  -\frac
{\vec{r}_{12}^{\,\,2}}{\vec{r}_{01}^{\,\,2}\vec{r}_{02}^{\,\,2}}\ln\left(
\frac{\vec{r}_{12}^{\,\,2}}{\tilde{\mu}^{2}}\right)  \right]  \right)
+(1\leftrightarrow3)+(2\leftrightarrow3)\right]
\]%
\[
-{\frac{9\alpha_{s}^{2}}{4\pi^{4}}}\!\int\!d\vec{r}_{0}d\vec{r}_{4}\left(
L_{12}^{C}+L_{13}^{C}+L_{23}^{C}-\frac{n_{f}}{54}(L_{12}^{q}+L_{13}^{q}%
+L_{23}^{q})\right)  (B_{044}+B_{004}-12)
\]%
\[
-{\frac{9\alpha_{s}^{2}}{64\pi^{3}}}\!\int\!d\vec{r}_{0}\left(  \tilde
{F}_{100}B_{100}^{+}+\tilde{F}_{230}B_{230}^{+}+(1\leftrightarrow
3)+(1\leftrightarrow2)\right)
\]%
\[
+{\frac{27\alpha_{s}^{2}}{4\pi^{2}}}\zeta(3)(3-\delta_{23}-\delta_{13}%
-\delta_{21})B_{123}^{+}%
\]%
\begin{equation}
-{\frac{9\alpha_{s}^{2}}{64\pi^{4}}}\!\int\!d\vec{r}_{0}d\vec{r}_{4}~\left(
\left\{  F_{140}+(0\leftrightarrow4)\right\}  B_{140}^{+}+(\text{all 5
permutations}\,1\leftrightarrow2\leftrightarrow3)\right)  .
\end{equation}
The linearized equations for C-odd composite 3QWL Green function is the
consequence of (\ref{CoddKnlo3g}) and (\ref{compositeOperator3g})%
\[
\frac{\partial B_{123}^{-conf}}{\partial\eta}\overset{\mathrm{3g}}{=}%
\frac{3\alpha_{s}\left(  \mu^{2}\right)  }{4\pi^{2}}\int d\vec{r}_{0}\left[
\left(  B_{100}^{-conf}+B_{320}^{-conf}+B_{200}^{-conf}+B_{310}^{-conf}%
\right.  \frac{{}}{{}}\right.
\]%
\[
\left.  -B_{300}^{-conf}-B_{210}^{-conf}-B_{123}^{-conf}\right)
\]%
\[
\!\left.  \times\left(  \frac{\vec{r}_{12}^{\,\,2}}{\vec{r}_{01}^{\,\,2}%
\vec{r}_{02}^{\,\,2}}-{\frac{3\alpha_{s}}{4\pi}}\beta\left[  \ln\left(
\frac{\vec{r}_{01}^{\,\,2}}{\vec{r}_{02}^{\,\,2}}\right)  \left(  \frac
{1}{\vec{r}_{02}^{\,\,2}}-\frac{1}{\vec{r}_{01}^{\,\,2}}\right)  -\frac
{\vec{r}_{12}^{\,\,2}}{\vec{r}_{01}^{\,\,2}\vec{r}_{02}^{\,\,2}}\ln\left(
\frac{\vec{r}_{12}^{\,\,2}}{\tilde{\mu}^{2}}\right)  \right]  \right)
+(1\leftrightarrow3)+(2\leftrightarrow3)\right]
\]%
\[
-{\frac{\alpha_{s}^{2}n_{f}}{24\pi^{4}}}\!\int\!d\vec{r}_{0}d\vec{r}%
_{4}~\left\{  \left(  2B_{014}^{-}-B_{001}^{-}-B_{144}^{-}\right)  (L_{12}%
^{q}+L_{13}^{q}-2L_{32}^{q})+(1\leftrightarrow3)+(1\leftrightarrow2)\right\}
\]%
\[
-{\frac{9\alpha_{s}^{2}}{64\pi^{3}}}\!\int\!d\vec{r}_{0}\left(  \tilde
{F}_{100}B_{100}^{-}+\tilde{F}_{230}B_{230}^{-}+(1\leftrightarrow
3)+(1\leftrightarrow2)\right)  +{\frac{27\alpha_{s}^{2}}{4\pi^{2}}}%
\zeta(3)(3-\delta_{23}-\delta_{13}-\delta_{21})B_{123}^{-}%
\]%
\begin{equation}
-{\frac{9\alpha_{s}^{2}}{64\pi^{4}}}\!\int\!d\vec{r}_{0}d\vec{r}_{4}~\left(
2F_{0}B_{040}^{-}+\left\{  F_{140}+(0\leftrightarrow4)\right\}  B_{140}%
^{-}+(\text{all 5 permutations}\,1\leftrightarrow2\leftrightarrow3)\right)  .
\end{equation}
From these expressions one can see that terms with $L_{ij},L_{ij}^{C}$, which
comprise the BFKL kernels contribute only to the evolution of C-even part of
the Green function while terms with $F_{0},\tilde{L}_{ij},\tilde{L}_{ij}^{C}$
contribute only to the evolution of the C-odd one.

The BK\ equation for the color dipole $B_{122}=2tr(U_{1}U_{2}^{\dag})$ in the
3-gluon approximation reads (see (\ref{BK3g}))%
\[
\frac{\partial B_{122}^{conf}}{\partial\eta}\overset{\mathrm{3g}}{=}%
\frac{3\alpha_{s}\left(  \mu^{2}\right)  }{2\pi^{2}}\int d\vec{r}_{0}%
(B_{100}^{conf}+B_{220}^{conf}-B_{122}^{conf}-6)
\]%
\[
\!\times\left(  \frac{\vec{r}_{12}^{\,\,2}}{\vec{r}_{01}^{\,\,2}\vec{r}%
_{02}^{\,\,2}}-{\frac{3\alpha_{s}}{4\pi}}\beta\left[  \ln\left(  \frac{\vec
{r}_{01}^{\,\,2}}{\vec{r}_{02}^{\,\,2}}\right)  \left(  \frac{1}{\vec{r}%
_{02}^{\,\,2}}-\frac{1}{\vec{r}_{01}^{\,\,2}}\right)  -\frac{\vec{r}%
_{12}^{\,\,2}}{\vec{r}_{01}^{\,\,2}\vec{r}_{02}^{\,\,2}}\ln\left(  \frac
{\vec{r}_{12}^{\,\,2}}{\tilde{\mu}^{2}}\right)  \right]  \right)
+{\frac{27\alpha_{s}^{2}}{2\pi^{2}}}\zeta(3)(B_{122}-6)
\]%
\[
-{\frac{9\alpha_{s}^{2}}{4\pi^{4}}}\!\int\!d\vec{r}_{0}d\vec{r}_{4}\left(
L_{12}^{C}-\frac{n_{f}}{54}L_{12}^{q}\right)  (B_{044}+B_{004}-12)-{\frac
{9\alpha_{s}^{2}}{4\pi^{4}}}\!\int\!d\vec{r}_{0}d\vec{r}_{4}~\tilde{L}%
_{12}^{C}(B_{044}-B_{040})
\]%
\begin{equation}
-{\frac{\alpha_{s}^{2}n_{f}}{12\pi^{4}}}\!\int\!d\vec{r}_{0}d\vec{r}%
_{4}~\left\{  \left(  2B_{014}-B_{001}-B_{144}\right)  -\left(  2B_{024}%
-B_{002}-B_{244}\right)  \right\}  L_{12}^{q}. \label{BK3G}%
\end{equation}
As is clear from the last line the evolution of the color dipole in the
3-gluon approximation depends on the 3QWL operators which have nondipole
structure. The BK\ equation for the C-even part of the color dipole operator
$B_{122}^{+}=2tr(U_{1}U_{2}^{\dag})+2tr(U_{1}^{\dag}U_{2})-6$ in the 3-gluon
approximation is the same as in the 2-gluon one (BFKL)%
\[
\frac{\partial B_{122}^{+conf}}{\partial\eta}\overset{\mathrm{3g}}{=}%
\frac{3\alpha_{s}\left(  \mu^{2}\right)  }{2\pi^{2}}\int d\vec{r}_{0}%
(B_{100}^{+conf}+B_{220}^{+conf}-B_{122}^{+conf})
\]%
\[
\!\times\left(  \frac{\vec{r}_{12}^{\,\,2}}{\vec{r}_{01}^{\,\,2}\vec{r}%
_{02}^{\,\,2}}-{\frac{3\alpha_{s}}{4\pi}}\beta\left[  \ln\left(  \frac{\vec
{r}_{01}^{\,\,2}}{\vec{r}_{02}^{\,\,2}}\right)  \left(  \frac{1}{\vec{r}%
_{02}^{\,\,2}}-\frac{1}{\vec{r}_{01}^{\,\,2}}\right)  -\frac{\vec{r}%
_{12}^{\,\,2}}{\vec{r}_{01}^{\,\,2}\vec{r}_{02}^{\,\,2}}\ln\left(  \frac
{\vec{r}_{12}^{\,\,2}}{\tilde{\mu}^{2}}\right)  \right]  \right)
\]%
\begin{equation}
-{\frac{9\alpha_{s}^{2}}{2\pi^{4}}}\!\int\!d\vec{r}_{0}d\vec{r}_{4}\left(
L_{12}^{C}-\frac{n_{f}}{54}L_{12}^{q}\right)  B_{044}^{+}+{\frac{27\alpha
_{s}^{2}}{2\pi^{2}}}\zeta(3)B_{122}^{+}.
\end{equation}
At the same time the BK\ equation for the C-odd part of the color dipole
operator $B_{122}^{-}=2tr(U_{1}U_{2}^{\dag})-2tr(U_{1}^{\dag}U_{2})$ in the
3-gluon approximation reads%
\[
\frac{\partial B_{122}^{-conf}}{\partial\eta}\overset{\mathrm{3g}}{=}%
\frac{3\alpha_{s}\left(  \mu^{2}\right)  }{2\pi^{2}}\int d\vec{r}_{0}%
(B_{100}^{-conf}+B_{220}^{-conf}-B_{122}^{-conf})
\]%
\[
\!\times\left(  \frac{\vec{r}_{12}^{\,\,2}}{\vec{r}_{01}^{\,\,2}\vec{r}%
_{02}^{\,\,2}}-{\frac{3\alpha_{s}}{4\pi}}\beta\left[  \ln\left(  \frac{\vec
{r}_{01}^{\,\,2}}{\vec{r}_{02}^{\,\,2}}\right)  \left(  \frac{1}{\vec{r}%
_{02}^{\,\,2}}-\frac{1}{\vec{r}_{01}^{\,\,2}}\right)  -\frac{\vec{r}%
_{12}^{\,\,2}}{\vec{r}_{01}^{\,\,2}\vec{r}_{02}^{\,\,2}}\ln\left(  \frac
{\vec{r}_{12}^{\,\,2}}{\tilde{\mu}^{2}}\right)  \right]  \right)
-{\frac{9\alpha_{s}^{2}}{2\pi^{4}}}\!\int\!d\vec{r}_{0}d\vec{r}_{4}~\tilde
{L}_{12}^{C}B_{044}^{-}%
\]%
\begin{equation}
+{\frac{27\alpha_{s}^{2}}{2\pi^{2}}}\zeta(3)B_{122}^{-}-{\frac{\alpha_{s}%
^{2}n_{f}}{12\pi^{4}}}\!\int\!d\vec{r}_{0}d\vec{r}_{4}~\left\{  \left(
2B_{014}^{-}-B_{001}^{-}-B_{144}^{-}\right)  -\left(  2B_{024}^{-}-B_{002}%
^{-}-B_{244}^{-}\right)  \right\}  L_{12}^{q}. \label{BK3G-}%
\end{equation}
This equation contains the nondipole 3QWL operators in its quark part.

\section{Conclusions}

In this paper we constructed the NLO evolution equation for the
\textquotedblleft color triple\textquotedblright\ - three-quark Wilson loop
operator $\varepsilon^{i^{\prime}j^{\prime}h^{\prime}}\varepsilon
_{ijh}U_{1i^{\prime}}^{i}U_{2j^{\prime}}^{j}U_{3h^{\prime}}^{h}$. As in the
case of color dipole evolution, for the \textquotedblleft rigid
cutoff\textquotedblright\ $Y<\eta$ of Wilson line the kernel of this equation
has non-conformal terms not related to renormalization. We have constructed
the composite 3QWL operator (\ref{anzatz})\ obeying the NLO evolution equation
with quasi-conformal kernel. We linearized the quasi-conformal equation in the
3-gluon approximation. It is worth noting that our results have correct dipole
limit in the case when the coordinates of two lines coincide. We also
constructed the 3-gluon approximation of the BK equation and showed that it
contains non-dipole 3QWL operators (\ref{BK3G}), (\ref{BK3G-}).

The 3QWL operator may have many phenomenological applications. First, it is a
natural $SU(3)$ model for a baryon Green function in the Regge limit. Also, it
is the irreducible operator describing C-odd (odderon) exchange. For example
as shown in the appendix \ref{quadrupole}, the odderon part of the quadrupole
operator $tr(U_{1}U_{2}^{\dag}U_{3}U_{4}^{\dag})$\ in the 3-gluon
approximation in $SU(3)$ can be decomposed into a sum of 3QWLs%
\[
2tr(U_{1}U_{2}^{\dag}U_{3}U_{4}^{\dag})-2tr(U_{4}U_{3}^{\dag}U_{2}U_{1}^{\dag
})\overset{\mathrm{3g}}{=}B_{144}^{-}+B_{322}^{-}-B_{433}^{-}-B_{211}%
^{-}+B_{124}^{-}+B_{234}^{-}-B_{123}^{-}-B_{134}^{-}.
\]
Moreover, even the NLO evolution equation for the dipole C-odd Green function
in the 3-gluon approximation (\ref{BK3G-})\ in QCD can not be written without
the introduction of the 3QWL operator.

The evolution equation for the C-odd part of the 3QWL operator is the
generalization of the BKP equation for odderon exchange to the saturation
regime. However, it is valid for the colorless object, i.e. for the function
$B_{ijk}^{-}=B^{-}\left(  \vec{r}_{i},\vec{r}_{j},\vec{r}_{k}\right)  ,$ which
vanishes as $\vec{r}_{i}=\vec{r}_{j}=\vec{r}_{k}.$ The linear approximation of
the equation for the C-odd part of the 3QWL should be equivalent to the NLO
BKP for odderon exchange acting in the space of such functions. One may try to
restore the full NLO BKP kernel from our result via the technique similar to
the one developed for the 2-point operators in \cite{Fadin:2011jg}.

\acknowledgments A. V. G. acknowledges support of the Russian Fund for Basic
Research grant 13-02-01023 and president grant MK-525.2013.2. The work by I.
B. was supported by contract DE-AC05-06OR23177 under which the Jefferson
Science Associates operate the Thomas Jefferson National Accelerator Facility.
The authors are grateful to S. Caron-Huot and M. Lublinsky for valuable discussions.

\appendix

\section{Notations}

\label{notation} First, let us describe our notations. We introduce the light
cone vectors $n_{1}$ and $n_{2}$%
\begin{equation}
n_{1}=\left(  1,0,0,1\right)  ,\quad n_{2}=\frac{1}{2}\left(  1,0,0,-1\right)
,\quad n_{1}^{+}=n_{2}^{-}=n_{1}n_{2}=1
\end{equation}
so that for any vector $p$ we have%
\begin{equation}
p^{+}=p_{-}=pn_{2}=\frac{1}{2}\left(  p^{0}+p^{3}\right)  ,\qquad p_{+}%
=p^{-}=pn_{1}=p^{0}-p^{3},
\end{equation}%
\begin{equation}
p=p^{+}n_{1}+p^{-}n_{2}+p_{\bot},\qquad p^{2}=2p^{+}p^{-}-\vec{p}^{\,2},
\end{equation}%
\begin{equation}
\quad p\,k=p^{\mu}k_{\mu}=p^{+}k^{-}+p^{-}k^{+}-\vec{p}\vec{k}=p_{+}%
k_{-}+p_{-}k_{+}-\vec{p}\vec{k}.
\end{equation}
The index convention is $a_{i}^{j}b_{j}^{k}=(ab)_{i}^{k}.$

Second, let us present the connection of our notation with the notation of
\cite{Balitsky:2013fea}. In that paper the quark and gluon coordinates were
denoted $z_{i}.$ So we have to change%
\begin{equation}
z_{1,2,3,4}\leftrightarrow\vec{r}_{1,2,3,4},\quad z_{5}\leftrightarrow\vec
{r}_{0}.
\end{equation}
Assuming this substitution for self-interaction we get
\begin{equation}
G_{3}={\frac{I_{1}}{2z_{45}^{2}}}\ln{\frac{z_{14}^{2}}{z_{15}^{2}}}-{\frac
{1}{z_{45}^{4}}},~~~~~G_{9}=-{\frac{(z_{14},z_{15})}{2z_{45}^{2}z_{14}%
^{2}z_{15}^{2}}}\ln{\frac{z_{14}^{2}}{z_{15}^{2}}.}%
\end{equation}
For pairwise interaction we obtain
\begin{equation}
8H_{1}=2\mathcal{J}_{1245}\ln{\frac{z_{14}^{2}}{z_{15}^{2}}},~~~~8H_{2}%
=-2(J_{1245}+J_{1254})\ln{\frac{z_{14}^{2}}{z_{15}^{2}}},
\end{equation}%
\begin{equation}
8H_{3}=2(J_{1245}-J_{1254})\ln{\frac{z_{14}^{2}}{z_{15}^{2}},}%
\end{equation}%
\begin{equation}
8H_{4}=\ln{\frac{z_{14}^{2}}{z_{15}^{2}}}~(2L-2J_{1245}+2J_{1254})-{\frac
{2}{z_{45}^{4}},}%
\end{equation}
or%
\begin{equation}
\hspace{-11mm}8(H_{3}+H_{4})=2L\ln{\frac{z_{14}^{2}}{z_{15}^{2}}}-{\frac
{2}{z_{45}^{4}}.}%
\end{equation}
Here $L$ is defined as
\begin{equation}
-2K-{\frac{8}{z_{45}^{4}}}+{\frac{2I_{1}}{z_{45}^{2}}}+{\frac{2I_{2}}%
{z_{45}^{2}}}=2L\ln{\frac{z_{14}^{2}}{z_{15}^{2}}}+(1\leftrightarrow
2)-{\frac{4}{z_{45}^{4}}}=8(H_{3}+H_{4}+1\leftrightarrow2).
\end{equation}
For triple interaction we have
\begin{equation}
H_{7}\equiv H_{5}+H_{6}={\frac{1}{2}}\mathcal{J}_{32145}\ln{\frac{z_{14}^{2}%
}{z_{15}^{2}}},~~~~~~~H_{8}\equiv H_{5}-H_{6}=-{\frac{1}{2}}\mathcal{J}%
_{32154}\ln{\frac{z_{14}^{2}}{z_{15}^{2}}}.
\end{equation}

\section{$SU(3)$ identities}

\label{identities}Here we present the list of SU(3) identities used in the
paper.%
\begin{equation}
U_{i}\cdot U_{j}\cdot U_{k}=(U_{i}U_{l}^{\dag})\cdot(U_{j}U_{l}^{\dag}%
)\cdot(U_{k}U_{l}^{\dag})=(U_{l}^{\dag}U_{i})\cdot(U_{l}^{\dag}U_{j}%
)\cdot(U_{l}^{\dag}U_{k}), \label{3qWlidentity}%
\end{equation}%
\begin{equation}
\varepsilon^{ijh}\varepsilon_{i^{\prime}j^{\prime}h^{\prime}}(U_{1}%
)_{i}^{i^{\prime}}(U_{1})_{j}^{j^{\prime}}=2(U_{1}^{\dag})_{h^{\prime}}%
^{h},\quad U_{1}\cdot U_{1}\cdot U_{3}=2tr(U_{1}^{\dag}U_{3}).
\label{su3-identity-1}%
\end{equation}
These identities follow from the definition of the group, namely from
unitarity and the fact that the determinant of $U$ is $1.$
\begin{equation}
\left(  U_{2}U_{4}^{\dag}U_{1}+U_{1}U_{4}^{\dag}U_{2}\right)  \cdot U_{4}\cdot
U_{3}=-B_{123}+\frac{1}{2}(B_{144}B_{324}+B_{244}B_{314}-B_{344}B_{214}).
\label{IDENTITY}%
\end{equation}
This identity can be checked using (\ref{3qWlidentity}) with $l=4$ and then
expanding the product of Levi-Civita symbols as%
\begin{equation}
\varepsilon_{ijh}\varepsilon^{i^{\prime}j^{\prime}h^{\prime}}=\left\vert
\begin{tabular}
[c]{lll}%
$\delta_{i}^{i^{\prime}}$ & $\delta_{i}^{j^{\prime}}$ & $\delta_{i}%
^{h^{\prime}}$\\
$\delta_{j}^{i^{\prime}}$ & $\delta_{j}^{j^{\prime}}$ & $\delta_{j}%
^{h^{\prime}}$\\
$\delta_{h}^{i^{\prime}}$ & $\delta_{h}^{j^{\prime}}$ & $\delta_{h}%
^{h^{\prime}}$%
\end{tabular}
\ \right\vert . \label{ee}%
\end{equation}%
\[
0=[\left(  U_{0}U_{4}{}^{\dag}U_{3}U_{0}{}^{\dag}U_{4}\right)  \cdot
U_{1}\cdot U_{2}-U_{1}\cdot U_{2}\cdot U_{4}tr\left(  U_{0}{}^{\dag}%
U_{3}\right)  tr\left(  U_{4}{}^{\dag}U_{0}\right)
\]%
\[
+tr\left(  U_{0}U_{4}{}^{\dag}\right)  \left(  U_{2}U_{0}{}^{\dag}U_{3}%
+U_{3}U_{0}{}^{\dag}U_{2}\right)  \cdot U_{1}\cdot U_{4}%
\]%
\begin{equation}
+\left(  U_{0}U_{4}{}^{\dag}U_{2}\right)  \cdot\left(  U_{3}U_{0}{}^{\dag
}U_{4}\right)  \cdot U_{1}+\left(  U_{0}U_{4}{}^{\dag}U_{3}\right)
\cdot\left(  U_{2}U_{0}{}^{\dag}U_{4}\right)  \cdot U_{1}+(1\leftrightarrow
2)]+(4\leftrightarrow0). \label{id1}%
\end{equation}
This identity relates the color structures in $\mathbf{G}_{12\langle3\rangle}%
$, $\mathbf{G}_{1\langle23\rangle}$ and $\mathbf{G}_{\langle13\rangle2}.$ By
$1\leftrightarrow2\leftrightarrow3$ transformation one can obtain 2 more
identities and totally eliminate 3 color structures from $\mathbf{G}%
_{12\langle3\rangle},$ $\mathbf{G}_{\langle1\rangle23}\mathbf{,}$ and
$\mathbf{G}_{1\langle2\rangle3}.$
\[
0=tr\left(  U_{0}{}^{\dag}U_{2}\right)  \left(  U_{0}U_{4}{}^{\dag}U_{3}%
+U_{3}U_{4}{}^{\dag}U_{0}\right)  \cdot U_{1}\cdot U_{4}%
\]%
\[
-\left(  U_{0}U_{4}{}^{\dag}U_{3}\right)  \cdot\left(  U_{2}U_{0}{}^{\dag
}U_{4}\right)  \cdot U_{1}-\left(  U_{3}U_{4}{}^{\dag}U_{0}\right)
\cdot\left(  U_{4}U_{0}{}^{\dag}U_{2}\right)  \cdot U_{1}%
\]%
\[
-tr\left(  U_{4}U_{0}{}^{\dag}\right)  \left(  U_{2}U_{4}{}^{\dag}U_{3}%
+U_{3}U_{4}{}^{\dag}U_{2}\right)  \cdot U_{0}\cdot U_{1}%
\]%
\[
+\left(  U_{3}U_{4}{}^{\dag}U_{2}U_{0}{}^{\dag}U_{4}\right)  \cdot U_{0}\cdot
U_{1}+\left(  U_{4}U_{0}{}^{\dag}U_{2}U_{4}{}^{\dag}U_{3}\right)  \cdot
U_{0}\cdot U_{1}%
\]%
\[
-\left(  U_{0}U_{4}{}^{\dag}U_{3}\right)  \cdot\left(  U_{2}U_{0}{}^{\dag
}U_{1}\right)  \cdot U_{4}-\left(  U_{1}U_{0}{}^{\dag}U_{2}\right)
\cdot\left(  U_{3}U_{4}{}^{\dag}U_{0}\right)  \cdot U_{4}%
\]%
\begin{equation}
+\left(  U_{1}U_{0}{}^{\dag}U_{4}\right)  \cdot\left(  U_{3}U_{4}{}^{\dag
}U_{2}\right)  \cdot U_{0}+\left(  U_{2}U_{4}{}^{\dag}U_{3}\right)
\cdot\left(  U_{4}U_{0}{}^{\dag}U_{1}\right)  \cdot U_{0}. \label{id2}%
\end{equation}
This identity relates all color structures in $\mathbf{G}_{1\langle23\rangle}$
and two structures in $\mathbf{G}_{\langle123\rangle}.$ It goes into 5
different identities after $1\leftrightarrow2\leftrightarrow3$ transformation,
which allows one to get rid of 6 structures.%
\[
0=tr\left(  U_{0}{}^{\dag}U_{2}\right)  \left(  U_{0}U_{4}{}^{\dag}U_{1}%
+U_{1}U_{4}{}^{\dag}U_{0}\right)  \cdot U_{3}\cdot U_{4}%
\]%
\[
+\left(  U_{0}U_{4}{}^{\dag}U_{2}U_{0}{}^{\dag}U_{1}+U_{1}U_{0}{}^{\dag}%
U_{2}U_{4}{}^{\dag}U_{0}\right)  \cdot U_{3}\cdot U_{4}%
\]%
\[
-tr\left(  U_{0}U_{4}{}^{\dag}\right)  \left(  U_{1}U_{0}{}^{\dag}U_{2}%
+U_{2}U_{0}{}^{\dag}U_{1}\right)  \cdot U_{3}\cdot U_{4}%
\]%
\[
-\left(  U_{0}U_{4}{}^{\dag}U_{1}\right)  \cdot\left(  U_{2}U_{0}{}^{\dag
}U_{4}\right)  \cdot U_{3}-\left(  U_{0}U_{4}{}^{\dag}U_{2}\right)
\cdot\left(  U_{1}U_{0}{}^{\dag}U_{4}\right)  \cdot U_{3}%
\]%
\[
-\left(  U_{1}U_{0}{}^{\dag}U_{4}\right)  \cdot\left(  U_{3}U_{4}{}^{\dag
}U_{2}\right)  \cdot U_{0}+\left(  U_{1}U_{4}{}^{\dag}U_{2}\right)
\cdot\left(  U_{3}U_{0}{}^{\dag}U_{4}\right)  \cdot U_{0}%
\]%
\begin{equation}
+\left(  U_{2}U_{4}{}^{\dag}U_{1}\right)  \cdot\left(  U_{4}U_{0}{}^{\dag
}U_{3}\right)  \cdot U_{0}-\left(  U_{2}U_{4}{}^{\dag}U_{3}\right)
\cdot\left(  U_{4}U_{0}{}^{\dag}U_{1}\right)  \cdot U_{0}+(4\leftrightarrow0).
\label{id3}%
\end{equation}
This identity relates 2 color structures in $\mathbf{G}_{\langle12\rangle3}$
and a structure in $\mathbf{G}_{\langle123\rangle}.$ It also goes into 5
different identities after $1\leftrightarrow2\leftrightarrow3$ transformation,
which allows one to get rid of 6 structures.%
\[
0=[U_{0}\cdot U_{1}\cdot U_{2}tr\left(  U_{0}{}^{\dag}U_{4}\right)  tr\left(
U_{4}{}^{\dag}U_{3}\right)
\]%
\[
-tr\left(  U_{4}U_{0}{}^{\dag}\right)  \left(  U_{1}U_{4}{}^{\dag}U_{3}%
+U_{3}U_{4}{}^{\dag}U_{1}\right)  \cdot U_{0}\cdot U_{2}+\left(  U_{0}U_{4}%
{}^{\dag}U_{1}\right)  \cdot\left(  U_{3}U_{0}{}^{\dag}U_{4}\right)  \cdot
U_{2}%
\]%
\begin{equation}
+\left(  U_{1}U_{4}{}^{\dag}U_{0}\right)  \cdot\left(  U_{4}U_{0}{}^{\dag
}U_{3}\right)  \cdot U_{2}+(1\leftrightarrow2)]-(4\leftrightarrow0).
\label{id4}%
\end{equation}
This identity relates 2 color structures in $\mathbf{G}_{\langle13\rangle2},$
2 color structures in $\mathbf{G}_{1\langle23\rangle}$ and a structure in
$\mathbf{G}_{12\langle3\rangle}.$ It goes into 2 different identities after
$1\leftrightarrow2\leftrightarrow3$ transformation, which allows one to get
rid of 3 more structures.%
\[
0=2tr\left(  U_{4}U_{0}{}^{\dag}\right)  \left(  U_{2}U_{4}{}^{\dag}%
U_{3}+U_{3}U_{4}{}^{\dag}U_{2}\right)  \cdot U_{0}\cdot U_{1}%
\]%
\[
+\left(  U_{0}U_{4}{}^{\dag}U_{1}+U_{1}U_{4}{}^{\dag}U_{0}\right)
\cdot\left(  U_{2}U_{0}{}^{\dag}U_{3}+U_{3}U_{0}{}^{\dag}U_{2}\right)  \cdot
U_{4}%
\]%
\[
+\left(  U_{0}U_{4}{}^{\dag}U_{2}-U_{2}U_{4}{}^{\dag}U_{0}\right)
\cdot\left(  U_{3}U_{0}{}^{\dag}U_{1}-U_{1}U_{0}{}^{\dag}U_{3}\right)  \cdot
U_{4}%
\]%
\begin{equation}
+\left(  U_{0}U_{4}{}^{\dag}U_{3}-U_{3}U_{4}{}^{\dag}U_{0}\right)
\cdot\left(  U_{2}U_{0}{}^{\dag}U_{1}-U_{1}U_{0}{}^{\dag}U_{2}\right)  \cdot
U_{4}-(4\leftrightarrow0). \label{id5}%
\end{equation}
This identity relates 3 color structures in $\mathbf{G}_{\langle132\rangle}$
and a color structure in $\mathbf{G}_{1\langle23\rangle}.$ It goes into 2
different identities after $1\leftrightarrow2\leftrightarrow3$ transformation,
which allows one to get rid of 3 more structures.

All these identities (\ref{id1}--\ref{id5}) can be checked using
(\ref{3qWlidentity}) with $l=1$ and then expanding the product of Levi-Civita
symbols via (\ref{ee}).
\[
0=2tr(U_{0}{}^{\dag}U_{3})\left(  U_{1}U_{4}{}^{\dag}U_{2}\right)  \cdot
U_{0}\cdot U_{4}-\left(  U_{1}U_{4}{}^{\dag}U_{2}\right)  \cdot\left(
U_{3}U_{0}{}^{\dag}U_{4}\right)  \cdot U_{0}%
\]%
\[
-\left(  U_{1}U_{4}{}^{\dag}U_{2}\right)  \cdot\left(  U_{4}U_{0}{}^{\dag
}U_{3}\right)  \cdot U_{0}-2\left(  U_{1}U_{4}{}^{\dag}U_{2}\right)  \cdot
U_{3}\cdot U_{4}%
\]%
\begin{equation}
-\left(  U_{1}U_{4}{}^{\dag}U_{2}U_{0}{}^{\dag}U_{3}\right)  \cdot U_{0}\cdot
U_{4}-\left(  U_{3}U_{0}{}^{\dag}U_{1}U_{4}{}^{\dag}U_{2}\right)  \cdot
U_{0}\cdot U_{4}+(1\leftrightarrow2). \label{simple_identity}%
\end{equation}
This identity can be proved directly using (\ref{IDENTITY}).%
\[
0=2tr\left(  U_{0}{}^{\dag}U_{4}\right)  \left(  U_{1}U_{4}{}^{\dag}%
U_{2}\right)  \cdot U_{0}\cdot U_{3}-tr\left(  U_{0}{}^{\dag}U_{1}\right)
\left(  U_{0}U_{4}{}^{\dag}U_{2}+U_{2}U_{4}{}^{\dag}U_{0}\right)  \cdot
U_{3}\cdot U_{4}%
\]%
\[
+\left(  U_{0}U_{4}{}^{\dag}U_{1}\right)  \cdot\left(  U_{2}U_{0}{}^{\dag
}U_{3}\right)  \cdot U_{4}+\left(  U_{0}U_{4}{}^{\dag}U_{1}\right)
\cdot\left(  U_{2}U_{0}{}^{\dag}U_{4}\right)  \cdot U_{3}%
\]%
\[
+\left(  U_{1}U_{4}{}^{\dag}U_{0}\right)  \cdot\left(  U_{3}U_{0}{}^{\dag
}U_{2}\right)  \cdot U_{4}+\left(  U_{1}U_{4}{}^{\dag}U_{0}\right)
\cdot\left(  U_{4}U_{0}{}^{\dag}U_{2}\right)  \cdot U_{3}%
\]%
\[
-\left(  U_{1}U_{4}{}^{\dag}U_{2}\right)  \cdot\left(  U_{3}U_{0}{}^{\dag
}U_{4}+U_{4}U_{0}{}^{\dag}U_{3}\right)  \cdot U_{0}%
\]%
\begin{equation}
-\left(  U_{1}U_{4}{}^{\dag}U_{2}U_{0}{}^{\dag}U_{4}\right)  \cdot U_{0}\cdot
U_{3}-\left(  U_{4}U_{0}{}^{\dag}U_{1}U_{4}{}^{\dag}U_{2}\right)  \cdot
U_{0}\cdot U_{3}+(1\leftrightarrow2). \label{id6}%
\end{equation}%
\[
0=tr\left(  U_{0}{}^{\dag}U_{2}\right)  \left(  U_{0}U_{4}{}^{\dag}U_{1}%
+U_{1}U_{4}{}^{\dag}U_{0}\right)  \cdot U_{3}\cdot U_{4}-2\left(  U_{1}U_{4}%
{}^{\dag}U_{2}+U_{2}U_{4}{}^{\dag}U_{1}\right)  \cdot U_{3}\cdot U_{4}%
\]%
\[
+U_{0}\cdot U_{3}\cdot U_{4}tr\left(  U_{0}{}^{\dag}U_{1}U_{4}{}^{\dag}%
U_{2}\right)  +U_{0}\cdot U_{3}\cdot U_{4}tr\left(  U_{0}{}^{\dag}U_{2}U_{4}%
{}^{\dag}U_{1}\right)
\]%
\[
-\left(  U_{1}U_{4}{}^{\dag}U_{2}U_{0}{}^{\dag}U_{3}\right)  \cdot U_{0}\cdot
U_{4}-\left(  U_{3}U_{0}{}^{\dag}U_{2}U_{4}{}^{\dag}U_{1}\right)  \cdot
U_{0}\cdot U_{4}%
\]%
\[
-\left(  U_{1}U_{4}{}^{\dag}U_{2}U_{0}{}^{\dag}U_{4}\right)  \cdot U_{0}\cdot
U_{3}-\left(  U_{4}U_{0}{}^{\dag}U_{2}U_{4}{}^{\dag}U_{1}\right)  \cdot
U_{0}\cdot U_{3}%
\]%
\[
-\left(  U_{0}U_{4}{}^{\dag}U_{1}\right)  \cdot\left(  U_{2}U_{0}{}^{\dag
}U_{3}\right)  \cdot U_{4}-\left(  U_{1}U_{4}{}^{\dag}U_{0}\right)
\cdot\left(  U_{3}U_{0}{}^{\dag}U_{2}\right)  \cdot U_{4}%
\]%
\begin{equation}
-\left(  U_{0}U_{4}{}^{\dag}U_{1}\right)  \cdot\left(  U_{2}U_{0}{}^{\dag
}U_{4}\right)  \cdot U_{3}-\left(  U_{1}U_{4}{}^{\dag}U_{0}\right)
\cdot\left(  U_{4}U_{0}{}^{\dag}U_{2}\right)  \cdot U_{3}. \label{id7}%
\end{equation}%
\[
0=tr\left(  U_{4}{}^{\dag}U_{1}\right)  \left(  U_{2}U_{0}{}^{\dag}U_{4}%
+U_{4}U_{0}{}^{\dag}U_{2}\right)  \cdot U_{0}\cdot U_{3}%
\]%
\[
+\left(  U_{0}U_{4}{}^{\dag}U_{1}\right)  \cdot\left(  U_{3}U_{0}{}^{\dag
}U_{2}\right)  \cdot U_{4}+\left(  U_{1}U_{4}{}^{\dag}U_{0}\right)
\cdot\left(  U_{2}U_{0}{}^{\dag}U_{3}\right)  \cdot U_{4}%
\]%
\begin{equation}
+\left(  U_{2}U_{0}{}^{\dag}U_{1}U_{4}{}^{\dag}U_{0}\right)  \cdot U_{3}\cdot
U_{4}-\left(  U_{4}U_{0}{}^{\dag}U_{2}U_{4}{}^{\dag}U_{1}\right)  \cdot
U_{0}\cdot U_{3}-(1\leftrightarrow2,0\leftrightarrow4). \label{id8}%
\end{equation}
These identities (\ref{id6}--\ref{id8}) also can be checked using
(\ref{3qWlidentity}) with $l=3$ and then expanding the product of Levi-Civita
symbols via (\ref{ee}).%
\[
0=\left(  U_{2}U_{4}{}^{\dag}U_{0}-U_{0}U_{4}{}^{\dag}U_{2}\right)
\cdot\left(  U_{3}U_{0}{}^{\dag}U_{1}-U_{1}U_{0}{}^{\dag}U_{3}\right)  \cdot
U_{4}%
\]%
\[
+\left(  U_{3}U_{4}{}^{\dag}U_{2}-U_{2}U_{4}{}^{\dag}U_{3}\right)
\cdot\left(  U_{4}U_{0}{}^{\dag}U_{1}-U_{1}U_{0}{}^{\dag}U_{4}\right)  \cdot
U_{0}%
\]%
\begin{equation}
+\left(  U_{1}U_{0}{}^{\dag}U_{2}-U_{2}U_{0}{}^{\dag}U_{1}\right)
\cdot\left(  U_{3}U_{4}{}^{\dag}U_{0}-U_{0}U_{4}{}^{\dag}U_{3}\right)  \cdot
U_{4}+(0\leftrightarrow4). \label{id9}%
\end{equation}
It can be proved using (\ref{3qWlidentity}) with $l=4$ and (\ref{ee}).
\[
0=2(U_{3}U_{0}{}^{\dag}U_{4}+U_{4}U_{0}{}^{\dag}U_{3})\cdot U_{1}\cdot
U_{2}-U_{1}\cdot U_{2}\cdot U_{3}tr(U_{0}{}^{\dag}U_{4})+(U_{1}U_{0}{}^{\dag
}U_{4}+U_{4}U_{0}{}^{\dag}U_{1})\cdot U_{2}\cdot U_{3}%
\]%
\begin{equation}
+3(U_{3}U_{0}{}^{\dag}U_{1}+U_{1}U_{0}{}^{\dag}U_{3})\cdot U_{2}\cdot
U_{4}-3U_{1}\cdot U_{2}\cdot U_{4}tr(U_{0}{}^{\dag}U_{3})+(1\leftrightarrow2).
\label{id10}%
\end{equation}
This identity is necessary for calculation of the quark contribution. It can
be proved using (\ref{3qWlidentity}) with $l=2$ and (\ref{ee}).

\section{Construction of conformal 4-point operator}

\label{4pointOperator}Here we derive the evolution equation for the operator
\begin{equation}
\left(  \left(  U_{2}U_{4}^{\dag}U_{1}+U_{1}U_{4}^{\dag}U_{2}\right)  \cdot
U_{4}\cdot U_{3}-2B_{123}\right)  =(-3B_{123}+\frac{1}{2}(B_{144}%
B_{324}+B_{244}B_{314}-B_{344}B_{214})).
\end{equation}
So one has to find the evolution of the operator $\left(  U_{1}U_{4}^{\dag
}U_{2}\right)  \cdot U_{4}\cdot U_{3}$ first. It reads
\[
\frac{\partial\left(  U_{1}U_{4}^{\dag}U_{2}\right)  \cdot U_{4}\cdot U_{3}%
}{\partial\eta}=\left(  U_{1}U_{4}^{\dag}U_{2}\right)  \cdot U_{4}\cdot
U_{3}\left(  -\frac{\alpha_{s}}{2\pi^{2}}\frac{4}{3}2\right)  \int d\vec
{z}_{0}\left(  \frac{1}{\vec{r}_{10}^{\,\,2}}+\frac{1}{\vec{r}_{20}^{\,\,2}%
}+\frac{1}{\vec{r}_{30}^{\,\,2}}+\frac{2}{\vec{r}_{40}^{\,\,2}}\right)
\]%
\[
-\frac{\alpha_{s}}{\pi^{2}}\left[  \left(  t^{c}U_{1}U_{4}^{\dag}t^{c}%
U_{2}\right)  \cdot U_{4}\cdot U_{3}+\left(  U_{1}t^{c}U_{4}^{\dag}U_{2}%
t^{c}\right)  \cdot U_{4}\cdot U_{3}\right]  \int d\vec{r}_{0}\frac{\left(
\vec{r}_{10}\vec{r}_{20}\right)  }{\vec{r}_{10}^{\,\,2}\vec{r}_{20}^{\,\,2}}%
\]%
\[
-\frac{\alpha_{s}}{\pi^{2}}\left[  \left(  t^{c}U_{1}U_{4}^{\dag}U_{2}\right)
\cdot U_{4}\cdot(t^{c}U_{3})+\left(  U_{1}t^{c}U_{4}^{\dag}U_{2}\right)  \cdot
U_{4}\cdot(U_{3}t^{c})\right]  \int d\vec{r}_{0}\frac{\left(  \vec{r}_{10}%
\vec{r}_{30}\right)  }{\vec{r}_{10}^{\,\,2}\vec{r}_{30}^{\,\,2}}%
\]%
\[
-\frac{\alpha_{s}}{\pi^{2}}\left[  \left(  U_{1}U_{4}^{\dag}t^{c}U_{2}\right)
\cdot U_{4}\cdot(t^{c}U_{3})+\left(  U_{1}U_{4}^{\dag}U_{2}t^{c}\right)  \cdot
U_{4}\cdot(U_{3}t^{c})\right]  \int d\vec{r}_{0}\frac{\left(  \vec{r}_{20}%
\vec{r}_{30}\right)  }{\vec{r}_{20}^{\,\,2}\vec{r}_{30}^{\,\,2}}%
\]%
\[
-\frac{\alpha_{s}}{\pi^{2}}\left[  \left(  t^{c}U_{1}U_{4}^{\dag}U_{2}\right)
\cdot(t^{c}U_{4})+\left(  U_{1}t^{c}U_{4}^{\dag}U_{2}\right)  \cdot(U_{4}%
t^{c})\right.
\]%
\[
\left.  -\left(  t^{c}U_{1}U_{4}^{\dag}t^{c}U_{2}\right)  \cdot U_{4}-\left(
U_{1}t^{c}t^{c}U_{4}^{\dag}U_{2}\right)  \cdot U_{4}\right]  \cdot U_{3}\int
d\vec{r}_{0}\frac{\left(  \vec{r}_{10}\vec{r}_{40}\right)  }{\vec{r}%
_{10}^{\,\,2}\vec{r}_{40}^{\,\,2}}%
\]%
\[
-\frac{\alpha_{s}}{\pi^{2}}\left[  \left(  U_{1}U_{4}^{\dag}t^{c}U_{2}\right)
\cdot(t^{c}U_{4})+\left(  U_{1}U_{4}^{\dag}U_{2}t^{c}\right)  \cdot(U_{4}%
t^{c})\right]
\]%
\[
\left.  -\left(  U_{1}U_{4}^{\dag}t^{c}t^{c}U_{2}\right)  \cdot U_{4}-\left(
U_{1}t^{c}U_{4}^{\dag}U_{2}t^{c}\right)  \cdot U_{4}\right]  \cdot U_{3}\int
d\vec{r}_{0}\frac{\left(  \vec{r}_{20}\vec{r}_{40}\right)  }{\vec{r}%
_{20}^{\,\,2}\vec{r}_{40}^{\,\,2}}%
\]%
\[
-\frac{\alpha_{s}}{\pi^{2}}\left[  \left(  U_{1}U_{4}^{\dag}U_{2}\right)
\cdot(t^{c}U_{4})\cdot(t^{c}U_{3})+\left(  U_{1}U_{4}^{\dag}U_{2}\right)
\cdot(U_{4}t^{c})\cdot(U_{3}t^{c})\right.
\]%
\[
\left.  -\left(  U_{1}U_{4}^{\dag}t^{c}U_{2}\right)  \cdot U_{4}\cdot
(t^{c}U_{3})-\left(  U_{1}t^{c}U_{4}^{\dag}U_{2}\right)  \cdot U_{4}%
\cdot(U_{3}t^{c})\right]  \int d\vec{r}_{0}\frac{\left(  \vec{r}_{30}\vec
{z}_{40}\right)  }{\vec{r}_{30}^{\,\,2}\vec{r}_{40}^{\,\,2}}%
\]%
\[
+\frac{\alpha_{s}}{\pi^{2}}\left[  \left(  U_{1}U_{4}^{\dag}t^{c}U_{2}\right)
\cdot(t^{c}U_{4})\cdot U_{3}+\left(  U_{1}t^{c}U_{4}^{\dag}U_{2}\right)
\cdot(U_{4}t^{c})\cdot U_{3}\right]  \int\frac{d\vec{r}_{0}}{\vec{r}%
_{40}^{\,\,2}}%
\]%
\[
+\frac{\alpha_{s}}{\pi^{2}}\int d\vec{r}_{0}U_{0}^{cd}\left(  \frac{\left(
U_{1}U_{4}^{\dag}U_{2}\right)  \cdot(t^{c}U_{4}t^{d})\cdot U_{3}}{\vec{r}%
_{04}^{\,\,2}}+\frac{\left(  U_{1}t^{d}U_{4}^{\dag}t^{c}U_{2}\right)  \cdot
U_{4}\cdot U_{3}}{\vec{r}_{04}^{\,\,2}}\right.
\]%
\[
\left.  +\frac{\left(  t^{c}U_{1}t^{d}U_{4}^{\dag}U_{2}\right)  \cdot
U_{4}\cdot U_{3}}{\vec{r}_{01}^{\,\,2}}+\frac{\left(  U_{1}U_{4}^{\dag}%
t^{c}U_{2}t^{d}\right)  \cdot U_{4}\cdot U_{3}}{\vec{r}_{02}^{\,\,2}}%
+\frac{\left(  U_{1}U_{4}^{\dag}U_{2}\right)  \cdot U_{4}\cdot(t^{c}U_{3}%
t^{d})}{\vec{r}_{03}^{\,\,2}}\right)
\]%
\[
+\frac{\alpha_{s}}{\pi^{2}}\int d\vec{r}_{0}\frac{\left(  \vec{r}_{03}\vec
{z}_{02}\right)  }{\vec{r}_{03}^{\,\,2}\vec{r}_{02}^{\,\,2}}U_{0}^{cd}\left(
\left(  U_{1}U_{4}^{\dag}t^{c}U_{2}\right)  \cdot U_{4}\cdot(U_{3}%
t^{d})+\left(  U_{1}U_{4}^{\dag}U_{2}t^{d}\right)  \cdot U_{4}\cdot(t^{c}%
U_{3})\right)
\]%
\[
+\frac{\alpha_{s}}{\pi^{2}}\int d\vec{r}_{0}\frac{\left(  \vec{r}_{01}\vec
{z}_{02}\right)  }{\vec{r}_{01}^{\,\,2}\vec{r}_{02}^{\,\,2}}U_{0}^{cd}\left(
\left(  U_{1}t^{d}U_{4}^{\dag}t^{c}U_{2}\right)  \cdot U_{4}\cdot
U_{3}+\left(  t^{c}U_{1}U_{4}^{\dag}U_{2}t^{d}\right)  \cdot U_{4}\cdot
U_{3}\right)
\]%
\[
+\frac{\alpha_{s}}{\pi^{2}}\int d\vec{r}_{0}\frac{\left(  \vec{r}_{01}\vec
{z}_{03}\right)  }{\vec{r}_{01}^{\,\,2}\vec{r}_{03}^{\,\,2}}U_{0}^{cd}\left(
\left(  U_{1}t^{d}U_{4}^{\dag}U_{2}\right)  \cdot U_{4}\cdot(t^{c}%
U_{3})+\left(  t^{c}U_{1}U_{4}^{\dag}U_{2}\right)  \cdot U_{4}\cdot(U_{3}%
t^{d})\right)
\]%
\[
+\frac{\alpha_{s}}{\pi^{2}}\int d\vec{r}_{0}\frac{\left(  \vec{r}_{04}\vec
{z}_{01}\right)  }{\vec{r}_{04}^{\,\,2}\vec{r}_{01}^{\,\,2}}U_{0}^{cd}\left(
\left(  U_{1}t^{d}U_{4}^{\dag}U_{2}\right)  \cdot(t^{c}U_{4})\cdot
U_{3}+\left(  t^{c}U_{1}U_{4}^{\dag}U_{2}\right)  \cdot(U_{4}t^{d})\cdot
U_{3}\frac{{}}{{}}\right.
\]%
\[
\left.  \frac{{}}{{}}-\left(  U_{1}t^{d}U_{4}^{\dag}t^{c}U_{2}\right)  \cdot
U_{4}\cdot U_{3}-\left(  t^{c}U_{1}t^{d}U_{4}^{\dag}U_{2}\right)  \cdot
U_{4}\cdot U_{3}\right)
\]%
\[
+\frac{\alpha_{s}}{\pi^{2}}\int d\vec{r}_{0}\frac{\left(  \vec{r}_{04}\vec
{z}_{02}\right)  }{\vec{r}_{04}^{\,\,2}\vec{r}_{02}^{\,\,2}}U_{0}^{cd}\left(
\left(  U_{1}U_{4}^{\dag}U_{2}t^{d}\right)  \cdot(t^{c}U_{4})\cdot
U_{3}+\left(  U_{1}U_{4}^{\dag}t^{c}U_{2}\right)  \cdot(U_{4}t^{d})\cdot
U_{3}\frac{{}}{{}}\right.
\]%
\[
\left.  \frac{{}}{{}}-\left(  U_{1}U_{4}^{\dag}t^{c}U_{2}t^{d}\right)  \cdot
U_{4}\cdot U_{3}-\left(  U_{1}t^{d}U_{4}^{\dag}t^{c}U_{2}\right)  \cdot
U_{4}\cdot U_{3}\right)
\]%
\[
+\frac{\alpha_{s}}{\pi^{2}}\int d\vec{r}_{0}\frac{\left(  \vec{r}_{04}\vec
{z}_{03}\right)  }{\vec{r}_{04}^{\,\,2}\vec{r}_{03}^{\,\,2}}U_{0}^{cd}\left(
\left(  U_{1}U_{4}^{\dag}U_{2}\right)  \cdot(t^{c}U_{4})\cdot(U_{3}%
t^{d})+\left(  U_{1}U_{4}^{\dag}U_{2}\right)  \cdot(U_{4}t^{d})\cdot
(t^{c}U_{3})\frac{{}}{{}}\right.
\]%
\[
\left.  \frac{{}}{{}}-\left(  U_{1}U_{4}^{\dag}t^{c}U_{2}\right)  \cdot
U_{4}\cdot(U_{3}t^{d})-\left(  U_{1}t^{d}U_{4}^{\dag}U_{2}\right)  \cdot
U_{4}\cdot(t^{c}U_{3})\right)
\]%
\begin{equation}
-\frac{\alpha_{s}}{\pi^{2}}\int\frac{d\vec{r}_{0}}{\vec{r}_{04}^{\,\,2}}%
U_{0}^{cd}\left(  \left(  U_{1}t^{d}U_{4}^{\dag}U_{2}\right)  \cdot(t^{c}%
U_{4})\cdot U_{3}+\left(  U_{1}U_{4}^{\dag}t^{c}U_{2}\right)  \cdot(U_{4}%
t^{d})\cdot U_{3}\right)  .
\end{equation}
Using (\ref{IDENTITY}) and (\ref{simple_identity}--\ref{id7}) after the
convolution one gets
\[
\frac{\partial}{\partial\eta}\left(  \left(  U_{2}U_{4}^{\dag}U_{1}+U_{1}%
U_{4}^{\dag}U_{2}\right)  \cdot U_{4}\cdot U_{3}-2B_{123}\right)
=\frac{\alpha_{s}}{4\pi^{2}}\int d\vec{r}_{0}%
\]%
\begin{equation}
\times\left(  A_{34}\frac{\vec{r}_{34}^{\,\,2}}{\vec{r}_{03}^{\,\,2}\vec
{r}_{04}^{\,\,2}}+A_{13}\frac{\vec{r}_{13}^{\,2}}{\vec{r}_{03}^{\,\,2}\vec
{r}_{01}^{\,\,2}}+A_{23}\frac{\vec{r}_{23}^{\,\,2}}{\vec{r}_{03}^{\,\,2}%
\vec{r}_{02}^{\,\,2}}+A_{14}\frac{\vec{r}_{14}^{\,\,2}}{\vec{r}_{01}%
^{\,\,2}\vec{r}_{04}^{\,\,2}}+A_{24}\frac{\vec{r}_{24}^{\,\,2}}{\vec{r}%
_{02}^{\,\,2}\vec{r}_{04}^{\,\,2}}+A_{12}\frac{\vec{r}_{12}^{\,\,2}}{\vec
{r}_{01}^{\,\,2}\vec{r}_{02}^{\,\,2}}\right)  .
\end{equation}
Here%
\[
A_{34}=-2\left(  U_{2}U_{4}^{\dag}U_{1}+U_{1}U_{4}^{\dag}U_{2}\right)  \cdot
U_{4}\cdot U_{3}+\left(  U_{3}U_{4}^{\dag}U_{1}+U_{1}U_{4}^{\dag}U_{3}\right)
\cdot U_{4}\cdot U_{2}%
\]%
\[
+\left(  U_{3}U_{4}^{\dag}U_{2}+U_{2}U_{4}^{\dag}U_{3}\right)  \cdot
U_{4}\cdot U_{1}+\left(  U_{2}U_{4}{}^{\dag}U_{1}+U_{1}U_{4}{}^{\dag}%
U_{2}\right)  \cdot\left(  U_{3}U_{0}{}^{\dag}U_{4}+U_{4}U_{0}{}^{\dag}%
U_{3}\right)  \cdot U_{0}%
\]%
\[
-\left(  U_{2}U_{4}{}^{\dag}U_{0}\right)  \cdot\left(  U_{3}U_{0}{}^{\dag
}U_{1}\right)  \cdot U_{4}-\left(  U_{0}U_{4}{}^{\dag}U_{2}\right)
\cdot\left(  U_{1}U_{0}{}^{\dag}U_{3}\right)  \cdot U_{4}%
\]%
\begin{equation}
-\left(  U_{0}U_{4}{}^{\dag}U_{1}\right)  \cdot\left(  U_{2}U_{0}{}^{\dag
}U_{3}\right)  \cdot U_{4}-\left(  U_{1}U_{4}{}^{\dag}U_{0}\right)
\cdot\left(  U_{3}U_{0}{}^{\dag}U_{2}\right)  \cdot U_{4}. \label{A34}%
\end{equation}%
\[
A_{13}=\left(  U_{0}U_{4}{}^{\dag}U_{2}\right)  \cdot\left(  U_{1}U_{0}%
{}^{\dag}U_{3}\right)  \cdot U_{4}+\left(  U_{2}U_{4}{}^{\dag}U_{0}\right)
\cdot\left(  U_{3}U_{0}{}^{\dag}U_{1}\right)  \cdot U_{4}%
\]%
\[
+\left(  U_{3}U_{0}{}^{\dag}U_{1}U_{4}{}^{\dag}U_{2}\right)  \cdot U_{0}\cdot
U_{4}+\left(  U_{2}U_{4}{}^{\dag}U_{1}U_{0}{}^{\dag}U_{3}\right)  \cdot
U_{0}\cdot U_{4}%
\]%
\[
-2\left(  U_{1}U_{0}{}^{\dag}U_{3}+U_{3}U_{0}{}^{\dag}U_{1}\right)  \cdot
U_{0}\cdot U_{2}-\left(  U_{3}U_{4}{}^{\dag}U_{2}+U_{2}U_{4}{}^{\dag}%
U_{3}\right)  \cdot U_{1}\cdot U_{4}%
\]%
\begin{equation}
-\left(  U_{1}U_{4}{}^{\dag}U_{2}+U_{2}U_{4}{}^{\dag}U_{1}\right)  \cdot
U_{3}\cdot U_{4}+4U_{1}\cdot U_{2}\cdot U_{3}.
\end{equation}%
\[
A_{14}=tr\left(  U_{0}{}^{\dag}U_{1}\right)  \left(  U_{2}U_{4}{}^{\dag}%
U_{0}+U_{0}U_{4}{}^{\dag}U_{2}\right)  \cdot U_{3}\cdot U_{4}+tr\left(
U_{4}{}^{\dag}U_{0}\right)  \left(  U_{1}U_{0}{}^{\dag}U_{2}+U_{2}U_{0}%
{}^{\dag}U_{1}\right)  \cdot U_{3}\cdot U_{4}%
\]%
\[
+\left(  U_{2}U_{4}{}^{\dag}U_{0}\right)  \cdot\left(  U_{4}U_{0}{}^{\dag
}U_{1}\right)  \cdot U_{3}+\left(  U_{0}U_{4}{}^{\dag}U_{2}\right)
\cdot\left(  U_{1}U_{0}{}^{\dag}U_{4}\right)  \cdot U_{3}%
\]%
\[
-2U_{4}\cdot U_{2}\cdot U_{3}tr\left(  U_{4}{}^{\dag}U_{1}\right)
-2U_{1}\cdot U_{2}\cdot U_{3}-4\left(  U_{1}U_{4}{}^{\dag}U_{2}+U_{2}U_{4}%
{}^{\dag}U_{1}\right)  \cdot U_{3}\cdot U_{4}%
\]%
\begin{equation}
+\left(  U_{4}U_{0}{}^{\dag}U_{1}U_{4}{}^{\dag}U_{2}\right)  \cdot U_{0}\cdot
U_{3}+\left(  U_{2}U_{4}{}^{\dag}U_{1}U_{0}{}^{\dag}U_{4}\right)  \cdot
U_{0}\cdot U_{3}.
\end{equation}%
\[
A_{12}=-2\left(  U_{1}U_{0}{}^{\dag}U_{2}+U_{2}U_{0}{}^{\dag}U_{1}\right)
\cdot U_{3}\cdot U_{0}-tr\left(  U_{4}{}^{\dag}U_{0}\right)  \left(
U_{1}U_{0}{}^{\dag}U_{2}+U_{2}U_{0}{}^{\dag}U_{1}\right)  \cdot U_{3}\cdot
U_{4}%
\]%
\[
+4U_{1}\cdot U_{2}\cdot U_{3}+2U_{4}\cdot U_{2}\cdot U_{3}tr\left(  U_{4}%
{}^{\dag}U_{1}\right)  +2U_{4}\cdot U_{1}\cdot U_{3}tr\left(  U_{4}{}^{\dag
}U_{2}\right)
\]%
\begin{equation}
-U_{0}\cdot U_{3}\cdot U_{4}\left(  tr\left(  U_{0}{}^{\dag}U_{1}U_{4}{}%
^{\dag}U_{2}\right)  +tr\left(  U_{0}{}^{\dag}U_{2}U_{4}{}^{\dag}U_{1}\right)
\right)  .
\end{equation}%
\begin{equation}
A_{23}=A_{13}|_{\vec{r}_{1}\leftrightarrow\vec{r}_{2},}\quad A_{24}%
=A_{14}|_{\vec{r}_{1}\leftrightarrow\vec{r}_{2}}. \label{A23}%
\end{equation}
Our prescription for the composite conformal operators reads
\cite{Balitsky:2009xg} (see also Ref. \cite{NLOJIMWLKonformal})%
\begin{equation}
O^{conf}=O+\frac{1}{2}\frac{\partial O}{\partial\eta}\left\vert _{\frac
{\vec{r}_{mn}^{\,\,2}}{\vec{r}_{im}^{\,\,2}\vec{r}_{in}^{\,\,2}}%
\rightarrow\frac{\vec{r}_{mn}^{\,\,2}}{\vec{r}_{im}^{\,\,2}\vec{r}%
_{in}^{\,\,2}}\ln\left(  \frac{\vec{r}_{mn}^{\,\,2}a}{\vec{r}_{im}^{\,\,2}%
\vec{r}_{in}^{\,\,2}}\right)  }\right.  ,
\end{equation}
where $a$ is an arbitrary constant. Thus%
\begin{equation}
B_{123}^{conf}=B_{123}+\frac{1}{2}\frac{\partial B_{123}}{\partial\eta
}\left\vert _{\frac{\vec{r}_{mn}^{\,\,2}}{\vec{r}_{im}^{\,\,2}\vec{r}%
_{in}^{\,\,2}}\rightarrow\frac{\vec{r}_{mn}^{\,\,2}}{\vec{r}_{im}^{\,\,2}%
\vec{r}_{in}^{\,\,2}}\ln\left(  \frac{\vec{r}_{mn}^{\,\,2}a}{\vec{r}%
_{im}^{\,\,2}\vec{r}_{in}^{\,\,2}}\right)  }\right.
\end{equation}%
\[
=B_{123}+\frac{\alpha_{s}3}{8\pi^{2}}\int d\vec{r}_{0}\left[  \frac{\vec
{r}_{12}^{\,\,2}}{\vec{r}_{01}^{\,\,2}\vec{r}_{02}^{\,\,2}}\ln\left(
\frac{a\vec{r}_{12}^{\,\,2}}{\vec{r}_{01}^{\,\,2}\vec{r}_{02}^{\,\,2}}\right)
\right.
\]%
\begin{equation}
\left.  \times(-B_{123}+\frac{1}{6}(B_{100}B_{320}+B_{200}B_{310}%
-B_{300}B_{210}))+(1\leftrightarrow3)+(2\leftrightarrow3)\right]  ,
\end{equation}%
\[
(-3B_{123}+\frac{1}{2}(B_{144}B_{324}+B_{244}B_{314}-B_{344}B_{214}))^{conf}%
\]%
\[
=(-3B_{123}+\frac{1}{2}(B_{144}B_{324}+B_{244}B_{314}-B_{344}B_{214}))
\]%
\begin{equation}
+\frac{1}{2}\frac{\partial}{\partial\eta}(-3B_{123}+\frac{1}{2}(B_{144}%
B_{324}+B_{244}B_{314}-B_{344}B_{214}))\left\vert _{\frac{\vec{r}_{mn}%
^{\,\,2}}{\vec{r}_{im}^{\,\,2}\vec{r}_{in}^{\,\,2}}\rightarrow\frac{\vec
{r}_{mn}^{\,\,2}}{\vec{r}_{im}^{\,\,2}\vec{r}_{in}^{\,\,2}}\ln\left(
\frac{\vec{r}_{mn}^{\,\,2}a}{\vec{r}_{im}^{\,\,2}\vec{r}_{in}^{\,\,2}}\right)
}\right.
\end{equation}%
\[
=(-3B_{123}+\frac{1}{2}(B_{144}B_{324}+B_{244}B_{314}-B_{344}B_{214}))
\]%
\[
+\frac{\alpha_{s}}{8\pi^{2}}\int d\vec{r}_{0}\left(  A_{34}\frac{\vec{r}%
_{34}^{\,\,2}}{\vec{r}_{03}^{\,\,2}\vec{r}_{04}^{\,\,2}}\ln\left(  \frac
{\vec{r}_{34}^{\,\,2}a}{\vec{r}_{03}^{\,\,2}\vec{r}_{04}^{\,\,2}}\right)
+A_{13}\frac{\vec{r}_{13}^{\,2}}{\vec{r}_{03}^{\,\,2}\vec{r}_{01}^{\,\,2}}%
\ln\left(  \frac{\vec{r}_{13}^{\,2}a}{\vec{r}_{03}^{\,\,2}\vec{r}_{01}%
^{\,\,2}}\right)  +A_{23}\frac{\vec{r}_{23}^{\,\,2}}{\vec{r}_{03}^{\,\,2}%
\vec{r}_{02}^{\,\,2}}\ln\left(  \frac{\vec{r}_{23}^{\,\,2}a}{\vec{r}%
_{03}^{\,\,2}\vec{r}_{02}^{\,\,2}}\right)  \right.
\]%
\begin{equation}
\left.  +A_{14}\frac{\vec{r}_{14}^{\,\,2}}{\vec{r}_{01}^{\,\,2}\vec{r}%
_{04}^{\,\,2}}\ln\left(  \frac{\vec{r}_{14}^{\,\,2}a}{\vec{r}_{01}^{\,\,2}%
\vec{r}_{04}^{\,\,2}}\right)  +A_{24}\frac{\vec{r}_{24}^{\,\,2}}{\vec{r}%
_{02}^{\,\,2}\vec{z}_{04}^{\,\,2}}\ln\left(  \frac{\vec{r}_{24}^{\,\,2}a}%
{\vec{r}_{02}^{\,\,2}\vec{z}_{04}^{\,\,2}}\right)  +A_{12}\frac{\vec{r}%
_{12}^{\,\,2}}{\vec{r}_{01}^{\,\,2}\vec{r}_{02}^{\,\,2}}\ln\left(  \frac
{\vec{r}_{12}^{\,\,2}a}{\vec{r}_{01}^{\,\,2}\vec{r}_{02}^{\,\,2}}\right)
\right)  .
\end{equation}
In the 3-gluon approximation%
\[
(-3B_{123}+\frac{1}{2}(B_{144}B_{324}+B_{244}B_{314}-B_{344}B_{214}))
\]%
\begin{equation}
\overset{\mathrm{3g}}{=}3(-B_{123}+B_{144}+B_{324}+B_{244}+B_{314}%
-B_{344}-B_{214}-6).
\end{equation}
Therefore%
\[
(-3B_{123}+\frac{1}{2}(B_{144}B_{324}+B_{244}B_{314}-B_{344}B_{214}))^{conf}%
\]%
\[
\overset{\mathrm{3g}}{=}3(-B_{123}+B_{144}+B_{324}+B_{244}+B_{314}%
-B_{344}-B_{214}-6)
\]%
\begin{equation}
+\frac{3}{2}\frac{\partial}{\partial\eta}(-B_{123}+B_{144}+B_{324}%
+B_{244}+B_{314}-B_{344}-B_{214}-6)\left\vert _{\frac{\vec{r}_{mn}^{\,\,2}%
}{\vec{r}_{im}^{\,\,2}\vec{r}_{in}^{\,\,2}}\rightarrow\frac{\vec{r}%
_{mn}^{\,\,2}}{\vec{r}_{im}^{\,\,2}\vec{r}_{in}^{\,\,2}}\ln\left(  \frac
{\vec{r}_{mn}^{\,\,2}a}{\vec{r}_{im}^{\,\,2}\vec{r}_{in}^{\,\,2}}\right)
}\right.
\end{equation}%
\begin{equation}
=3(-B_{123}^{conf}+B_{144}^{conf}+B_{324}^{conf}+B_{244}^{conf}+B_{314}%
^{conf}-B_{344}^{conf}-B_{214}^{conf}-6). \label{compositeOperator3g}%
\end{equation}

\section{Integrals}

\label{integrals}Here we describe the calculation of integral (\ref{Z}). It
reads%
\begin{equation}
\int d\vec{r}_{4}Z_{12}=J_{12}-(1\leftrightarrow3).
\end{equation}%
\[
J_{12}=\frac{\vec{r}_{12}{}^{2}}{8\vec{r}_{01}{}^{2}\vec{r}_{02}{}^{2}}%
\int\!d\vec{r}_{4}\left[  \left(  {\frac{\vec{r}_{03}^{\,\,2}}{\vec{r}%
_{04}^{\,\,2}\vec{r}_{34}^{\,\,2}}}-{\frac{\vec{r}_{02}^{\,\,2}}{\vec{r}%
_{04}^{\,\,2}\vec{r}_{24}^{\,\,2}}}\right)  \ln\left(  {\frac{\vec{r}_{02}%
{}^{2}\vec{r}_{14}{}^{2}}{\vec{r}_{04}{}^{2}\vec{r}_{12}{}^{2}}}\right)
\right.
\]%
\begin{equation}
\left.  +{\frac{\vec{r}_{01}{}^{2}}{\vec{r}_{04}{}^{2}\vec{r}_{14}{}^{2}}}%
\ln\left(  {\frac{\vec{r}_{02}{}^{2}\vec{r}_{34}{}^{2}}{\vec{r}_{03}{}^{2}%
\vec{r}_{24}{}^{2}}}\right)  +{\frac{\vec{r}_{13}{}^{2}}{\vec{r}_{14}{}%
^{2}\vec{r}_{34}{}^{2}}}\ln\left(  {\frac{\vec{r}_{03}{}^{2}\vec{r}_{12}{}%
^{2}}{\vec{r}_{02}{}^{2}\vec{r}_{13}{}^{2}}}\right)  \right]  .
\end{equation}
Since the integral is conformally invariant, one can set $\vec{r}_{0}=0$ and
make the inversion, then calculate the integral and then again make the
inversion and restore $\vec{r}_{0}.$
\[
J_{12}\overset{\vec{r}_{0}=0}{\rightarrow}\frac{\vec{r}_{12}{}^{2}}{8\vec
{r}_{1}{}^{2}\vec{r}_{2}{}^{2}}\int\!dr_{4}\left[  \left(  {\frac{\vec{r}%
_{3}{}^{2}}{\vec{r}_{4}{}^{2}\vec{r}_{34}{}^{2}}}-{\frac{\vec{r}_{2}{}^{2}%
}{\vec{r}_{4}{}^{2}\vec{r}_{24}{}^{2}}}\right)  \ln\left(  {\frac{\vec{r}%
_{2}{}^{2}\vec{r}_{14}{}^{2}}{\vec{r}_{4}{}^{2}\vec{r}_{12}{}^{2}}}\right)
\right.
\]%
\begin{equation}
\left.  +{\frac{\vec{r}_{1}{}^{2}}{\vec{r}_{4}{}^{2}\vec{r}_{14}{}^{2}}}%
\ln\left(  {\frac{\vec{r}_{2}{}^{2}\vec{r}_{34}{}^{2}}{\vec{r}_{3}{}^{2}%
\vec{r}_{24}{}^{2}}}\right)  +{\frac{\vec{r}_{13}{}^{2}}{\vec{r}_{14}{}%
^{2}\vec{r}_{34}{}^{2}}}\ln\left(  {\frac{\vec{r}_{3}{}^{2}\vec{r}_{12}{}^{2}%
}{\vec{r}_{2}{}^{2}\vec{r}_{13}{}^{2}}}\right)  \right]
\end{equation}%
\begin{equation}
\overset{\mathrm{inversion}}{\rightarrow}\frac{r_{12}^{2}}{8}\int%
\!dr_{4}\left[  \left(  {\frac{1}{\vec{r}_{34}{}^{2}}}-{\frac{1}{\vec{r}%
_{24}{}^{2}}}\right)  \ln\left(  {\frac{\vec{r}_{14}{}^{2}}{\vec{r}_{12}{}%
^{2}}}\right)  +{\frac{1}{\vec{r}_{14}{}^{2}}}\ln\left(  {\frac{\vec{r}_{34}%
{}^{2}}{\vec{r}_{24}{}^{2}}}\right)  +{\frac{\vec{r}_{13}{}^{2}}{\vec{r}%
_{14}{}^{2}\vec{r}_{34}{}^{2}}}\ln\left(  {\frac{\vec{r}_{12}{}^{2}}{\vec
{r}_{13}{}^{2}}}\right)  \right]  .
\end{equation}
Using the integrals from appendix A in \cite{Fadin:2009za} we have
\begin{equation}
J_{12}\rightarrow-\pi\frac{r_{12}^{2}}{8}\ln^{2}\left(  {\frac{\vec{r}%
_{13}^{\,\,2}}{\vec{r}_{12}^{\,\,2}}}\right)  .
\end{equation}
After inversion and restoring of $\vec{r}_{0}$ we get
\begin{equation}
J_{12}=~-\pi\frac{\vec{r}_{12}{}^{2}}{8\vec{r}_{01}{}^{2}\vec{r}_{02}{}^{2}%
}\ln{}^{2}\left(  {\frac{\vec{r}_{12}{}^{2}\vec{r}_{30}{}^{2}}{\vec{r}_{13}%
{}^{2}\vec{r}_{20}{}^{2}}}\right)  {.}%
\end{equation}
Therefore
\begin{equation}
\int\frac{d\vec{r}_{4}}{\pi}Z_{12}={\frac{\vec{r}_{32}{}^{2}}{8\vec{r}_{03}%
{}^{2}\vec{r}_{02}{}^{2}}}\ln^{2}\left(  {\frac{\vec{r}_{32}{}^{2}\vec{r}%
_{10}{}^{2}}{\vec{r}_{13}{}^{2}\vec{r}_{20}{}^{2}}}\right)  -{\frac{\vec
{r}_{12}{}^{2}}{8\vec{r}_{01}{}^{2}\vec{r}_{02}{}^{2}}}\ln^{2}\left(
{\frac{\vec{r}_{12}{}^{2}\vec{r}_{30}{}^{2}}{\vec{r}_{13}{}^{2}\vec{r}_{20}%
{}^{2}}}\right)  {.}%
\end{equation}
Now we will integrate $F_{100}$ (\ref{F100}) w.r.t. $\vec{r}_{4}.$ Again we
set $\vec{r}_{0}=0$, do inversion, and calculate the integral in the
$d=2+2\epsilon$ dimensional space using the integrals from appendix A in
\cite{Fadin:2009za}\ and
\begin{equation}
\int\frac{d^{2+2\epsilon}r_{14}}{\pi^{1+\epsilon}\Gamma\left(  1-\epsilon
\right)  }\frac{r_{34}{}^{2}}{r_{14}{}^{2}r_{24}{}^{2}}=\frac{r_{13}%
^{2}+r_{23}^{2}-r_{12}^{2}}{r_{12}^{2}}\left(  \frac{1}{\epsilon}+\ln\left(
r_{12}^{2}\right)  \right)  +O\left(  \epsilon\right)  . \label{int1}%
\end{equation}
We get%
\[
\int\frac{d\vec{r}_{4}}{\pi}F_{100}+(2\leftrightarrow3)\rightarrow\int%
\frac{d^{d}r_{4}}{\pi}\left(  \frac{r_{12}{}^{2}}{r_{24}{}^{2}}\ln\left(
\frac{r_{14}{}^{2}}{r_{12}{}^{2}}\right)  +\frac{r_{23}{}^{2}r_{12}{}^{2}%
}{2r_{14}{}^{2}r_{24}{}^{2}}\ln\left(  \frac{r_{14}{}^{2}r_{23}{}^{2}}%
{r_{12}{}^{2}r_{24}{}^{2}}\right)  \right.
\]%
\[
+\frac{r_{13}{}^{2}r_{12}{}^{2}}{r_{14}{}^{2}r_{24}{}^{2}}\ln\left(
\frac{r_{13}{}^{2}r_{24}{}^{2}}{r_{12}{}^{2}r_{14}{}^{2}}\right)
-\frac{r_{13}{}^{2}}{r_{24}{}^{2}}\ln\left(  \frac{r_{13}{}^{2}r_{24}{}^{2}%
}{r_{14}{}^{4}{}}\right)  +\frac{r_{23}{}^{2}}{2r_{24}{}^{2}}\ln\left(
\frac{r_{23}{}^{2}r_{24}{}^{4}{}r_{34}{}^{2}}{r_{14}{}^{8}{}}\right)
\]%
\begin{equation}
\left.  +\frac{r_{23}{}^{2}}{2r_{14}{}^{2}}\ln\left(  \frac{r_{24}{}^{2}%
r_{34}{}^{2}}{r_{14}{}^{2}r_{23}{}^{2}}\right)  -\frac{r_{34}{}^{2}r_{12}%
{}^{2}}{2r_{14}{}^{2}r_{24}{}^{2}}\ln\left(  \frac{r_{24}{}^{2}r_{34}{}^{6}{}%
}{r_{12}{}^{6}{}r_{14}{}^{2}}\right)  -\frac{r_{12}{}^{2}}{2r_{14}{}^{2}}%
\ln\left(  \frac{r_{34}{}^{4}{}}{r_{12}{}^{2}r_{14}{}^{2}}\right)  \right)
+(2\leftrightarrow3)
\end{equation}%
\[
\overset{d\rightarrow2}{\rightarrow}\left(  \frac{3\left(  r_{13}{}^{2}%
-r_{12}{}^{2}\right)  }{4}-\frac{r_{23}{}^{2}}{2}\right)  \ln^{2}\left(
\frac{r_{12}{}^{2}}{r_{23}{}^{2}}\right)  +\left(  \frac{3\left(  r_{12}{}%
^{2}-r_{13}{}^{2}\right)  }{4}-\frac{r_{23}{}^{2}}{2}\right)  \ln^{2}\left(
\frac{r_{13}{}^{2}}{r_{23}{}^{2}}\right)
\]%
\begin{equation}
+\left(  \frac{3}{4}r_{23}{}^{2}-r_{12}{}^{2}-r_{13}{}^{2}\right)  \ln
^{2}\left(  \frac{r_{12}{}^{2}}{r_{13}{}^{2}}\right)  +\frac{3}{2}%
S_{123}I\left(  r_{12}{}^{2},r_{13}{}^{2},r_{23}{}^{2}\right)  .
\end{equation}
Here%
\begin{equation}
S_{123}=r_{12}{}^{4}{}+r_{13}{}^{4}{}+r_{23}{}^{4}-2r_{13}{}^{2}r_{12}{}%
^{2}-2r_{23}{}^{2}r_{12}{}^{2}{}-2r_{13}{}^{2}r_{23}{}^{2}%
\end{equation}
is the Cayley-Menger determinant proportional to the squared area of the
triangle with the corners at $r=r_{1,2,3},$ and%
\begin{equation}
I(a,b,c)=\int_{0}^{1}\frac{dx}{a(1-x)+bx-cx(1-x)}\ln\left(  \frac
{a(1-x)+bx}{cx(1-x)}\right)  ~ \label{integral I}%
\end{equation}%
\begin{equation}
=\int_{0}^{1}\int_{0}^{1}\int_{0}^{1}\frac{dx_{1}dx_{2}dx_{3}\delta
(1-x_{1}-x_{2}-x_{3})}{(ax_{1}+bx_{2}+cx_{3})(x_{1}x_{2}+x_{1}x_{3}+x_{2}%
x_{3})}~
\end{equation}%
\begin{equation}
=\int_{0}^{1}dx\int_{0}^{1}{dz}\;\frac{1}{cx(1-x)z+(b(1-x)+ax)(1-z)}~.
\end{equation}
is symmetric w.r.t. interchange of its arguments function defined in
\cite{Fadin:2002hz}. Performing inversion and restoring $r_{0},$ we get%
\[
\int\frac{d\vec{r}_{4}}{\pi}F_{100}+(2\leftrightarrow3)=\left(  \frac{3\vec
{r}_{23}{}^{2}}{4\vec{r}_{02}{}^{2}\vec{r}_{03}{}^{2}}-\frac{\vec{r}_{12}%
{}^{2}}{\vec{r}_{01}{}^{2}\vec{r}_{02}{}^{2}}-\frac{\vec{r}_{13}{}^{2}}%
{\vec{r}_{01}{}^{2}\vec{r}_{03}{}^{2}}\right)  \ln^{2}\left(  \frac{\vec
{r}_{03}{}^{2}\vec{r}_{12}{}^{2}}{\vec{r}_{02}{}^{2}\vec{r}_{13}{}^{2}%
}\right)
\]%
\[
+\left(  \frac{3\vec{r}_{12}{}^{2}}{4\vec{r}_{01}{}^{2}\vec{r}_{02}{}^{2}%
}-\frac{3\vec{r}_{13}{}^{2}}{4\vec{r}_{01}{}^{2}\vec{r}_{03}{}^{2}}-\frac
{\vec{r}_{23}{}^{2}}{2\vec{r}_{02}{}^{2}\vec{r}_{03}{}^{2}}\right)  \ln
^{2}\left(  \frac{\vec{r}_{02}{}^{2}\vec{r}_{13}{}^{2}}{\vec{r}_{01}{}^{2}%
\vec{r}_{23}{}^{2}}\right)
\]%
\[
+\left(  \frac{3\vec{r}_{13}{}^{2}}{4\vec{r}_{01}{}^{2}\vec{r}_{03}{}^{2}%
}-\frac{3\vec{r}_{12}{}^{2}}{4\vec{r}_{01}{}^{2}\vec{r}_{02}{}^{2}}-\frac
{\vec{r}_{23}{}^{2}}{2\vec{r}_{02}{}^{2}\vec{r}_{03}{}^{2}}\right)  \ln
^{2}\left(  \frac{\vec{r}_{03}{}^{2}\vec{r}_{12}{}^{2}}{\vec{r}_{01}{}^{2}%
\vec{r}_{23}{}^{2}}\right)
\]%
\[
+\frac{3}{2}\tilde{S}_{123}I\left(  \frac{\vec{r}_{12}{}^{2}}{\vec{r}_{01}%
{}^{2}\vec{r}_{02}{}^{2}},\frac{\vec{r}_{13}{}^{2}}{\vec{r}_{01}{}^{2}\vec
{r}_{03}{}^{2}},\frac{\vec{r}_{23}{}^{2}}{\vec{r}_{02}{}^{2}\vec{r}_{03}{}%
^{2}}\right)
\]%
\[
+X\left(  \frac{\vec{r}_{02}{}^{2}\vec{r}_{13}{}^{2}}{\vec{r}_{03}{}^{2}%
\vec{r}_{12}{}^{2}},\frac{\vec{r}_{02}{}^{2}\vec{r}_{13}{}^{2}}{\vec{r}_{01}%
{}^{2}\vec{r}_{23}{}^{2}}\right)  \delta(\vec{r}_{20})+X\left(  \frac{\vec
{r}_{03}{}^{2}\vec{r}_{12}{}^{2}}{\vec{r}_{02}{}^{2}\vec{r}_{13}{}^{2}}%
,\frac{\vec{r}_{03}{}^{2}\vec{r}_{12}{}^{2}}{\vec{r}_{01}{}^{2}\vec{r}_{23}%
{}^{2}}\right)  \delta(\vec{r}_{30})
\]%
\begin{equation}
+Y\left(  \frac{\vec{r}_{01}{}^{2}\vec{r}_{23}{}^{2}}{\vec{r}_{03}{}^{2}%
\vec{r}_{12}{}^{2}},\frac{\vec{r}_{01}{}^{2}\vec{r}_{23}{}^{2}}{\vec{r}_{02}%
{}^{2}\vec{r}_{13}{}^{2}}\right)  \delta\left(  \vec{r}_{10}\right)  .
\end{equation}
Here%
\begin{equation}
\tilde{S}_{123}=\left(  \frac{\vec{r}_{12}{}^{4}{}}{\vec{r}_{01}{}^{4}{}%
\vec{r}_{02}{}^{4}{}}+\frac{\vec{r}_{13}{}^{4}{}}{\vec{r}_{01}{}^{4}{}\vec
{r}_{03}{}^{4}{}}+\frac{\vec{r}_{23}{}^{4}{}}{\vec{r}_{02}{}^{4}{}\vec{r}%
_{03}{}^{4}{}}-\frac{2\vec{r}_{13}{}^{2}\vec{r}_{12}{}^{2}}{\vec{r}_{01}{}%
^{4}{}\vec{r}_{02}{}^{2}\vec{r}_{03}{}^{2}}-\frac{2\vec{r}_{23}{}^{2}\vec
{r}_{12}{}^{2}}{\vec{r}_{01}{}^{2}\vec{r}_{02}{}^{4}{}\vec{r}_{03}{}^{2}%
}-\frac{2\vec{r}_{13}{}^{2}\vec{r}_{23}{}^{2}}{\vec{r}_{01}{}^{2}\vec{r}%
_{02}{}^{2}\vec{r}_{03}{}^{4}{}}\right)  \label{Sinv}%
\end{equation}
and we added the delta-functional contributions, which may be lost via
inversion. Thanks to conformal invariance of the integral such contributions
may depend only on conformally invariant ratios. We can find the values of the
unknown functions $X$ and $Y$ at $\vec{r}_{2}=\vec{r}_{3},\vec{r}_{2}=\vec
{r}_{1},\vec{r}_{1}=\vec{r}_{3}$. Using (\ref{int_L12C}) we have%
\[
\int\frac{d\vec{r}_{4}}{\pi}F_{100}+(2\leftrightarrow3)|_{\vec{r}_{2}=\vec
{r}_{3}}=16\int\frac{d\vec{r}_{4}}{\pi}\tilde{L}_{12}^{C}=24\pi\zeta
(3)[\delta(\vec{r}_{10})-\delta(\vec{r}_{20})]
\]%
\begin{equation}
=2X\left(  1,\infty\right)  \delta(\vec{r}_{20})+Y\left(  0,0\right)
\delta\left(  \vec{r}_{10}\right)  .
\end{equation}
Therefore
\begin{equation}
X\left(  1,\infty\right)  =-12\pi\zeta\left(  3\right)  ,\quad Y\left(
0,0\right)  =24\pi\zeta(3).
\end{equation}%
\[
\int\frac{d\vec{r}_{4}}{\pi}F_{100}+(2\leftrightarrow3)|_{\vec{r}_{1}=\vec
{r}_{3}}=-4\int\frac{d\vec{r}_{4}}{\pi}\tilde{L}_{12}^{C}=-6\pi\zeta
(3)[\delta(\vec{r}_{10})-\delta(\vec{r}_{20})]
\]%
\begin{equation}
=X\left(  0,0\right)  \delta(\vec{r}_{20})+(Y\left(  1,\infty\right)
+X\left(  \infty,1\right)  )\delta\left(  \vec{r}_{10}\right)  .
\end{equation}
Here again we used (\ref{int_L12C}). Therefore
\begin{equation}
X\left(  0,0\right)  =6\pi\zeta\left(  3\right)  ,\quad Y\left(
1,\infty\right)  +X\left(  \infty,1\right)  =-6\pi\zeta(3).
\end{equation}
If $\vec{r}_{2}\neq\vec{r}_{3},\vec{r}_{2}\neq\vec{r}_{1},\vec{r}_{1}\neq
\vec{r}_{3}$ then the arguments of $X$ and $Y$ are fixed by the integration
w.r.t. $\vec{r}_{0}$%
\begin{equation}
X\left(  \frac{\vec{r}_{02}{}^{2}\vec{r}_{13}{}^{2}}{\vec{r}_{03}{}^{2}\vec
{r}_{12}{}^{2}},\frac{\vec{r}_{02}{}^{2}\vec{r}_{13}{}^{2}}{\vec{r}_{01}{}%
^{2}\vec{r}_{23}{}^{2}}\right)  \delta(\vec{r}_{20})=X\left(  0,0\right)
\delta(\vec{r}_{20})=6\pi\zeta\left(  3\right)  \delta(\vec{r}_{20}),
\end{equation}%
\begin{equation}
Y\left(  \frac{\vec{r}_{01}{}^{2}\vec{r}_{23}{}^{2}}{\vec{r}_{03}{}^{2}\vec
{r}_{12}{}^{2}},\frac{\vec{r}_{01}{}^{2}\vec{r}_{23}{}^{2}}{\vec{r}_{02}{}%
^{2}\vec{r}_{13}{}^{2}}\right)  \delta\left(  \vec{r}_{10}\right)  =Y\left(
0,0\right)  \delta\left(  \vec{r}_{10}\right)  =24\pi\zeta(3)\delta\left(
\vec{r}_{10}\right)  .
\end{equation}
As a result, one can write%
\[
\int\frac{d\vec{r}_{4}}{\pi}F_{100}+(2\leftrightarrow3)=\left(  \frac{3\vec
{r}_{23}{}^{2}}{4\vec{r}_{02}{}^{2}\vec{r}_{03}{}^{2}}-\frac{\vec{r}_{12}%
{}^{2}}{\vec{r}_{01}{}^{2}\vec{r}_{02}{}^{2}}-\frac{\vec{r}_{13}{}^{2}}%
{\vec{r}_{01}{}^{2}\vec{r}_{03}{}^{2}}\right)  \ln^{2}\left(  \frac{\vec
{r}_{03}{}^{2}\vec{r}_{12}{}^{2}}{\vec{r}_{02}{}^{2}\vec{r}_{13}{}^{2}%
}\right)
\]%
\[
+\left(  \frac{3\vec{r}_{12}{}^{2}}{4\vec{r}_{01}{}^{2}\vec{r}_{02}{}^{2}%
}-\frac{3\vec{r}_{13}{}^{2}}{4\vec{r}_{01}{}^{2}\vec{r}_{03}{}^{2}}-\frac
{\vec{r}_{23}{}^{2}}{2\vec{r}_{02}{}^{2}\vec{r}_{03}{}^{2}}\right)  \ln
^{2}\left(  \frac{\vec{r}_{02}{}^{2}\vec{r}_{13}{}^{2}}{\vec{r}_{01}{}^{2}%
\vec{r}_{23}{}^{2}}\right)
\]%
\[
+\left(  \frac{3\vec{r}_{13}{}^{2}}{4\vec{r}_{01}{}^{2}\vec{r}_{03}{}^{2}%
}-\frac{3\vec{r}_{12}{}^{2}}{4\vec{r}_{01}{}^{2}\vec{r}_{02}{}^{2}}-\frac
{\vec{r}_{23}{}^{2}}{2\vec{r}_{02}{}^{2}\vec{r}_{03}{}^{2}}\right)  \ln
^{2}\left(  \frac{\vec{r}_{03}{}^{2}\vec{r}_{12}{}^{2}}{\vec{r}_{01}{}^{2}%
\vec{r}_{23}{}^{2}}\right)
\]%
\[
+\frac{3}{2}\tilde{S}_{123}I\left(  \frac{\vec{r}_{12}{}^{2}}{\vec{r}_{01}%
{}^{2}\vec{r}_{02}{}^{2}},\frac{\vec{r}_{13}{}^{2}}{\vec{r}_{01}{}^{2}\vec
{r}_{03}{}^{2}},\frac{\vec{r}_{23}{}^{2}}{\vec{r}_{02}{}^{2}\vec{r}_{03}{}%
^{2}}\right)
\]%
\[
+6\pi\zeta\left(  3\right)  \left(  \delta(\vec{r}_{20})+\delta(\vec{r}%
_{30})\right)  +24\pi\zeta(3)\delta\left(  \vec{r}_{10}\right)
\]%
\begin{equation}
-36\pi\zeta\left(  3\right)  \delta_{23}\delta(\vec{r}_{20})-36\pi
\zeta(3)(\delta_{13}+\delta_{12})\delta\left(  \vec{r}_{10}\right)
+72\pi\zeta(3)\delta_{13}\delta_{12}\delta\left(  \vec{r}_{10}\right)  .
\label{F100integrated}%
\end{equation}
Here $\delta_{ij}=1,$ if $\vec{r}_{i}=\vec{r}_{j}$ and $\delta_{ij}=0$
otherwise.\ The last term is added since the total contribution at $\vec
{r}_{1}=\vec{r}_{2}=\vec{r}_{3}$ is 0.

Now we will integrate $F_{230}$ (\ref{F230}) w.r.t. $\vec{r}_{4}.$ Again we
set $\vec{r}_{0}=0$, do inversion, and calculate the integral in the
$d$-dimensional space using the integrals from appendix A in
\cite{Fadin:2009za} and (\ref{int1}). We get%
\[
\int\frac{d\vec{r}_{4}}{\pi}F_{230}+(2\leftrightarrow3)\rightarrow\int%
\frac{d^{d}r_{4}}{\pi}\left(  \frac{r_{34}{}^{2}r_{12}{}^{2}}{2r_{14}{}%
^{2}r_{24}{}^{2}}\ln\left(  \frac{r_{14}{}^{2}r_{34}{}^{6}{}}{{}r_{24}{}%
^{2}r_{12}{}^{6}}\right)  +\frac{r_{12}{}^{2}}{2r_{14}{}^{2}}\ln\left(
\frac{r_{12}{}^{2}r_{34}{}^{8}{}}{r_{14}{}^{4}{}r_{24}{}^{6}{}}\right)
\right.
\]%
\[
-\frac{r_{12}{}^{2}}{2r_{24}{}^{2}}\ln\left(  \frac{r_{14}{}^{2}}{r_{12}{}%
^{2}}\right)  -\frac{r_{23}{}^{2}r_{12}{}^{2}}{r_{14}{}^{2}r_{24}{}^{2}}%
\ln\left(  \frac{r_{14}{}^{2}r_{23}{}^{2}}{r_{12}{}^{2}r_{24}{}^{2}}\right)
-\frac{r_{13}{}^{2}r_{12}{}^{2}}{2r_{14}{}^{2}r_{24}{}^{2}}\ln\left(
\frac{r_{13}{}^{2}r_{24}{}^{2}}{r_{12}{}^{2}r_{14}{}^{2}}\right)
\]%
\begin{equation}
\left.  -\frac{r_{23}{}^{2}}{r_{14}{}^{2}}\ln\left(  \frac{r_{24}{}^{2}%
r_{34}{}^{2}}{r_{14}{}^{2}r_{23}{}^{2}}\right)  +\frac{r_{13}{}^{2}}{2r_{24}%
{}^{2}}\ln\left(  \frac{r_{13}{}^{2}r_{24}{}^{2}}{r_{14}{}^{4}{}}\right)
-\frac{r_{23}{}^{2}}{2r_{24}{}^{2}}\ln\left(  \frac{r_{23}{}^{4}{}r_{24}{}%
^{2}}{r_{14}{}^{4}{}r_{34}{}^{2}}\right)  \right)  +(2\leftrightarrow3)
\end{equation}%
\[
\overset{d\rightarrow2}{\rightarrow}\left(  \frac{3\left(  r_{12}{}^{2}%
-r_{13}{}^{2}\right)  }{4}+r_{23}{}^{2}\right)  \ln^{2}\left(  \frac{r_{12}%
{}^{2}}{r_{23}{}^{2}}\right)  +\left(  \frac{3\left(  r_{13}{}^{2}-r_{12}%
{}^{2}\right)  }{4}+r_{23}{}^{2}\right)  \ln^{2}\left(  \frac{r_{13}{}^{2}%
}{r_{23}{}^{2}}\right)
\]%
\begin{equation}
+\left(  \frac{r_{12}{}^{2}+r_{13}{}^{2}}{2}-\frac{3}{4}r_{23}{}^{2}\right)
\ln^{2}\left(  \frac{r_{12}{}^{2}}{r_{13}{}^{2}}\right)  -\frac{3}{2}%
S_{123}I\left(  r_{12}{}^{2},r_{13}{}^{2},r_{23}{}^{2}\right)  .
\end{equation}
Again, inverting and restoring $r_{0}$ we have%
\[
\int\frac{d\vec{r}_{4}}{\pi}F_{230}+(2\leftrightarrow3)=\left(  \frac{\vec
{r}_{12}{}^{2}}{2\vec{r}_{01}{}^{2}\vec{r}_{02}{}^{2}}+\frac{\vec{r}_{13}%
{}^{2}}{2\vec{r}_{01}{}^{2}\vec{r}_{03}{}^{2}}-\frac{3\vec{r}_{23}{}^{2}%
}{4\vec{r}_{02}{}^{2}\vec{r}_{03}{}^{2}}\right)  \ln^{2}\left(  \frac{\vec
{r}_{03}{}^{2}\vec{r}_{12}{}^{2}}{\vec{r}_{02}{}^{2}\vec{r}_{13}{}^{2}%
}\right)
\]%
\[
+\left(  \frac{3\vec{r}_{13}{}^{2}}{4\vec{r}_{01}{}^{2}\vec{r}_{03}{}^{2}%
}-\frac{3\vec{r}_{12}{}^{2}}{4\vec{r}_{01}{}^{2}\vec{r}_{02}{}^{2}}+\frac
{\vec{r}_{23}{}^{2}}{\vec{r}_{02}{}^{2}\vec{r}_{03}{}^{2}}\right)  \ln
^{2}\left(  \frac{\vec{r}_{02}{}^{2}\vec{r}_{13}{}^{2}}{\vec{r}_{01}{}^{2}%
\vec{r}_{23}{}^{2}}\right)
\]%
\[
+\left(  \frac{3\vec{r}_{12}{}^{2}}{4\vec{r}_{01}{}^{2}\vec{r}_{02}{}^{2}%
}-\frac{3\vec{r}_{13}{}^{2}}{4\vec{r}_{01}{}^{2}\vec{r}_{03}{}^{2}}+\frac
{\vec{r}_{23}{}^{2}}{\vec{r}_{02}{}^{2}\vec{r}_{03}{}^{2}}\right)  \ln
^{2}\left(  \frac{\vec{r}_{03}{}^{2}\vec{r}_{12}{}^{2}}{\vec{r}_{01}{}^{2}%
\vec{r}_{23}{}^{2}}\right)
\]%
\[
-\frac{3}{2}\tilde{S}_{123}I\left(  \frac{\vec{r}_{12}{}^{2}}{\vec{r}_{01}%
{}^{2}\vec{r}_{02}{}^{2}},\frac{\vec{r}_{13}{}^{2}}{\vec{r}_{01}{}^{2}\vec
{r}_{03}{}^{2}},\frac{\vec{r}_{23}{}^{2}}{\vec{r}_{02}{}^{2}\vec{r}_{03}{}%
^{2}}\right)
\]%
\[
+\tilde{X}\left(  \frac{\vec{r}_{02}{}^{2}\vec{r}_{13}{}^{2}}{\vec{r}_{03}%
{}^{2}\vec{r}_{12}{}^{2}},\frac{\vec{r}_{02}{}^{2}\vec{r}_{13}{}^{2}}{\vec
{r}_{01}{}^{2}\vec{r}_{23}{}^{2}}\right)  \delta(\vec{r}_{20})+\tilde
{X}\left(  \frac{\vec{r}_{03}{}^{2}\vec{r}_{12}{}^{2}}{\vec{r}_{02}{}^{2}%
\vec{r}_{13}{}^{2}},\frac{\vec{r}_{03}{}^{2}\vec{r}_{12}{}^{2}}{\vec{r}_{01}%
{}^{2}\vec{r}_{23}{}^{2}}\right)  \delta(\vec{r}_{30})
\]%
\begin{equation}
+\tilde{Y}\left(  \frac{\vec{r}_{01}{}^{2}\vec{r}_{23}{}^{2}}{\vec{r}_{03}%
{}^{2}\vec{r}_{12}{}^{2}},\frac{\vec{r}_{01}{}^{2}\vec{r}_{23}{}^{2}}{\vec
{r}_{02}{}^{2}\vec{r}_{13}{}^{2}}\right)  \delta\left(  \vec{r}_{10}\right)  .
\end{equation}
Again, we can find the values of $\tilde{X}$ and $\tilde{Y}$ putting $\vec
{r}_{2}=\vec{r}_{3},\vec{r}_{2}=\vec{r}_{1},\vec{r}_{1}=\vec{r}_{3}$ in this
equation. Indeed via (\ref{int_L12C}) we have,%
\[
\int\frac{d\vec{r}_{4}}{\pi}F_{230}+(2\leftrightarrow3)|_{\vec{r}_{2}=\vec
{r}_{3}}=-8\int\frac{d\vec{r}_{4}}{\pi}\tilde{L}_{12}^{C}=-12\pi
\zeta(3)[\delta(\vec{r}_{10})-\delta(\vec{r}_{20})]
\]%
\begin{equation}
=2\tilde{X}\left(  1,\infty\right)  \delta(\vec{r}_{20})+\tilde{Y}\left(
0,0\right)  \delta\left(  \vec{r}_{10}\right)  .
\end{equation}
Therefore
\begin{equation}
\tilde{X}\left(  1,\infty\right)  =6\pi\zeta\left(  3\right)  ,\quad\tilde
{Y}\left(  0,0\right)  =-12\pi\zeta(3).
\end{equation}
Using (\ref{int_L12C}) again, we get%
\[
\int\frac{d\vec{r}_{4}}{\pi}F_{230}+(2\leftrightarrow3)|_{\vec{r}_{1}=\vec
{r}_{3}}=8\int\frac{d\vec{r}_{4}}{\pi}\tilde{L}_{12}^{C}=12\pi\zeta
(3)[\delta(\vec{r}_{10})-\delta(\vec{r}_{20})]
\]%
\begin{equation}
=\tilde{X}\left(  0,0\right)  \delta(\vec{r}_{20})+\left(  \tilde{X}\left(
\infty,1\right)  +\tilde{Y}\left(  1,\infty\right)  \right)  \delta\left(
\vec{r}_{10}\right)  .
\end{equation}
Therefore
\begin{equation}
\tilde{X}\left(  0,0\right)  =-12\pi\zeta\left(  3\right)  ,\quad\tilde
{Y}\left(  1,\infty\right)  +\tilde{X}\left(  \infty,1\right)  =12\pi\zeta(3).
\end{equation}
If $\vec{r}_{2}\neq\vec{r}_{3},\vec{r}_{2}\neq\vec{r}_{1},\vec{r}_{1}\neq
\vec{r}_{3}$ then the arguments of $\tilde{X}$ and $\tilde{Y}$ are fixed by
the integration w.r.t. $\vec{r}_{0}$%
\begin{equation}
\tilde{X}\left(  \frac{\vec{r}_{02}{}^{2}\vec{r}_{13}{}^{2}}{\vec{r}_{03}%
{}^{2}\vec{r}_{12}{}^{2}},\frac{\vec{r}_{02}{}^{2}\vec{r}_{13}{}^{2}}{\vec
{r}_{01}{}^{2}\vec{r}_{23}{}^{2}}\right)  \delta(\vec{r}_{20})=\tilde
{X}\left(  0,0\right)  \delta(\vec{r}_{20})=-12\pi\zeta\left(  3\right)
\delta(\vec{r}_{20}),
\end{equation}%
\begin{equation}
\tilde{Y}\left(  \frac{\vec{r}_{01}{}^{2}\vec{r}_{23}{}^{2}}{\vec{r}_{03}%
{}^{2}\vec{r}_{12}{}^{2}},\frac{\vec{r}_{01}{}^{2}\vec{r}_{23}{}^{2}}{\vec
{r}_{02}{}^{2}\vec{r}_{13}{}^{2}}\right)  \delta\left(  \vec{r}_{10}\right)
=\tilde{Y}\left(  0,0\right)  \delta\left(  \vec{r}_{10}\right)  =-12\pi
\zeta(3)\delta\left(  \vec{r}_{10}\right)  .
\end{equation}
Finally,%
\[
\int\frac{d\vec{r}_{4}}{\pi}F_{230}+(2\leftrightarrow3)=\left(  \frac{\vec
{r}_{12}{}^{2}}{2\vec{r}_{01}{}^{2}\vec{r}_{02}{}^{2}}+\frac{\vec{r}_{13}%
{}^{2}}{2\vec{r}_{01}{}^{2}\vec{r}_{03}{}^{2}}-\frac{3\vec{r}_{23}{}^{2}%
}{4\vec{r}_{02}{}^{2}\vec{r}_{03}{}^{2}}\right)  \ln^{2}\left(  \frac{\vec
{r}_{03}{}^{2}\vec{r}_{12}{}^{2}}{\vec{r}_{02}{}^{2}\vec{r}_{13}{}^{2}%
}\right)
\]%
\[
+\left(  \frac{3\vec{r}_{13}{}^{2}}{4\vec{r}_{01}{}^{2}\vec{r}_{03}{}^{2}%
}-\frac{3\vec{r}_{12}{}^{2}}{4\vec{r}_{01}{}^{2}\vec{r}_{02}{}^{2}}+\frac
{\vec{r}_{23}{}^{2}}{\vec{r}_{02}{}^{2}\vec{r}_{03}{}^{2}}\right)  \ln
^{2}\left(  \frac{\vec{r}_{02}{}^{2}\vec{r}_{13}{}^{2}}{\vec{r}_{01}{}^{2}%
\vec{r}_{23}{}^{2}}\right)
\]%
\[
+\left(  \frac{3\vec{r}_{12}{}^{2}}{4\vec{r}_{01}{}^{2}\vec{r}_{02}{}^{2}%
}-\frac{3\vec{r}_{13}{}^{2}}{4\vec{r}_{01}{}^{2}\vec{r}_{03}{}^{2}}+\frac
{\vec{r}_{23}{}^{2}}{\vec{r}_{02}{}^{2}\vec{r}_{03}{}^{2}}\right)  \ln
^{2}\left(  \frac{\vec{r}_{03}{}^{2}\vec{r}_{12}{}^{2}}{\vec{r}_{01}{}^{2}%
\vec{r}_{23}{}^{2}}\right)
\]%
\[
-\frac{3}{2}\tilde{S}_{123}I\left(  \frac{\vec{r}_{12}{}^{2}}{\vec{r}_{01}%
{}^{2}\vec{r}_{02}{}^{2}},\frac{\vec{r}_{13}{}^{2}}{\vec{r}_{01}{}^{2}\vec
{r}_{03}{}^{2}},\frac{\vec{r}_{23}{}^{2}}{\vec{r}_{02}{}^{2}\vec{r}_{03}{}%
^{2}}\right)
\]%
\[
-12\pi\zeta\left(  3\right)  (\delta(\vec{r}_{20})+\delta(\vec{r}_{30}%
)+\delta\left(  \vec{r}_{10}\right)  )
\]%
\begin{equation}
+36\pi\zeta\left(  3\right)  \delta_{23}\delta(\vec{r}_{20})+36\pi
\zeta(3)(\delta_{13}+\delta_{12})\delta\left(  \vec{r}_{10}\right)
-72\pi\zeta(3)\delta_{13}\delta_{12}\delta\left(  \vec{r}_{10}\right)  .
\label{F230integrated}%
\end{equation}
Now we will integrate (\ref{F140}) and prove equality (\ref{constraint3-1}).
Again we set $\vec{r}_{0}=0$, do inversion, and calculate the integral in the
$d$-dimensional space using the integrals from appendix A in
\cite{Fadin:2009za} and (\ref{int1}). We get%
\[
\int\frac{d\vec{r}_{4}}{\pi}~\left(  \left\{  F_{140}+(0\leftrightarrow
4)\right\}  +(2\leftrightarrow3)\right)  \rightarrow\int\frac{d^{d}r_{4}}{\pi
}\left(  \frac{r_{12}{}^{2}}{r_{14}{}^{2}}\ln\left(  \frac{r_{12}{}^{2}%
r_{34}{}^{4}{}}{r_{14}{}^{2}r_{24}{}^{4}{}}\right)  \right.
\]%
\[
+\frac{r_{12}{}^{2}}{r_{24}{}^{2}}\ln\left(  \frac{r_{12}{}^{2}r_{24}{}^{2}%
}{r_{34}{}^{4}{}}\right)  -\frac{r_{23}{}^{2}r_{12}{}^{2}}{r_{14}{}^{2}%
r_{24}{}^{2}}\ln\left(  \frac{r_{14}{}^{2}r_{23}{}^{2}}{r_{12}{}^{2}r_{24}%
{}^{2}}\right)  -\frac{r_{23}{}^{2}r_{12}{}^{2}}{r_{24}{}^{2}r_{34}{}^{2}}%
\ln\left(  \frac{r_{23}{}^{2}r_{24}{}^{2}}{r_{12}{}^{2}r_{34}{}^{2}}\right)
\]%
\[
-\frac{r_{23}{}^{2}}{r_{14}{}^{2}}\ln\left(  \frac{r_{24}{}^{2}r_{34}{}^{2}%
}{r_{14}{}^{2}r_{23}{}^{2}}\right)  +\frac{r_{13}{}^{2}}{r_{24}{}^{2}}%
\ln\left(  \frac{r_{34}{}^{4}{}}{r_{13}{}^{2}r_{24}{}^{2}}\right)
+\frac{r_{23}{}^{2}}{r_{24}{}^{2}}\ln\left(  \frac{r_{34}{}^{2}}{r_{23}{}^{2}%
}\right)
\]%
\begin{equation}
\left.  +\frac{r_{23}{}^{2}}{r_{34}{}^{2}}\ln\left(  \frac{r_{24}{}^{2}%
}{r_{23}{}^{2}}\right)  +\frac{r_{13}{}^{2}r_{24}{}^{2}}{r_{14}{}^{2}r_{34}%
{}^{2}}\ln\left(  \frac{r_{14}{}^{2}r_{24}{}^{2}}{r_{13}{}^{2}r_{34}{}^{2}%
}\right)  -\frac{r_{14}{}^{2}r_{23}{}^{2}}{r_{24}{}^{2}r_{34}{}^{2}}\ln\left(
\frac{r_{14}{}^{2}}{r_{23}{}^{2}}\right)  \right)  +(2\leftrightarrow3)
\end{equation}%
\begin{equation}
\overset{d\rightarrow2}{\rightarrow}-\frac{r_{13}{}^{2}+r_{12}{}^{2}}{2}%
\ln^{2}\left(  \frac{r_{12}{}^{2}}{r_{13}{}^{2}}\right)  +\frac{1}{2}r_{23}%
{}^{2}\left(  \ln^{2}\left(  \frac{r_{12}{}^{2}}{r_{23}{}^{2}}\right)
+\ln^{2}\left(  \frac{r_{13}{}^{2}}{r_{23}{}^{2}}\right)  \right)  .
\end{equation}
Inverting and restoring $\vec{r}_{0},$ we get
\[
\int\frac{d\vec{r}_{4}}{\pi}~\left(  \left\{  F_{140}+(0\leftrightarrow
4)\right\}  +(2\leftrightarrow3)\right)  =-\frac{1}{2}\left(  \frac{\vec
{r}_{12}{}^{2}}{\vec{r}_{01}{}^{2}\vec{r}_{02}{}^{2}}+\frac{\vec{r}_{13}{}%
^{2}}{\vec{r}_{01}{}^{2}\vec{r}_{03}{}^{2}}\right)  \ln^{2}\left(  \frac
{\vec{r}_{03}{}^{2}\vec{r}_{12}{}^{2}}{\vec{r}_{02}{}^{2}\vec{r}_{13}{}^{2}%
}\right)
\]%
\begin{equation}
+\frac{\vec{r}_{23}{}^{2}}{2\vec{r}_{02}{}^{2}\vec{r}_{03}{}^{2}}\ln
^{2}\left(  \frac{\vec{r}_{03}{}^{2}\vec{r}_{12}{}^{2}}{\vec{r}_{01}{}^{2}%
\vec{r}_{23}{}^{2}}\right)  +\frac{\vec{r}_{23}{}^{2}}{2\vec{r}_{02}{}^{2}%
\vec{r}_{03}{}^{2}}\ln^{2}\left(  \frac{\vec{r}_{02}{}^{2}\vec{r}_{13}{}^{2}%
}{\vec{r}_{01}{}^{2}\vec{r}_{23}{}^{2}}\right)  .
\end{equation}
This integral has no delta functional contributions since it equals 0 at
$\vec{r}_{1}=\vec{r}_{2},\vec{r}_{1}=\vec{r}_{3},\vec{r}_{3}=\vec{r}_{2}.$

\section{Decomposition of C-odd quadrupole operator}

\label{quadrupole}Here we demonstrate that the C-odd part of the quadrupole
operator $tr(U_{1}U_{2}^{\dag}U_{3}U_{4}^{\dag})$\ in the 3-gluon
approximation in $SU(3)$ can be decomposed into a sum of 3QWLs. Indeed%
\[
2tr(U_{1}U_{2}^{\dag}U_{3}U_{4}^{\dag})=\left(  (U_{1}-U_{2})(U_{2}^{\dag
}-U_{3}^{\dag})(U_{3}-U_{4})\right)  \cdot U_{4}\cdot U_{4}%
\]%
\begin{equation}
-B_{133}+B_{233}+B_{144}-B_{244}+B_{344}+B_{122}-6
\end{equation}%
\begin{equation}
\overset{\mathrm{3g}}{=}-(U_{1}-U_{2})(U_{2}-U_{3})(U_{3}-U_{4})\cdot E\cdot
E-B_{133}+B_{233}+B_{144}-B_{244}+B_{344}+B_{122}-6.
\end{equation}
Therefore%
\[
2tr(U_{1}U_{2}^{\dag}U_{3}U_{4}^{\dag})-2tr(U_{4}U_{3}^{\dag}U_{2}U_{1}^{\dag
})\overset{\mathrm{3g}}{=}-B_{133}^{-}+B_{233}^{-}+B_{144}^{-}-B_{244}%
^{-}+B_{344}^{-}+B_{122}^{-}%
\]%
\[
-((U_{1}-U_{2})(U_{2}-U_{3})(U_{3}-U_{4})+(U_{3}-U_{4})(U_{2}-U_{3}%
)(U_{1}-U_{2}))\cdot E\cdot E
\]%
\[
\overset{\mathrm{3g}}{=}-B_{133}^{-}+B_{233}^{-}+B_{144}^{-}-B_{244}%
^{-}+B_{344}^{-}+B_{122}^{-}-2(U_{1}-U_{2})\cdot(U_{2}-U_{3})\cdot(U_{3}%
-U_{4})
\]%
\[
\overset{\mathrm{3g}}{=}-B_{133}^{-}+B_{233}^{-}+B_{144}^{-}-B_{244}%
^{-}+B_{344}^{-}+B_{122}^{-}%
\]%
\[
-(U_{1}-U_{2})\cdot(U_{2}-U_{3})\cdot(U_{3}-U_{4})+(U_{1}^{\dag}-U_{2}^{\dag
})\cdot(U_{2}^{\dag}-U_{3}^{\dag})\cdot(U_{3}^{\dag}-U_{4}^{\dag})
\]%
\begin{equation}
=B_{144}^{-}+B_{322}^{-}-B_{433}^{-}-B_{211}^{-}+B_{124}^{-}+B_{234}%
^{-}-B_{123}^{-}-B_{134}^{-}.
\end{equation}

\end{document}